\documentclass[letterpaper,11pt,leqno]{article}
\usepackage{paper}
\usepackage{threeparttable}  
\usepackage{booktabs}        

\usepackage{booktabs}
\usepackage{threeparttable}
\usepackage{adjustbox}
\usepackage{pdflscape}
\usepackage{longtable}
\usepackage{makecell}
\usepackage{graphicx}
\usepackage{float}
\hypersetup{pdftitle={Minimalist LaTeX Template for Academic Papers}}

\newcommand{\bib}{paper.bib}

\begin{document}

\title{Recession Detection in Japan using Labor Market Data}

\author{Neha Sikand and Rongjin Zhang
%
\thanks{Department of Economics, University of California, Santa Cruz. \\We are deeply indebted to Pascal Michaillat for his invaluable advice, continuous guidance, encouragement, and support. We are extremely grateful to Kenneth Kletzer and Alonso Villacorta who contributed valuable comments and suggestions at different stages of this project. The paper has also benefited from suggestions made at the macroeconomics workshop at UCSC. We are also very grateful to Yuta Takahashi for all his help and guidance on locating Japanese data.}}

\date{May 2026}   


\begin{titlepage}
\maketitle

Recession indicators are often viewed as U.S.-specific, raising the question of whether labor market–based rules such as the Sahm Rule and the Michez Rule can reliably detect recessions in other countries. To answer this, we evaluate whether such rules can be adapted to Japan by calibrating thresholds and smoothing parameters to Japanese labor market data. We construct a large set of 95,832 recession indicators combining unemployment and vacancy data. The selected classifiers are statistically perfect as they identify all 11 historical recessions in the 1970–2021 training period without generating any false positives. Among these, 193 classifiers lie on the anticipation--precision frontier. Restricting attention to the high-precision segment yields six classifiers with a standard deviation of detection errors below 3 months. The selected classifier ensemble signals recessions, on average, 0.06 months after their true onset. Overall, these findings suggest that slack-based labor market rules provide a general framework for improving real time recession detection across countries.

\end{titlepage}

\section{Introduction}\label{s:introduction}


As emphasized by \cite{hamilton2011calling}, policymakers, firms, and households are often more concerned with identifying when the economy is turning than with measuring the precise magnitude of GDP growth. Economic turning points mark regime changes in the overall state of the economy, signaling transitions between expansion and recession. These transitions are associated with broad declines in output, weakening labor market conditions, reduced demand faced by firms, and rising economic uncertainty. 
Because of this importance, the study of recession detection has been extensively developed in the United States. 

This paper aims to extend these labor-market based recession detection methods to a non-U.S. context by asking the question: \textit{can labor-market based recession detection rules be effectively applied to Japan?} While many recession detection methods are developed and evaluated primarily in the U.S. context, it remains unclear whether such methods generalize well to economies with different labor market structures, institutional settings, and business cycle dynamics. Japan provides an especially informative setting because it combines a large advanced economy with detailed and consistently measured labor market data spanning several decades. At the same time, Japan exhibits distinct labor market characteristics and unusually long delays in official recession dating by the Economic and Social Research Institute (ESRI), making timely recession detection particularly informative.

The timely identification of recessions remains a central challenge in Japan. Using the business cycle chronology from the ESRI, we compare recession start dates (business cycle peaks) to their official announcement dates and document large and systematic delays. On average, recessions are announced 26.5 months after their onset, with a standard deviation of about 10 months (Table~\ref{tab:ESRI_Recession_Start_Announcments}). As a result, policymakers and markets often operate for extended periods without reliable confirmation that the economy has entered a recession, limiting the scope for timely policy responses. This issue is particularly important in Japan, where three decades of secular stagnation and relatively muted fluctuations in unemployment make recession detection both more difficult and more consequential. At the same time, Japan remains exposed to recession risk. In 2025, Japan narrowly avoided a technical recession, commonly defined as two consecutive quarters of negative GDP growth, highlighting both the difficulty of real time assessment and the economy’s continuing vulnerability to downturns.

To illustrate the economic meaning of the Japanese recession dates, Figure 1 plots HP-filtered real GDP together with the official recession dates announced by the ESRI. The figure shows that the recession dates generally coincide with sharp declines in detrended economic activity. This is important because some audiences may be unfamiliar with the Japanese recession dating system or uncertain about what these recession dates are intended to measure. The HP-filtered GDP series demonstrates that the official recession dates correspond to meaningful turning points in aggregate economic activity rather than arbitrary classifications. In other words, the recession dates capture periods in which the Japanese economy experiences broad and significant slowdowns relative to trend.

A natural benchmark for real time recession detection is the Sahm Rule, which signals a recession when the three-month moving average of the unemployment rate rises sufficiently above its recent minimum. While the rule has performed well in the United States, it relies solely on unemployment as an indicator of labor market conditions. This may present limitations in Japan, where unemployment fluctuations are substantially more muted than in more volatile economies such as the United States. As shown in Figures 2 and 3, U.S. recessions are characterized by sharp increases in unemployment and declines in vacancies, whereas Japanese recessions exhibit much smaller movements in unemployment. In contrast, vacancy rates in Japan display stronger cyclical variation and may better capture fluctuations in labor demand that are not fully reflected in unemployment data.

Recent work by \cite{michaillatsaez2025} proposes incorporating both unemployment and vacancy rates into recession detection through a combined unemployment–vacancy indicator, referred to as the Michez Rule. By integrating information from both labor supply and labor demand conditions, the approach aims to improve the timeliness and accuracy of recession detection.

Building on \cite{michaillat2025}, this paper shows that labor-market recession rules can be successfully adapted to Japan. We utilize monthly Japanese unemployment data from FRED, monthly job vacancy rate data from Japan’s Ministry of Health, Labour and Welfare, and monthly official recession dates from the ESRI. We construct 95,832 recession indicators using combinations of unemployment and vacancy data with alternative smoothing methods, transformations, and thresholds. Among these, 193 classifiers are perfect classifiers that lie on the anticipation--precision frontier. They produce no false negatives (actual recessions missed by the classifiers) and no false positives (non-recessions mistakenly detected by the classifiers). The selected classifier ensemble signals recessions on average 0.06 months after their true onset with a standard deviation of detection errors of 2.72 months. The classifier ensemble correctly identifies all 11 Japanese recessions between 1970 and 2021 without any false positives. These findings are consistent with the evidence for the United States reported by \citet{michaillat2025}, where the classifier ensemble detected recession starts on average 1.2 months after onset between 1979 and 2021, compared to 6.3 months for the National Bureau of Economic Research (NBER) Business Cycle Dating Committee. Overall, the results suggest that slack-based labor-market indicators provide a robust framework for real time recession detection outside the United States.

\begin{figure}[b!]
    \centering
    \includegraphics[
        width=\linewidth,
    ]{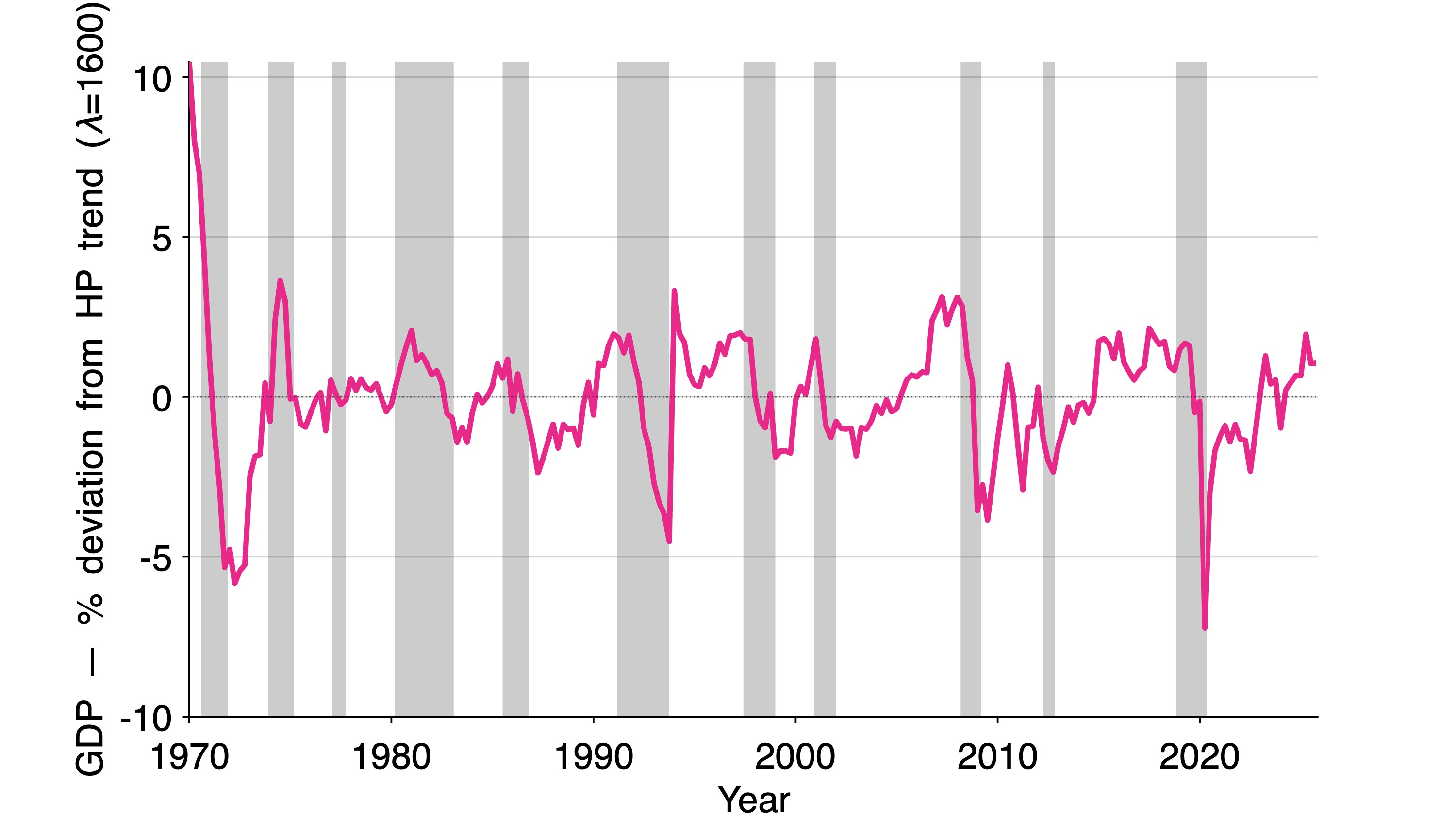}
    \caption{Japanese real GDP deviations from trend, 1970--2025}
    \label{fig:japan_gdp_recession}
    
    \vspace{0.3em}
    \begin{minipage}{0.95\linewidth}
    \footnotesize
    \textit{Notes:} The figure plots the percentage deviation of Japanese real GDP from its Hodrick--Prescott (HP) trend with smoothing parameter $\lambda = 1600$. Grey shaded regions indicate recessions dated by the Economic and Social Research Institute (ESRI). Peaks and troughs in GDP relative to trend closely align with ESRI business-cycle turning points, highlighting the cyclical fluctuations associated with Japanese recessions.
    \end{minipage}
\end{figure}

\begin{table}[H]
\centering
\begin{threeparttable}
\caption{ESRI Recession Start Announcements}
\label{tab:ESRI_Recession_Start_Announcments}

\begin{tabular}{cccc}
\toprule
\textbf{Cycle} & \textbf{Peak (Start Date)} & \textbf{Announcement Date} & \textbf{Delay (months)} \\
\midrule
11th & Feb 1991 & Oct 1993 & $\sim$32 \\
12th & May 1997 & Jul 2000 & $\sim$38 \\
13th & Nov 2000 & May 2002 & $\sim$18 \\
14th & Feb 2008 & Jul 2009 & $\sim$17 \\
15th & Mar 2012 & Aug 2013 & $\sim$17 \\
16th & Oct 2018 & Nov 9, 2021 & $\sim$37 \\
\midrule
\textbf{Mean} &  &  & \textbf{26.5} \\
\textbf{Std. Dev.} &  &  & \textbf{10.25} \\
\bottomrule
\end{tabular}

\begin{tablenotes}[flushleft]
\footnotesize
\item Notes: The table reports the delay between the official start dates of Japanese recessions, defined by ESRI business-cycle peaks, and the dates on which ESRI publicly announced these turning points. Delays are measured in months between the recession start date and the corresponding announcement date. The results highlight the substantial lag in official recession dating, motivating the need for timely recession detection methods.
\end{tablenotes}

\end{threeparttable}
\end{table}

\section{Literature Review}

This paper contributes to several strands of the recession-detection literature. First, it contributes to the literature on real time recession dating and business-cycle turning points. \cite{hamilton2011calling} studies the difficulty of identifying recession turning points in real time, emphasizing challenges such as data revisions, structural change, and noisy macroeconomic signals. The paper also highlights the trade-off between using simple and robust recession indicators versus incorporating large information sets to improve turning-point detection. This trade-off motivates our approach, which focuses on extracting recession signals from a small set of labor-market indicators rather than relying on large macroeconomic datasets.

A large part of the literature attempts to improve recession detection by incorporating increasingly large sets of macroeconomic indicators. These approaches typically rely on dynamic factor models or large collections of macroeconomic indicators. For example, \cite{stock2014estimating} use large information sets to estimate business-cycle turning points, while \cite{chen2011forecasting} and \cite{fornaro2016forecasting} similarly rely on many macroeconomic series to improve recession prediction. In contrast, another strand of the literature emphasizes that relatively simple recession rules can perform remarkably well in practice. \cite{piger2019turning} shows that narrow sets of carefully selected variables can provide strong classification performance, while \cite{sahm2019direct} proposes the Sahm Rule, a simple labor-market-based recession indicator using unemployment dynamics.

More recent work focuses on improving recession detection by combining multiple measures of labor-market slack and systematically searching across large model spaces to identify robust recession indicators. \cite{michaillatsaez2025} introduce the Michez Rule, which combines unemployment and vacancy data to reduce noise in recession signals and improve real time recession detection in U.S. data. Relatedly, \cite{michaillat2025} shows that recession detection can be improved by systematically searching across alternative smoothing methods, transformations, and thresholds to construct optimized recession classifiers from large model spaces.

This paper contributes to this literature by examining whether labor-market-based recession rules generalize outside the United States. Japan provides a useful setting because unemployment fluctuations are relatively muted while vacancy rates display stronger cyclical movements around business-cycle turning points. In addition, Japanese vacancy data may provide a more representative measure of labor-market conditions than vacancy data available for the United States, making Japan a particularly informative environment for evaluating vacancy-based recession indicators. We therefore assess whether vacancy-augmented recession indicators retain predictive power in a labor market where unemployment alone may provide weaker recession signals. More broadly, the paper studies whether labor-market-based recession rules provide a general framework for real time turning-point detection across countries.

\section{Data}

This section presents the data on Japanese recessions, unemployment, and job vacancies used in the paper. The data coverage is from January 1970 to December 2025.

\subsection{Japan Time Series Data}
Our analysis uses monthly unemployment, vacancy, employment, and business-cycle dating data for Japan spanning 1970 to 2025. The unemployment data are obtained from the Labour Force Survey conducted by the Statistics Bureau of Japan, while vacancy and employment data are taken from Labour Market Indicators (General, including Part-time). Business-cycle reference dates and GDP data are provided by the Economic and Social Research Institute (ESRI). Using the vacancy and employment data, we construct a monthly vacancy rate series.

\subsection{Business Cycle Dating}
Table~\ref{tab:japan_recession_start_1970} reports the recession start dates based on the chronology provided by the Economic and Social Research Institute. ESRI recession dates are widely treated as the official business-cycle chronology in Japan and are routinely referenced in financial press coverage discussing recession risk (\cite{esri_about,japantimes_recession_2025}). Over our sample period, this chronology identifies 11 recession episodes. These dates serve as the ground truth for evaluating our recession indicators. Our objective is to assess how early and how accurately these indicators detect recession onsets in real time.

To fix ideas, ESRI’s role in Japan is broadly analogous to that of the \cite{nber_business_cycles_2023} (NBER) in the United States. Similar to the NBER, ESRI identifies business-cycle turning points using a broad range of economic indicators rather than relying on a single variable. ESRI constructs a monthly coincident composite index designed to track overall economic activity using indicators related to production, employment, consumption, investment, retail sales, exports, and other measures of economic conditions. ESRI then determines business-cycle peaks and troughs by examining sustained movements in this composite index together with broader economic conditions (\cite{esri_about}).

Although ESRI’s approach is conceptually similar to the NBER’s business-cycle dating methodology, the two institutions differ in implementation. ESRI places greater emphasis on a formally constructed coincident composite index, whereas the NBER evaluates several macroeconomic indicators individually when determining turning points. Recent reporting has similarly emphasized the central role of the coincident index in ESRI’s recession dating process (\cite{japantimes_recession_2025}).

\begin{table}[t!]
\centering
\begin{threeparttable}
\caption{Start Dates of Japanese Recessions (1970-2025)}
\label{tab:japan_recession_start_1970}

\begin{tabular}{ccl}
\hline
\textbf{No.} & \textbf{Episode} & \textbf{Start Month} \\
\hline
1  & Recession 1  & Jul.\ 1970 \\
2  & Recession 2  & Nov.\ 1973 \\
3  & Recession 3  & Jan.\ 1977 \\
4  & Recession 4  & Feb.\ 1980 \\
5  & Recession 5  & Jun.\ 1985 \\
6  & Recession 6  & Feb.\ 1991 \\
7  & Recession 7  & May\ 1997 \\
8  & Recession 8  & Nov.\ 2000 \\
9  & Recession 9  & Feb.\ 2008 \\
10 & Recession 10 & Mar.\ 2012 \\
11 & Recession 11 & Oct.\ 2018 \\
\hline
\end{tabular}

\begin{tablenotes}[flushleft]
\footnotesize
\item\textit{Notes:} The table reports the start dates of Japanese recessions identified by the ESRI. ESRI determines business-cycle turning points using a broad range of macroeconomic indicators and a coincident composite index designed to track overall economic activity. By convention, recession start dates correspond to business-cycle peaks.
\end{tablenotes}

\end{threeparttable}
\end{table}

\subsection{GDP Data}
Japan’s official GDP data are produced by the ESRI through the Quarterly Estimates of GDP. The GDP series are reported at the quarterly frequency and include nominal and real measures under the expenditure approach to national accounts. In this paper, GDP data are used as a broad measure of aggregate economic activity and for comparison with officially dated business-cycle turning points.
\subsection{Unemployment and Vacancy Data}
The unemployment rate data are obtained from the OECD Main Economic Indicators database through the Federal Reserve Economic Data (FRED) system and are based on Japan’s Labour Force Survey conducted by the Statistics Bureau of Japan. The Labour Force Survey is a monthly household survey designed to measure employment and unemployment conditions in Japan. Following OECD and International Labour Organization (ILO) definitions, individuals are classified as unemployed if they are without work, available for work, and actively seeking employment. The unemployment rate is measured as the number of unemployed persons divided by the labor force and is reported as a monthly seasonally adjusted series for individuals aged 15--64.

In Japan, the Ministry of Health, Labour and Welfare compiles monthly statistics on job openings, job seekers, and employment placements through the Public Employment Security Offices (Hello Work) system, and publishes indicators such as the job openings-to-applicants ratio as part of the General Employment Placement Situation statistics. In this paper, we use seasonally adjusted monthly data on the total number of active job openings (monthly active vacancies) obtained from the official e-Stat database. We also obtain seasonally adjusted monthly labor force data from e-Stat. The vacancy rate is then constructed as the total number of active job openings divided by the total labor force.

\subsection{Labor Market Dynamics}
\begin{figure}[H]
    \centering
    \includegraphics[width=1\linewidth]{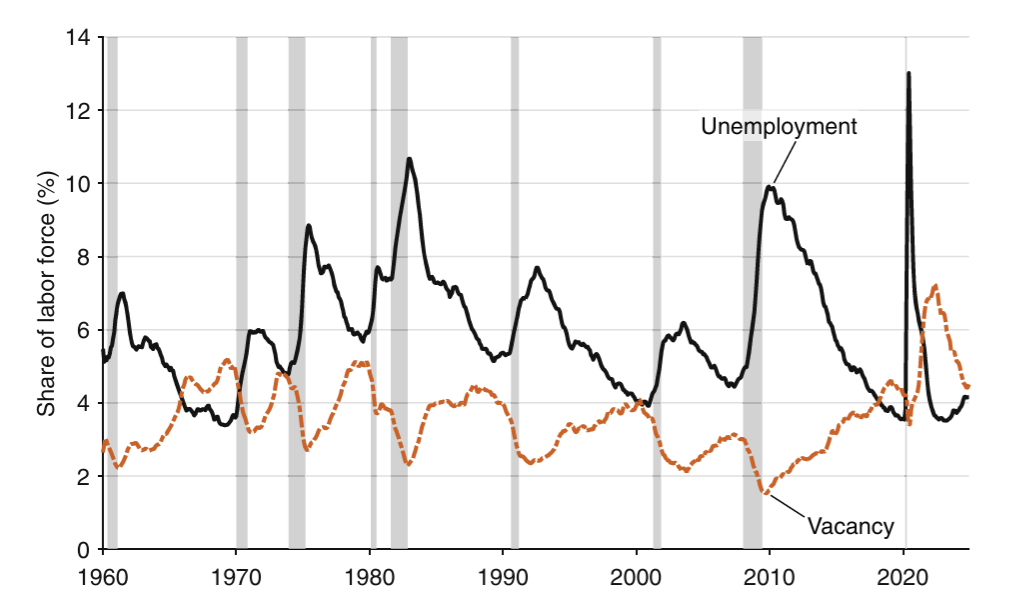}
    \caption{United States: Unemployment and Vacancy Rate}
    \label{fig:US_unemployement_vacancy_data}

    \vspace{0.3em}
    \begin{minipage}{0.95\linewidth}
    \footnotesize
    \textit{Notes:} U.S. unemployment and vacancy rates, January 1960--December 2024. This figure is reproduced from \cite{michaillat2025}. The unemployment rate (solid black line) is based on BLS data, and the vacancy rate (orange dashed line) combines estimates from \cite{barnichon2010composite} and the BLS. Both series are shown as three-month moving averages of monthly data. Shaded areas indicate recessions as dated by the NBER. We use this figure as a benchmark for comparison with our analysis.
    \end{minipage}
\end{figure}

To provide a reference point for the Japanese analysis, we first examine labor market dynamics in the United States, where recession detection rules such as the Sahm and Michez rules were originally developed and evaluated. Figure~\ref{fig:US_unemployement_vacancy_data} plots the unemployment rate and the vacancy rate over time, with shaded regions indicating recessions dated by the National Bureau of Economic Research. A clear pattern emerges: unemployment rises during recessions, while vacancies decline, implying that the two series move in opposite directions over the business cycle.

\begin{figure}[H]
    \centering
    \includegraphics[width=1\linewidth]{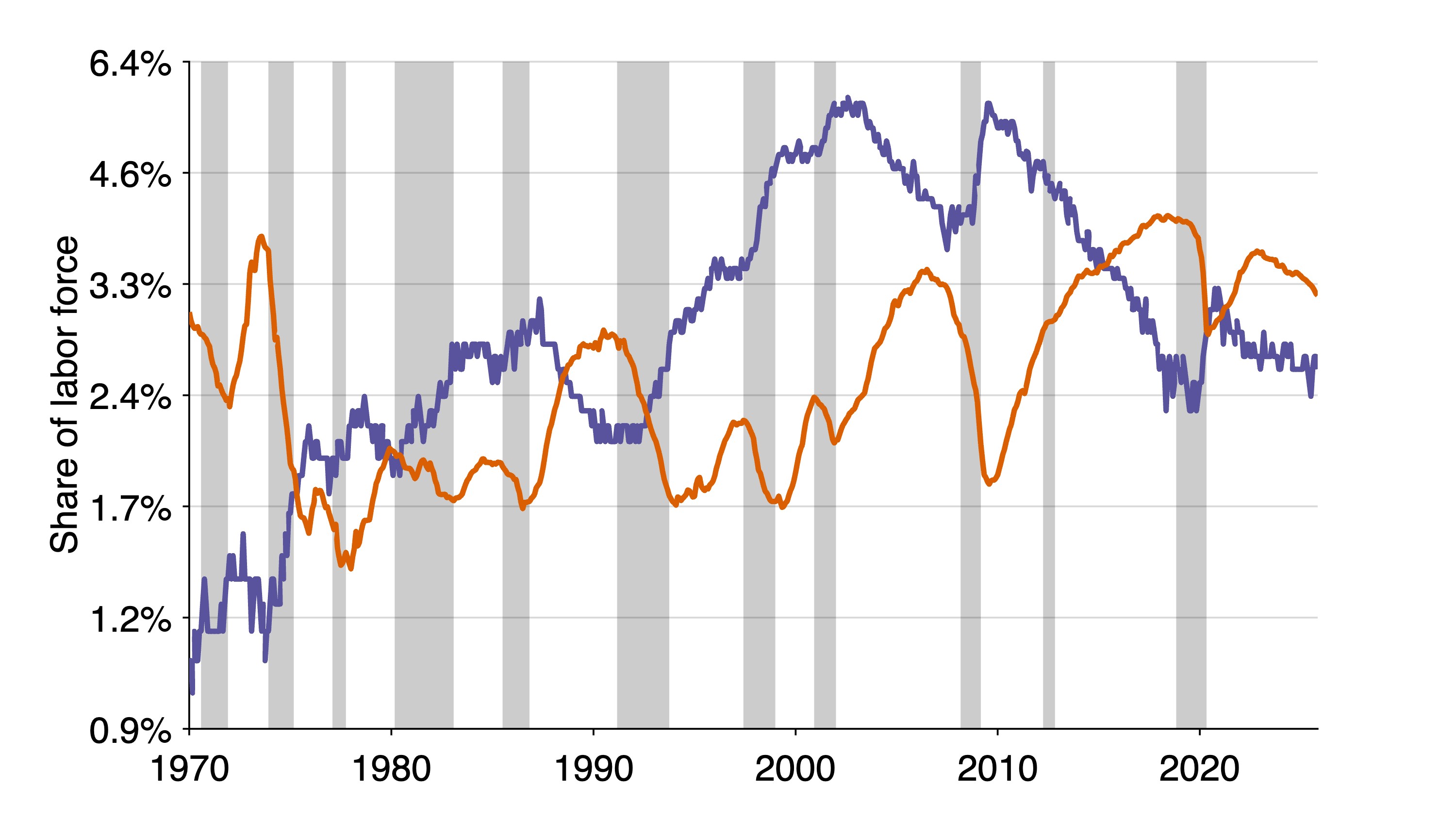}
    \caption{Japan: Unemployment Rate and Vacancy Rate}
    \label{fig:japan_unemployment_vacancy_rate}

    \vspace{0.3em}
    \begin{minipage}{0.95\linewidth}
    \footnotesize
    \textit{Notes:} Monthly Japanese unemployment and vacancy rates from January 1970 to December 2025. The unemployment rate (solid blue line) is constructed using data from FRED, while the vacancy rate (orange line) is computed using Japan’s labor force survey. Both series are expressed as three-month moving averages of monthly observations. Shaded areas indicate recessions dated by the Economic and Social Research Institute (ESRI).
    \end{minipage}
\end{figure}

We next examine labor market dynamics in Japan. Figure~\ref{fig:japan_unemployment_vacancy_rate} plots the unemployment rate and the vacancy rate over time, with shaded regions indicating recessions dated by the Economic and Social Research Institute. As in the United States, unemployment tends to rise during recessions, while vacancies decline, implying that the two series move in opposite directions over the business cycle. However, these patterns are less pronounced in Japan. In particular, fluctuations in unemployment are more gradual, and signals appear less sharp around some recession episodes. This attenuation suggests that recession detection using labor market data alone may be more challenging in the Japanese context. At the same time, the underlying economic intuition remains: vacancies reflect labor demand, while unemployment captures labor market slack. Accordingly, combining these two indicators may yield a more informative signal for recession detection in Japan.

\section{Background: Sahm Rule and Michez Rule in Japan}
This section presents the empirical results for recession detection in Japan using labor-market based indicators. We first implement the Sahm and Michez rules in the Japanese context. While the Sahm rule tracks increases in unemployment relative to recent lows, the Michez rule combines unemployment and vacancy data to construct a less noisy recession indicator. We then evaluate the ability of these indicators to identify historical recession episodes in real time.
\subsection{Sahm Rule in Japan}
First, we smooth the unemployment rate $u(t)$ by taking a three-month moving average:
\begin{equation}
\bar{u}(t) = \frac{u(t) + u(t-1) + u(t-2)}{3}.
\end{equation}

This step reduces short-term fluctuations and isolates persistent movements in unemployment.

Second, we measure the increase in the smoothed unemployment rate relative to its lowest value over the preceding 12 months:
\begin{equation}
\tilde{u}(t) = \bar{u}(t) - \min_{0 \leq s \leq 12} \{\bar{u}(t-s)\}.
\end{equation}

This transformation captures how much unemployment has risen from its recent trough. By construction, the indicator remains near zero when unemployment is stable or declining, and becomes positive when unemployment begins to increase.

A recession is identified when the indicator exceeds 0.5 percentage points:
\begin{equation}
\tilde{u}(t-1) < 0.5 \quad \text{and} \quad \tilde{u}(t) \geq 0.5
\end{equation}

\begin{figure}[H]
    \centering

    \includegraphics[width=\linewidth]{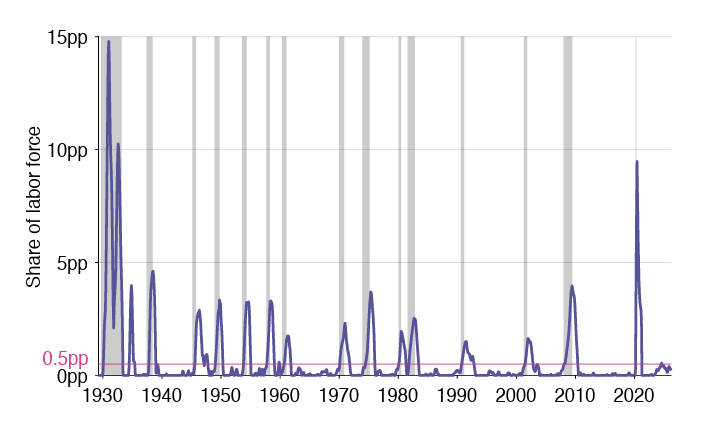}
    \caption{Sahm Rule Results for the United States}
    \label{fig:Sahm_Rule_US}
    \vspace{1em}

    \includegraphics[width=\linewidth]{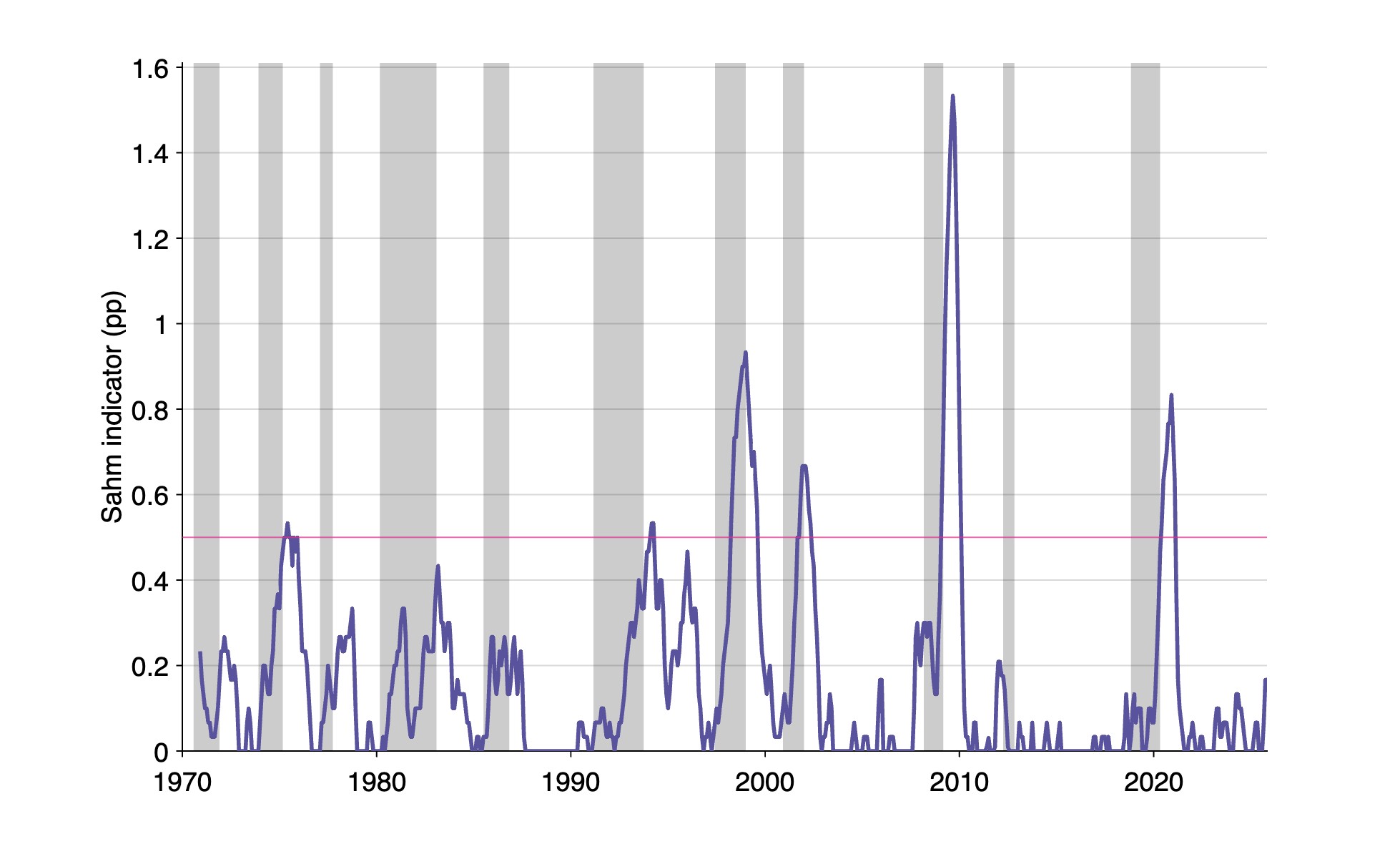}

    \caption{Sahm Rule Results for Japan}
    \label{fig:sahm_rule_comparison}

    \vspace{0.3em}
    \begin{minipage}{0.95\linewidth}
    \footnotesize
    \textit{Notes:} The top panel plots the Sahm indicator for the United States, defined as the increase in the three-month moving average of the unemployment rate relative to its twelve-month minimum, using data from the Bureau of Labor Statistics from 1930 to the present. Shaded areas denote recessions dated by the NBER. The recession threshold is 0.5 percentage points.

    \vspace{0.3em}

    The bottom panel plots the Sahm indicator for Japan defined as the increase in the three-month moving average of the unemployment rate relative to its twelve-month minimum, using unemployment data from FRED spanning 1970--2025. Shaded areas denote recessions dated by the ESRI. The recession threshold is 0.5 percentage points.
    \end{minipage}
\end{figure}

Figure~\ref{fig:Sahm_Rule_US} shows that the Sahm indicator performs well in the United States. The horizontal line marks the 0.5 percentage point threshold used to signal a recession. The indicator rises sharply during recessions and typically crosses the threshold in these periods, successfully identifying downturns. However, because the rule relies entirely on unemployment, the signal can be noisy and occasionally misleading. In particular, the indicator generates false positives in several historical episodes, including 1934, when it peaked at 3.98 percentage points despite no recession, as well as in 1947 and 1959, when it crossed the threshold without an associated downturn. Even in more recent data, smaller fluctuations, such as those around 2003, do not correspond to recessions. These patterns suggest that, while the Sahm Rule is simple and informative, reliance on a single labor market indicator limits precision around turning points.

The Sahm indicator performs poorly in Japan (Figure ~\ref{fig:sahm_rule_comparison}). It misses several recessions prior to 1990, detects the 1990s downturn with a delay, and fails to capture episodes in the late 1970s and around 2012. Overall, it identifies only 4 of 11 recessions, less than half, indicating that a direct application of the Sahm rule does not translate well to the Japanese context.

\clearpage


\subsection{Michez Rule in Japan}
We implement the Michez rule, which combines unemployment and vacancy data to produce a less noisy recession indicator.

First, we construct the vacancy-based component. Let $v(t)$ denote the vacancy rate and define its three-month moving average:
\begin{equation}
\bar{v}(t) = \frac{v(t) + v(t-1) + v(t-2)}{3}.
\end{equation}

We then measure the decline in the smoothed vacancy rate relative to its maximum over the preceding 12 months:
\begin{equation}
\tilde{v}(t) = \max_{0 \leq s \leq 12} \bar{v}(t - s) - \bar{v}(t).
\end{equation}

Let $\tilde{u}(t)$ denote the unemployment-based indicator defined in Section 4.1.

\begin{figure}[t!]
    \centering
    \includegraphics[width=1\linewidth]{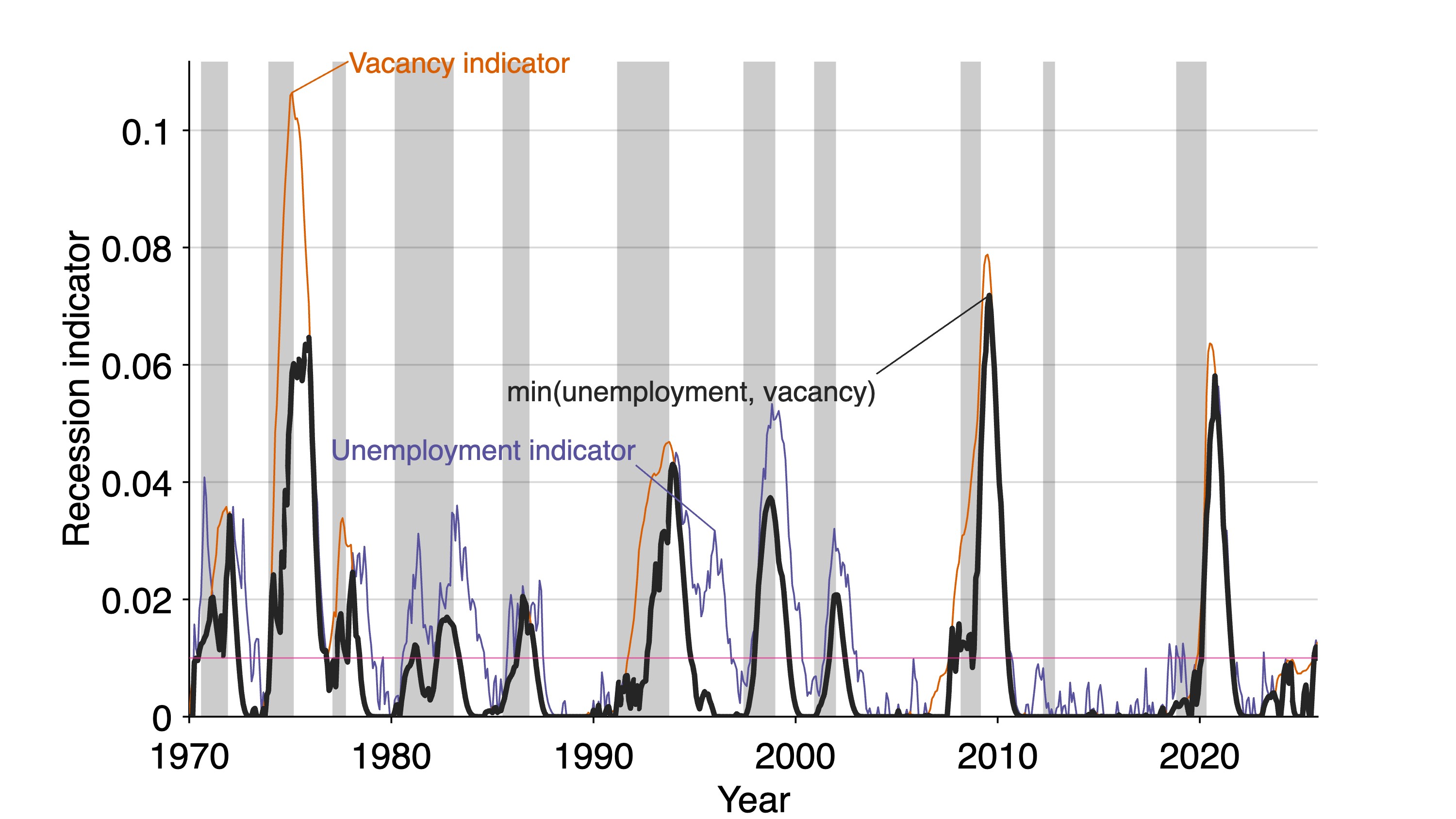}
    
    \caption{Michez Rule: Minimum of Two Indicators}
    \label{fig:michez_rule_min}

    \vspace{0.3em}
    \begin{minipage}{0.95\linewidth}
    \footnotesize
    \textit{Notes:} The figure plots three recession indicators for Japan from 1970 onward. The blue line shows the unemployment-based indicator, measured as the increase in the unemployment rate relative to its recent minimum. The orange dashed line shows the vacancy-based indicator, measured as the decline in the vacancy rate relative to its recent maximum. The black line represents the Michez rule, defined as the minimum of the two indicators. Gray shaded regions indicate recessions dated by the Economic and Social Research Institute (ESRI).
    \end{minipage}
\end{figure}

The Michez rule combines these two indicators by taking their minimum as shown in Figure ~\ref{fig:michez_rule_min}:
\begin{equation}
m(t) = \min \{ \tilde{u}(t), \tilde{v}(t) \}.
\end{equation}

\begin{figure}[p!]
    \centering
    \includegraphics[width=1\linewidth]{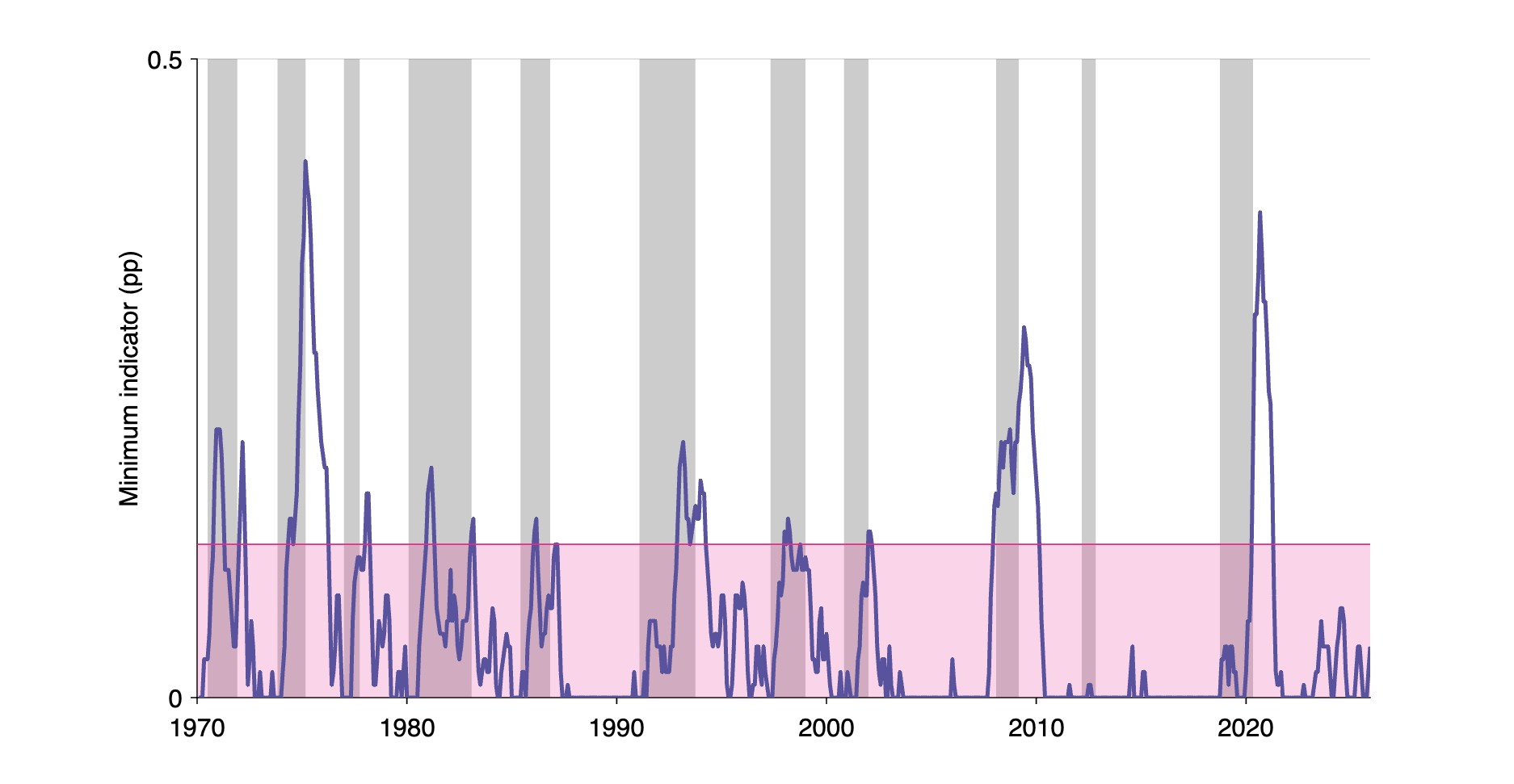}
    
    \caption{Michez Rule Results}
    \label{fig:michez_rule}

    \vspace{0.3em}
    \begin{minipage}{0.95\linewidth}
    \footnotesize
    \textit{Notes:} The figure plots the Michez minimum indicator over time using Japanese labor-market data. Gray shaded regions indicate recessions dated by the Economic and Social Research Institute (ESRI). The horizontal line represents the recession threshold of 0.29. The Michez rule combines unemployment and vacancy indicators by taking their minimum, following Michaillat and Saez (2025). Labor-market data are obtained from the Labor Force Survey of the Statistics Bureau of Japan.
    \end{minipage}
\end{figure}

The Michez rule further improves recession detection by incorporating vacancy data
alongside unemployment. Following the original U.S. application, we implement the standard Michez rule using a fixed threshold of 0.29, such that recessions are detected whenever the indicator crosses the threshold. As shown in Figure~\ref{fig:michez_rule}, the minimum indicator
closely tracks recession periods, with threshold crossings occurring near the onset
of most downturns. Relative to the Sahm rule, the Michez rule generates fewer false
positives, with the indicator remaining near zero during expansions and exhibiting
fewer spurious spikes outside recession periods.

This improvement reflects the ability of the Michez rule to filter out transient labor market fluctuations by requiring both rising unemployment and declining vacancies. Incorporating vacancy data yields a more accurate and less noisy recession signal, improving the tradeoff between false positives and false negatives. 

In our baseline specification, this fixed-threshold rule serves as a benchmark.
We then extend the analysis by replacing the ad hoc threshold with the probabilistic
classifier selection and aggregation procedure described above.



\section{Construction of the Recession Indicators}
Using unemployment and vacancy data, we construct 95,832 recession indicators in this section. The construction process consists of smoothing the data, identifying turning points, scaling the resulting movements, and combining the single-variable measures into composite indicators. In the next section, we pair these indicators with thresholds to detect recession episodes.
\subsection{Smoothing the Data}
To construct the recession indicators, we first smooth the unemployment and vacancy series using both simple moving averages and exponentially weighted moving averages, applying the approach in \citet{michaillat2025}. We begin by defining trailing moving averages for the unemployment and vacancy rate. In both equations $\alpha=0,1,\dots,11$ represents the smoothing window. For example, if $\alpha = 0$, the series is unchanged; if $\alpha = 11$, the formula produces a 12-month trailing average. The standard Sahm and Michez rule specifications correspond to $\alpha=2$, which yields a 3-month moving average.
\begin{equation}
\bar{u}(t)=\frac{\sum_{k=0}^{\alpha}u(t-k)}{\alpha+1},
\end{equation}

\begin{equation}
\bar{v}(t)=\frac{\sum_{k=0}^{\alpha}v(t-k)}{\alpha+1}.
\end{equation}

In addition, we examine exponentially weighted moving averages for the unemployment and vacancy rate where $\alpha \in [0.1,1]$ determines the degree of persistence in the series. A value of $\alpha=1$ implies no smoothing, while lower values of $\alpha$ generate smoother series. 

\begin{equation}
\bar{u}(t)=\alpha u(t)+(1-\alpha)\bar{u}(t-1),
\label{eq:trailing_min_unemployment}
\end{equation}

\begin{equation}
\bar{v}(t)=\alpha v(t)+(1-\alpha)\bar{v}(t-1).
\end{equation}

\subsection{Detecting Turning Points}
To identify turning points in labor market conditions, we compare current unemployment and vacancy rates to their recent extrema over a trailing window, following \citet{michaillat2025}. We define the trailing minimum unemployment rate and trailing maximum vacancy rate as:

\begin{equation}
u^{\min}(t)=\min_{0\leq k \leq \beta}\bar{u}(t-k),
\label{eq:trailing_min_vacancy}
\end{equation}

\begin{equation}
v^{\max}(t)=\max_{0\leq k \leq \beta}\bar{v}(t-k),
\label{eq:trailing_max_vacancy}
\end{equation}

where $\beta=1,2,\dots,18$ specifies the length of the trailing window. We then construct the unemployment and vacancy indicators as

\begin{equation}
\tilde{u}(t)=\bar{u}(t)-u^{\min}(t),
\label{eq:min_unemployment}
\end{equation}

\begin{equation}
\tilde{v}(t)=v^{\max}(t)-\bar{v}(t).
\label{eq:max_vacancy}
\end{equation}

The unemployment indicator measures how much unemployment has increased relative to recent lows, while the vacancy indicator measures how much vacancies have fallen relative to recent highs. These transformations are motivated by the fact that recessions are typically associated with rising unemployment and declining vacancies. Larger values of $\tilde{u}(t)$ and $\tilde{v}(t)$ therefore correspond to greater labor market deterioration.

\subsection{Scaling Variations}

Following \citet{michaillat2025}, we consider alternative ways of measuring changes in labor market conditions. Since recessions may depend on either level or proportional changes in unemployment and vacancy rates, we apply a Box--Cox transformation to construct alternative indicators.

For unemployment, we define

\begin{equation}
\hat{u}(t)=
\frac{
[u(t)]^{\gamma}-[u^{\min}(t)]^{\gamma}
}{
\gamma
}
\simeq
\frac{\tilde{u}(t)}{u^{\min}(t)^{1-\gamma}},
\label{eq:boxcox_unemployment}
\end{equation}

where $\gamma \in \{0,0.1,\dots,1\}$ controls the scaling of the transformation. The parameter $\gamma$ determines whether the indicator emphasizes absolute or proportional changes in unemployment. When $\gamma=1$, the transformation reduces to level changes in unemployment, while $\gamma \to 0$ approximates percentage or log changes. Intermediate values of $\gamma$ generate indicators that lie between these two cases.

As $\gamma \to 0$, the transformation becomes

\begin{equation}
\hat{u}(t)
=
\log\left(
\frac{u(t)}{u^{\min}(t)}
\right)
\simeq
\frac{\tilde{u}(t)}{u^{\min}(t)},
\label{eq:log_unemployment}
\end{equation}

which captures proportional changes in unemployment. When $\gamma=1$, the transformation simplifies to

\begin{equation}
\hat{u}(t)=\tilde{u}(t),
\label{eq:level_unemployment}
\end{equation}

which measures level changes in unemployment.

Similarly, for vacancies, we define
\begin{equation}
\hat{v}(t)=
\frac{
[v^{\max}(t)]^{\gamma}-[\bar{v}(t)]^{\gamma}
}{
\gamma
}
\simeq
\frac{\tilde{v}(t)}{[v^{\max}(t)]^{1-\gamma}}
\label{eq:boxcox_vacancy}
\end{equation}

where $\gamma$ has the same interpretation as in the unemployment transformation. The vacancy specification differs slightly from the unemployment specification because recessions are associated with declines in vacancies relative to a recent maximum, rather than increases in unemployment relative to a recent minimum. Lower values of $\gamma$ place greater emphasis on proportional declines in vacancies, while $\gamma=1$ corresponds to level declines.
\begin{figure}[p!]
    \centering
    
    \begin{subfigure}{\linewidth}
        \centering
        \includegraphics[width=1\linewidth]{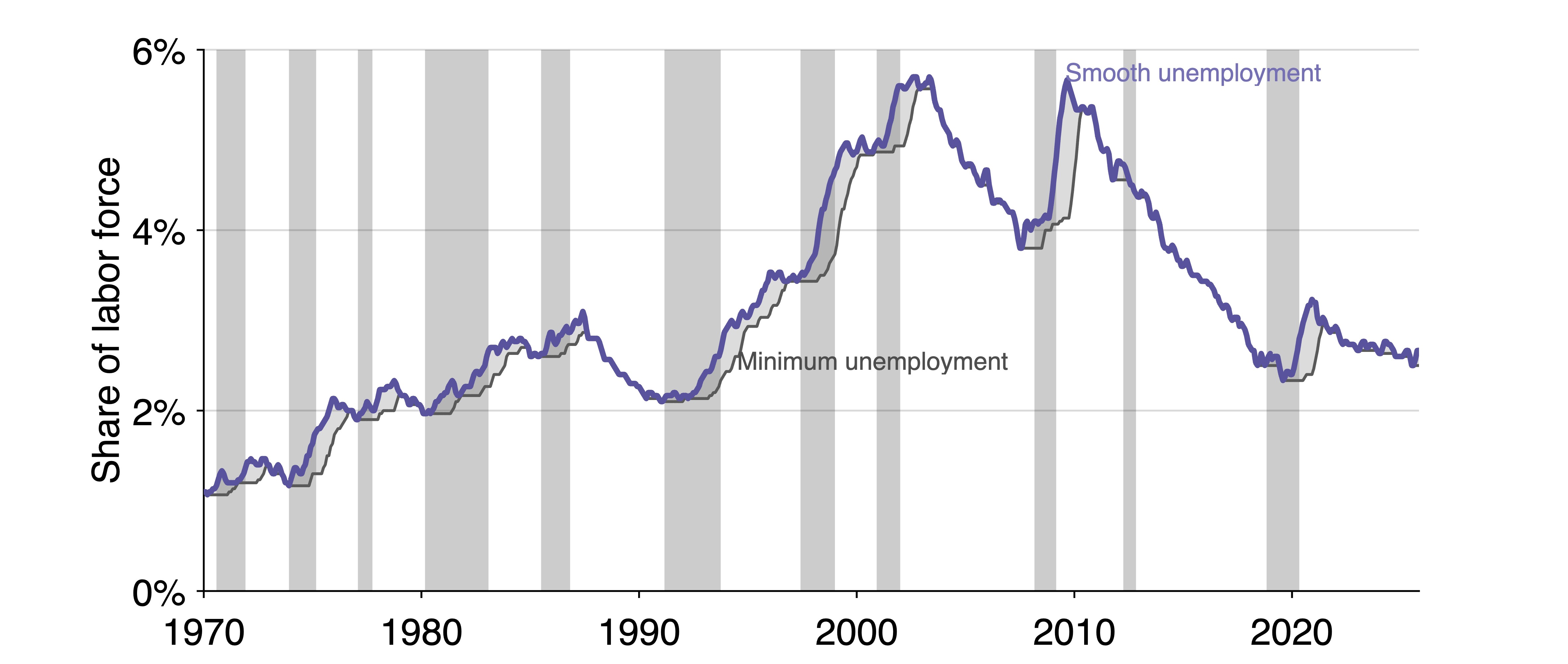}
        \caption{Trailing Minimum of the Unemployment Rate}
        \label{fig:trailing_minimum}
    \end{subfigure}
    
    \vspace{0.5cm}
    
    \begin{subfigure}{\linewidth}
        \centering
        \includegraphics[width=1\linewidth]{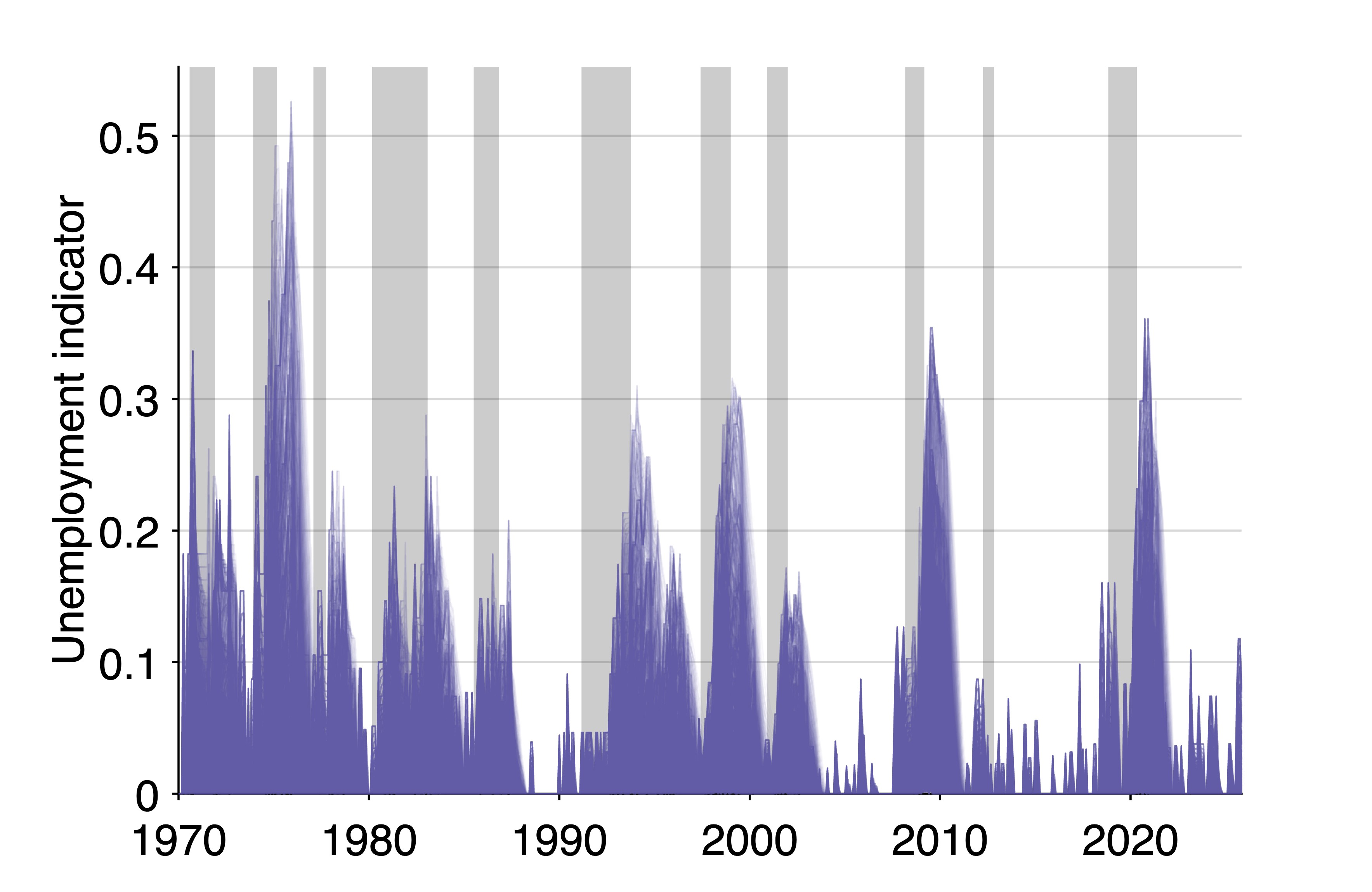}
        \caption{Increases in the Unemployment Rate}
        \label{fig:figure_unemployment}
    \end{subfigure}
    
    \caption{Increase in the Japanese Unemployment Rate (1970--2025)}
    \label{fig:combined_unemployment_figures}

    \vspace{0.3em}
    \begin{minipage}{0.95\linewidth}
    \footnotesize
    \textit{Notes:} Panel (a) plots the trailing minimum of the smoothed unemployment rate computed using Equation (\ref{eq:trailing_min_unemployment}). Panel (b) plots the unemployment increase defined in Equation (\ref{eq:min_unemployment}) as the difference between the smoothed unemployment rate and its recent minimum. Gray shaded regions indicate recessions dated by the ESRI.
    \end{minipage}
\end{figure}
\begin{figure}[p!]
    \centering
    
    \begin{subfigure}{\linewidth}
        \centering
        \includegraphics[width=\linewidth]{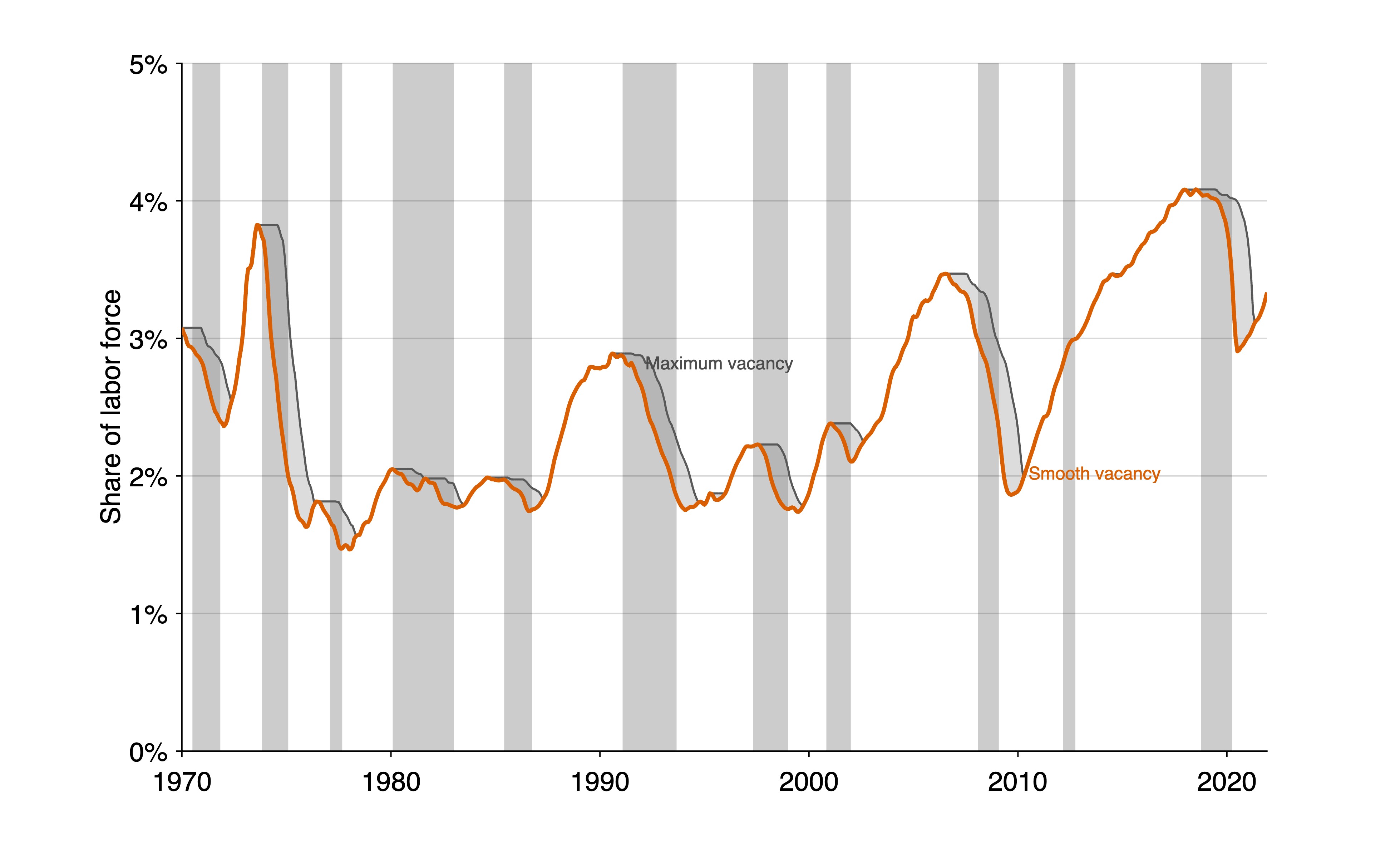}
        \caption{Trailing Maximum of the Vacancy Rate}
        \label{fig:vacancy_trailing_maximum}
    \end{subfigure}
    
    \vspace{0.5cm}
    
    \begin{subfigure}{\linewidth}
        \centering
        \includegraphics[width=\linewidth]{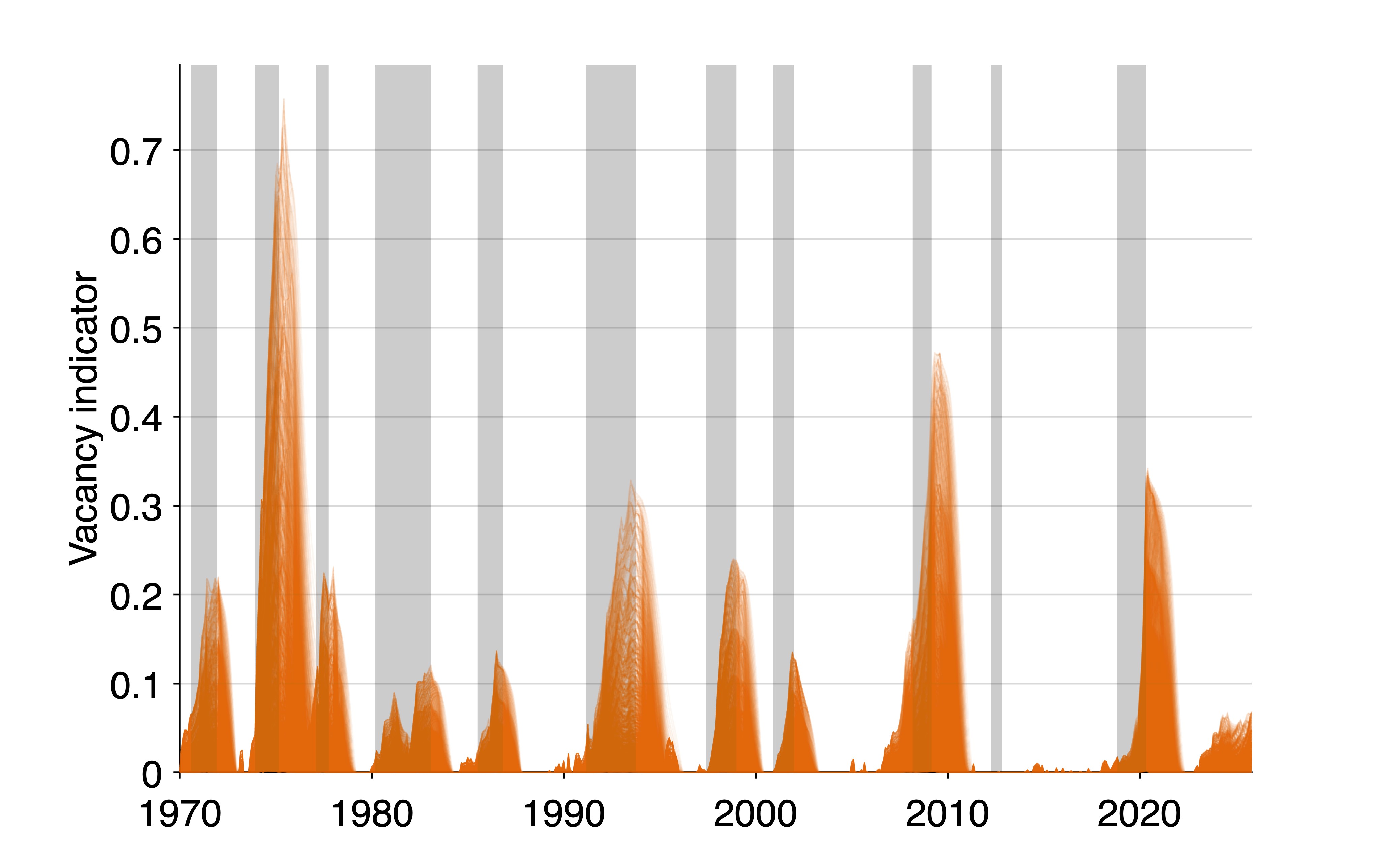}
        \caption{Decreases in the Vacancy Rate}
        \label{fig:figure_vacancy_indicator_pp}
    \end{subfigure}
    
    \caption{Decrease in the Japanese Vacancy Rate (1970--2025)}
    \label{fig:vacancy_indicators}

    \vspace{0.3em}
    \begin{minipage}{0.95\linewidth}
    \footnotesize
    \textit{Notes:} Panel (a) plots the trailing maximum of the smoothed vacancy rate computed using Equation (\ref{eq:trailing_max_vacancy}). Panel (b) plots the vacancy decrease defined in Equation (\ref{eq:max_vacancy}) as the difference between the smoothed vacancy rate and its recent maximum. Gray shaded regions indicate recessions dated by the ESRI.
    \end{minipage}
\end{figure}

    
        
    
    
        
    
    

\subsection{Combining Single-Variable Indicators}

Finally, we consider recession indicators based on unemployment alone, vacancies alone, and weighted combinations of the two measures constructed using alternative weighting schemes. $\delta=0,0.1,\dots,1$ determines the weight placed on the unemployment indicator. The Sahm Rule uses the unemployment indicator, corresponding to $\delta = 1$.

\begin{equation}
i(t)=\delta \hat{u}(t)+(1-\delta)\hat{v}(t),
\end{equation}

Following \citet{michaillatsaez2025}, we also construct indicators based on the minimum and maximum of the unemployment and vacancy signals:

\begin{equation}
i(t)=\delta \min(\hat{u}(t),\hat{v}(t))+(1-\delta)\max(\hat{u}(t),\hat{v}(t)),
\end{equation}

where $\delta=0,0.1,\dots,1$ determines the weight on the minimum indicator. The Michez rule uses the minimum indicator, corresponding to $\delta=1$.


Rather than imposing a specific functional form, we consider a broad class of recession indicators constructed from unemployment and vacancy data. These indicators allow for different ways of measuring changes in labor market conditions, ranging from level changes to percentage changes through Box–Cox transformations. We further combine unemployment and vacancy signals using both linear combinations and nonlinear transformations, such as the minimum and maximum of the two series. This approach generates a large set of candidate indicators that differ in their sensitivity and robustness, allowing the data to determine which specifications best capture recession dynamics.

Figures ~\ref{fig:figure_unemployment} and ~\ref{fig:figure_vacancy_indicator_pp} plot the unemployment- and vacancy-based indicators for Japan. The unemployment-based indicators capture increases relative to their recent minimum, while the vacancy-based indicators capture declines relative to their recent maximum. Both increase during recession periods, but they differ in their dynamics. The unemployment-based indicators evolve smoothly and adjust gradually, whereas the vacancy-based indicators exhibit sharper movements and align more closely with the onset of recessions. These patterns suggest that vacancy-based measures provide a more responsive signal of labor market conditions, while unemployment-based measures reflect slower adjustment. Taken together, the two series provide complementary information, motivating their combination to improve recession detection.

\begin{figure}[t!]
    \centering
    \includegraphics[width=1\linewidth]{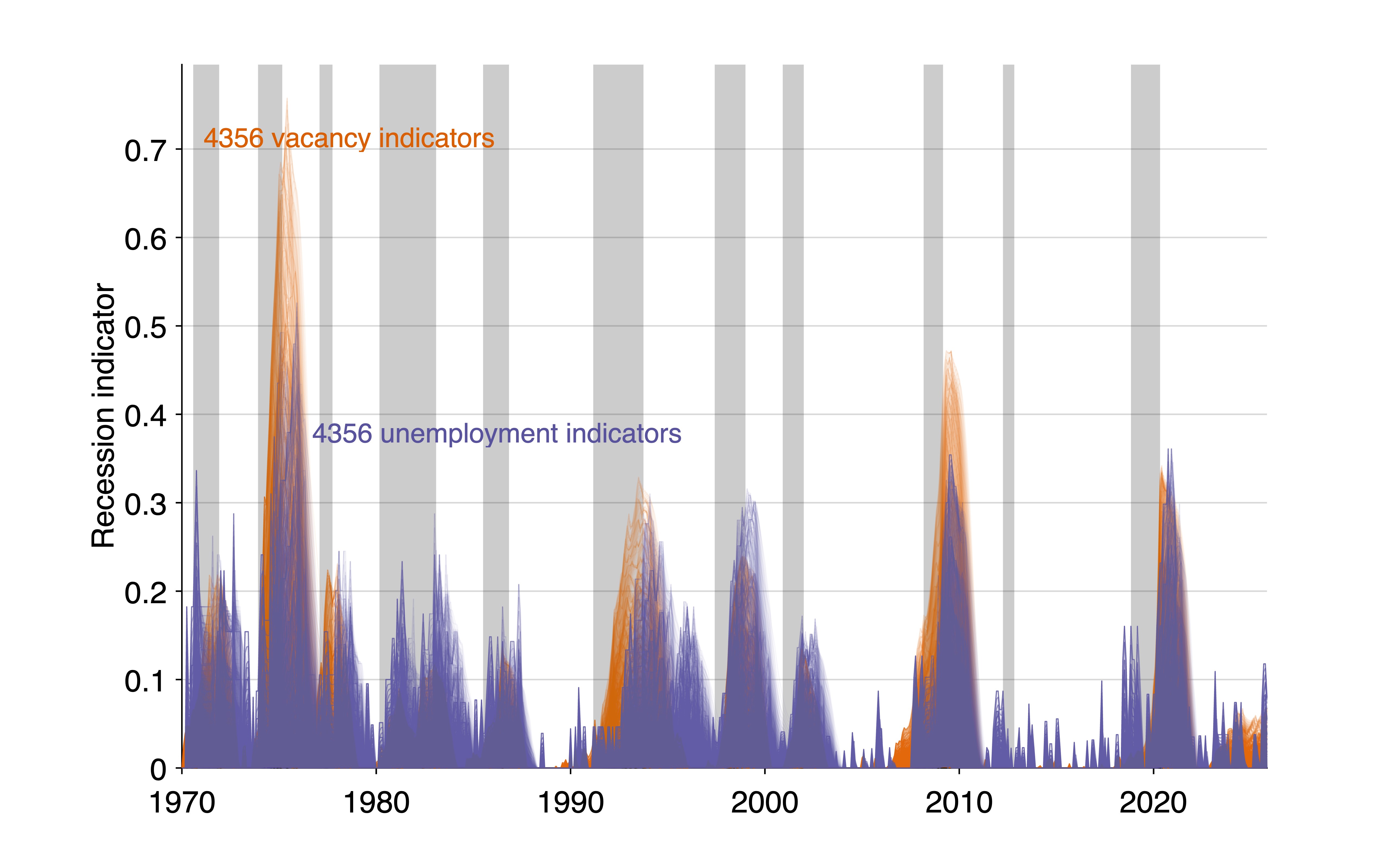}
    \caption{4,356 Unemployment Indicators and 4,356 Vacancy Indicators}
    \label{fig:placeholder-1}

    \vspace{0.3em}
    \begin{minipage}{0.95\linewidth}
    \footnotesize
    \textit{Notes:} The unemployment indicators are constructed by applying the scaling transformation in Equation (\ref{eq:boxcox_unemployment}) to the unemployment increases shown in Figure~\ref{fig:figure_unemployment}. Similarly, the vacancy indicators are constructed by applying the scaling transformation in Equation (\ref{eq:boxcox_vacancy}) to the vacancy decreases shown in Figure~\ref{fig:figure_vacancy_indicator_pp}. Gray shaded regions indicate recessions dated by the Economic and Social Research Institute (ESRI).
    \end{minipage}
\end{figure}

\subsection{Summary}
Using the unemployment rate, we generate a total of $(12+10)\times18\times11 = 4{,}356$ recession indicators by considering alternative smoothing methods, trailing windows, and scaling transformations. Applying the same procedure to the vacancy rate yields an additional 4,356 indicators. 

Next, we combine the unemployment and vacancy indicators using linear combinations, producing $4{,}356 \times 11 = 47{,}916$ composite indicators. We then construct an additional 47,916 indicators by taking the minimum and maximum across these combined measures. Altogether, this process generates $47{,}916 + 47{,}916 = 95{,}832$ recession indicators, which serve as the basis for constructing recession classifiers.

\section{Results}

Following the framework proposed by \cite{michaillat2025}, described in Section 5, we construct recession classifiers as combinations of labor-market indicators and threshold values. We then evaluate their performance using the anticipation--precision frontier. Among 193 frontier classifiers, we select six classifiers whose standard deviations of detection errors are below three months, which form the final high-precision set used in the analysis.

\subsection{Detection Methodology}

After obtaining 95,832 recession indicators, we generate hundreds of millions of recession classifiers by applying alternative thresholds to each indicator. As shown in Figure~\ref{fig:michez_rule}, the indicator can cross the threshold multiple times during recoveries, generating spurious recession signals. This pattern motivates the reset criterion requiring the indicator to return to zero before a new recession can be identified.

Therefore, a recession begins when the indicator crosses the threshold from below while the economy is currently in expansion. To avoid false recession signals caused by temporary movements around the threshold, the methodology keeps track of whether the economy is already in a recession or expansion state. In particular, the economy can only enter a new recession if it was previously in expansion, and it only returns to expansion once the indicator falls back to zero. This prevents short-lived rebounds in the indicator during recoveries from being incorrectly treated as new recessions. Although more computationally intensive than simple threshold rules, the procedure produces a more stable and economically intuitive classification of business-cycle turning points.

\subsection{Selecting Perfect Classifiers}
Using 95,832 recession indicators together with 2,500 alternative thresholds ranging from 0.0001 to 0.25, we generate 1,844,564 perfect recession classifiers. For each classifier, we evaluate its ability to identify the official Japanese recession episodes over the sample period. We retain only classifiers that correctly identify all recessions without generating false positives or false negatives. A false negative occurs when a classifier fails to identify a true recession, while a false positive occurs when a classifier incorrectly signals a recession during an expansion. After restricting attention to the high-precision segment, we compare the six classifiers’ mean and standard deviation of detection errors in order to identify recession rules that provide the strongest real time turning-point performance.

\subsection{Evaluating Perfect Classifiers}
To select the classifiers included in the ensemble, we again follow \citet{michaillat2025} exactly. In principle, a policymaker with mean--variance preferences over detection error would choose the frontier classifier $k$ that minimizes
\[
\mu(k) + \lambda \sigma(k),
\]
where $\mu(k)$ and $\sigma(k)$ denote the mean and standard deviation of the detection error, and $\lambda > 0$ captures the weight placed on precision relative to anticipation. Because $\lambda$ is not observed, we do not select a single classifier. Instead, as in \citet{michaillat2025}, we retain all classifiers on the anticipation--precision frontier whose standard deviation of detection errors satisfies
\[
\sigma(k) < 3 \text{ months}.
\]
Under the normal approximation for detection errors, this restriction implies that the associated 95\% confidence interval for the recession start date has width
\[
4\sigma(k) < 12 \text{ months}.
\]
We therefore define the ensemble as
\[
\mathcal{K}^{*} = \{ k \in \mathcal{F} : \sigma(k) < 3 \},
\]
where $\mathcal{F}$ denotes the set of classifiers on the anticipation--precision frontier. This is the same selection rule used in \citet{michaillat2025}: it focuses attention on the high-precision segment of the frontier and excludes classifiers whose implied timing uncertainty exceeds one year.

\subsection{Finding the Anticipation-Precision Frontier}
We then select classifiers along the anticipation–precision frontier, defined as those that minimize both the mean detection delay and its variability. This frontier highlights classifiers that achieve the best tradeoff between timely and reliable recession detection.

\begin{figure}[t!]
    \centering

    \begin{subfigure}[t]{0.9\linewidth}
        \centering
        \includegraphics[width=\linewidth]{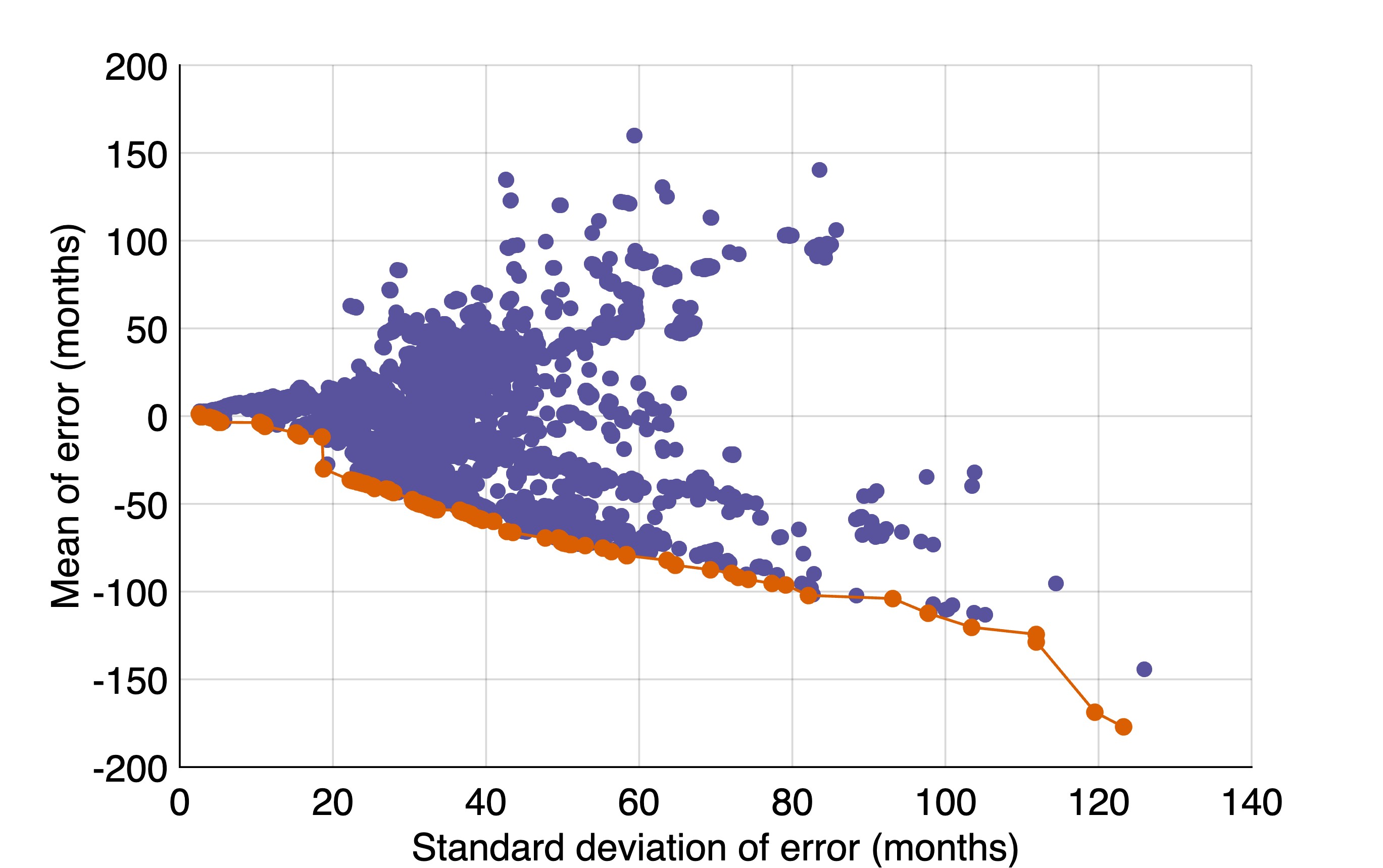}
        \caption{193 Perfect Classifiers on the Anticipation--Precision Frontier}
        \label{fig:figure_frontier}
    \end{subfigure}

    \vspace{0.4cm}

    \begin{subfigure}[t]{0.9\linewidth}
        \centering
        \includegraphics[width=\linewidth]{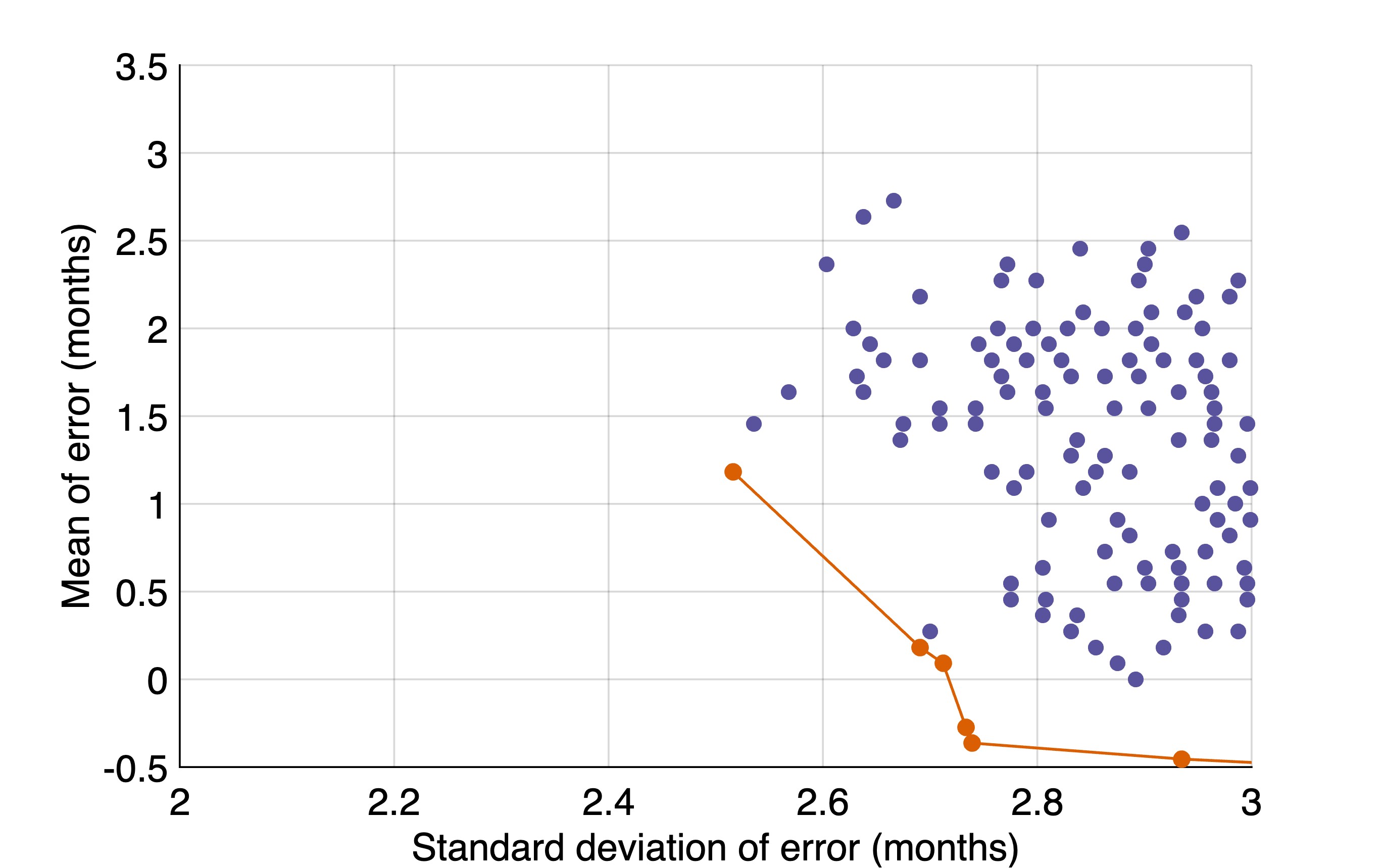}
        \caption{6 Perfect Classifiers satisfying $\sigma(k) < 3$ on the Anticipation--Precision Frontier}
        \label{fig:figure_frontier_precision}
    \end{subfigure}

    \caption{Anticipation and Precision of 1,844,564 Perfect Recession Classifiers in Japan, January 1970--December 2021}
    
    \label{fig:frontier_combined}

    \vspace{0.3em}
    \begin{minipage}{0.95\linewidth}
    \footnotesize
    \textit{Notes:} The figures plot the mean detection error against the standard deviation of detection errors for recession classifiers constructed using alternative indicators and thresholds. Each point represents a classifier. The orange frontier traces classifiers that maximize anticipation for a given level of dispersion, or equivalently maximize precision for a given level of anticipation. Panel (a) shows the full anticipation--precision frontier across all classifiers, while panel (b) focuses on the high-precision region satisfying $\sigma(k) < 3$. Detection errors are measured relative to recession start dates dated by the Economic and Social Research Institute (ESRI).
    \end{minipage}
\end{figure}

Figure ~\ref{fig:figure_frontier} plots the mean and standard deviation of detection errors across a large set of recession classifiers. The anticipation–precision frontier identifies the set of optimal classifiers that minimize detection delay for a given level of dispersion, or equivalently, minimize dispersion for a given level of delay. The downward-sloping frontier reveals a clear tradeoff: classifiers that achieve greater anticipation (more negative mean errors) exhibit substantially higher variability in detection timing, while classifiers with low dispersion detect recessions more consistently but with a modest delay.

The frontier contains 193 classifiers out of the 1,844,564 perfect classifiers. The left-most classifier, which achieves the highest precision, has a mean detection error of approximately 0.06 months and a has a standard deviation of detection errors of 2.7 months, indicating very accurate timing with minimal delay. In contrast, the right-most classifier exhibits substantial anticipation, with a mean detection error of -208.5 months and a standard deviation detection error of 158 months, reflecting a large loss in precision. \citet{michaillat2025} conducts this analysis for the United States from 1979–2021. The classifier ensemble in the United States detects recessions only 1.2 months after the recession has started. In Japan, we are able to detect recessions in 0.06 months after onset, while on average the Japanese ESRI committee detects recessions 26.5 months later, as shown in Table~\ref{tab:ESRI_Recession_Start_Announcments}. This shows that the algorithm is applicable to different labor markets worldwide and may perform even better in the Japanese context. Despite major differences between the United States and Japan in labor market structure and business cycle dynamics, the classifier ensemble is still able to detect recessions quickly and accurately. These findings suggest that combining unemployment and vacancy information provides a robust and internationally applicable recession indicator.

Table~\ref{tab:mainensemblethresholds1} reports episode-level recession detection results for the selected ensemble under alternative thresholds. For each threshold, the table lists the official start date, the detected start date, and the corresponding detection error for all 11 recession episodes, allowing us to examine how detection timing varies with the choice of threshold. Several patterns emerge.

\begin{table}[!htbp]
\centering
\begin{threeparttable}

\caption{Detected Recession Starts by Threshold for the Main Ensemble}
\label{tab:mainensemblethresholds1}

\scriptsize
\setlength{\tabcolsep}{12pt}
\renewcommand{\arraystretch}{1.2}

\begin{tabular}{ccccccc}
\toprule
\makecell{Ensemble\\Member} & Threshold & Recession & \makecell{Official\\Start} & \makecell{Official\\Trough} & \makecell{Detection\\Start} & \makecell{Error\\(Months)} \\
\midrule

\multicolumn{7}{c}{\textit{Threshold = 0.0005}} \\
1 & 0.0005 & 1  & 1970.58 & 1971.92 & 1970.25 & -4 \\
1 & 0.0005 & 2  & 1973.92 & 1975.17 & 1974.08 & 2 \\
1 & 0.0005 & 3  & 1977.08 & 1977.75 & 1977.33 & 3 \\
1 & 0.0005 & 4  & 1980.17 & 1983.08 & 1980.58 & 5 \\
1 & 0.0005 & 5  & 1985.50 & 1986.83 & 1985.75 & 3 \\
1 & 0.0005 & 6  & 1991.17 & 1993.75 & 1991.42 & 3 \\
1 & 0.0005 & 7  & 1997.42 & 1999.00 & 1997.50 & 1 \\
1 & 0.0005 & 8  & 2000.92 & 2002.00 & 2001.17 & 3 \\
1 & 0.0005 & 9  & 2008.17 & 2009.17 & 2008.42 & 3 \\
1 & 0.0005 & 10 & 2012.25 & 2012.83 & 2012.50 & 3 \\
1 & 0.0005 & 11 & 2018.83 & 2020.33 & 2019.08 & 3 \\

\midrule
\multicolumn{7}{c}{\textit{Threshold = 0.0056}} \\
2 & 0.0056 & 1  & 1970.58 & 1971.92 & 1970.25 & -4 \\
2 & 0.0056 & 2  & 1973.92 & 1975.17 & 1973.67 & -3 \\
2 & 0.0056 & 3  & 1977.08 & 1977.75 & 1976.83 & -3 \\
2 & 0.0056 & 4  & 1980.17 & 1983.08 & 1979.92 & -3 \\
2 & 0.0056 & 5  & 1985.50 & 1986.83 & 1985.75 & 3 \\
2 & 0.0056 & 6  & 1991.17 & 1993.75 & 1991.17 & 0 \\
2 & 0.0056 & 7  & 1997.42 & 1999.00 & 1997.42 & 0 \\
2 & 0.0056 & 8  & 2000.92 & 2002.00 & 2000.25 & -8 \\
2 & 0.0056 & 9  & 2008.17 & 2009.17 & 2008.00 & -2 \\
2 & 0.0056 & 10 & 2012.25 & 2012.83 & 2012.25 & 0 \\
2 & 0.0056 & 11 & 2018.83 & 2020.33 & 2018.92 & 1 \\

\midrule
\multicolumn{7}{c}{\textit{Threshold = 0.0081}} \\
3 & 0.0081 & 1  & 1970.58 & 1971.92 & 1970.25 & -4 \\
3 & 0.0081 & 2  & 1973.92 & 1975.17 & 1973.58 & -4 \\
3 & 0.0081 & 3  & 1977.08 & 1977.75 & 1976.75 & -4 \\
3 & 0.0081 & 4  & 1980.17 & 1983.08 & 1979.83 & -4 \\
3 & 0.0081 & 5  & 1985.50 & 1986.83 & 1985.75 & 3 \\
3 & 0.0081 & 6  & 1991.17 & 1993.75 & 1990.75 & -5 \\
3 & 0.0081 & 7  & 1997.42 & 1999.00 & 1997.42 & 0 \\
3 & 0.0081 & 8  & 2000.92 & 2002.00 & 2000.17 & -9 \\
3 & 0.0081 & 9  & 2008.17 & 2009.17 & 2007.92 & -3 \\
3 & 0.0081 & 10 & 2012.25 & 2012.83 & 2012.25 & 0 \\
3 & 0.0081 & 11 & 2018.83 & 2020.33 & 2018.83 & 0 \\

\bottomrule
\end{tabular}

\begin{tablenotes}[flushleft]
\scriptsize
\item \textit{Notes:} The table reports recession detection dates for Japan under alternative ensemble thresholds. Official start dates correspond to the benchmark recession chronology dated by the Economic and Social Research Institute (ESRI). Detection start denotes the first month in which the recession indicator crosses the specified threshold. Detection errors are measured as the difference, in months, between the detected and official recession start dates. Negative values indicate early detection, while positive values indicate delayed detection.
\end{tablenotes}

\end{threeparttable}
\end{table}

First, lower thresholds tend to detect recessions earlier, often generating smaller or negative detection errors. In many episodes, detection occurs either at the official start date or slightly in advance, although some variation across recessions remains.

Second, higher thresholds generally produce later detection relative to lower thresholds, with more frequent delays across episodes. However, detection timing is not uniformly later in all cases, and variation across recessions persists.

Third, the intermediate threshold  does not consistently improve performance across episodes. It produces a mix of early and delayed detections and does not uniformly reduce variation in detection timing relative to other thresholds.

Finally, the table reveals substantial heterogeneity across recession episodes. Some downturns are detected at similar times across thresholds, while others exhibit larger differences in detection timing depending on the threshold. Overall, the episode-level results demonstrate that the anticipation–precision tradeoff is not an artifact of aggregation, but arises consistently across recessions and varies with the choice of threshold.

\subsection{Selecting the Frontier Classifiers with Highest Precision}
The classifier ensemble is composed of the specifications on the anticipation–precision frontier with a standard deviation of detection errors below 3 months (Table~\ref{tab:classifier-ensemble}). All selected classifiers use simple moving average (SMA) smoothing, with smoothing parameters $\alpha$ between 6 and 9 months and turning horizons $\beta$ between 5 and 7 months. The Box–Cox transformation parameter $\gamma$ varies across specifications, while the indicators are combined linearly with high weights ($\delta \in [0.8, 0.9]$). The associated thresholds are small, ranging from 0.0005 to 0.0073.

The selected classifiers detect all recession episodes in the sample without generating false positives. The mean detection error ranges from $-0.45$ to $1.18$ months, with an average of $0.06$ months, indicating that, on average, the classifiers detect recessions slightly after onset. The standard deviation of detection errors ranges from 2.5 to 2.9 months, with an average of 2.72 months, implying that detection timing remains tightly clustered across episodes. 

Overall, the ensemble delivers recession signals that are timely and consistent, reflecting the restriction to the high-precision segment of the anticipation–precision frontier.

\begin{table}[H]
    \centering
    \begin{threeparttable}
    \caption{Classifier Ensemble Selected from the Anticipation--Precision Frontier}
    \label{tab:classifier-ensemble}
    \footnotesize
    \setlength{\tabcolsep}{6pt}
    \renewcommand{\arraystretch}{1.12}

    \begin{tabular}{ccccccccc}
        \toprule
        Smoothing &
        Smoothing &
        Curving &
        Turning &
        Mixing &
        Mixing &
        Threshold &
        Standard &
        Mean \\
        method &
        parameter &
        parameter &
        parameter &
        method &
        parameter &
         &
        error &
        error \\
        \midrule
        SMA & 6 & 0.8 & 8 & linear & 0.9 & 0.0005 & 2.52 & 1.18 \\
        SMA & 5 & 0.1 & 9 & linear & 0.8 & 0.0056 & 2.69 & 0.18 \\
        SMA & 5 & 0.0 & 9 & linear & 0.9 & 0.0081 & 2.71 & 0.09 \\
        SMA & 5 & 0.1 & 9 & linear & 0.8 & 0.0054 & 2.73 & -0.27 \\
        SMA & 5 & 0.0 & 9 & linear & 0.8 & 0.0073 & 2.74 & -0.36 \\
        SMA & 5 & 0.0 & 9 & linear & 0.8 & 0.0072 & 2.93 & -0.45 \\
        \midrule
        \textbf{Average} & \textbf{5.17} & \textbf{0.17} & \textbf{8.83} & \textbf{--} & \textbf{0.83} & \textbf{0.0057} & \textbf{2.72} & \textbf{0.06} \\
        \bottomrule
    \end{tabular}

    \begin{tablenotes}[flushleft]
        \footnotesize
        \item \textit{Notes:} Detection-error statistics are reported in months. \texttt{SMA} denotes simple moving average smoothing. The table reports classifiers selected from the anticipation--precision frontier subject to the restriction $\sigma(k) < 3$ months. The Box--Cox parameter $\gamma$ governs the transformation of the recession indicator, while the turning parameter $\beta$ determines the horizon used to compute turning points. Indicators are combined linearly using weight $\delta$. A recession is identified when the resulting indicator crosses the threshold $\zeta$ from below. Detection errors are defined as the difference, in months, between estimated and official recession start dates dated by the Economic and Social Research Institute (ESRI).
    \end{tablenotes}

    \end{threeparttable}
\end{table}

\clearpage
\begin{figure}[H]
    \centering
    \includegraphics[width=0.8\linewidth]{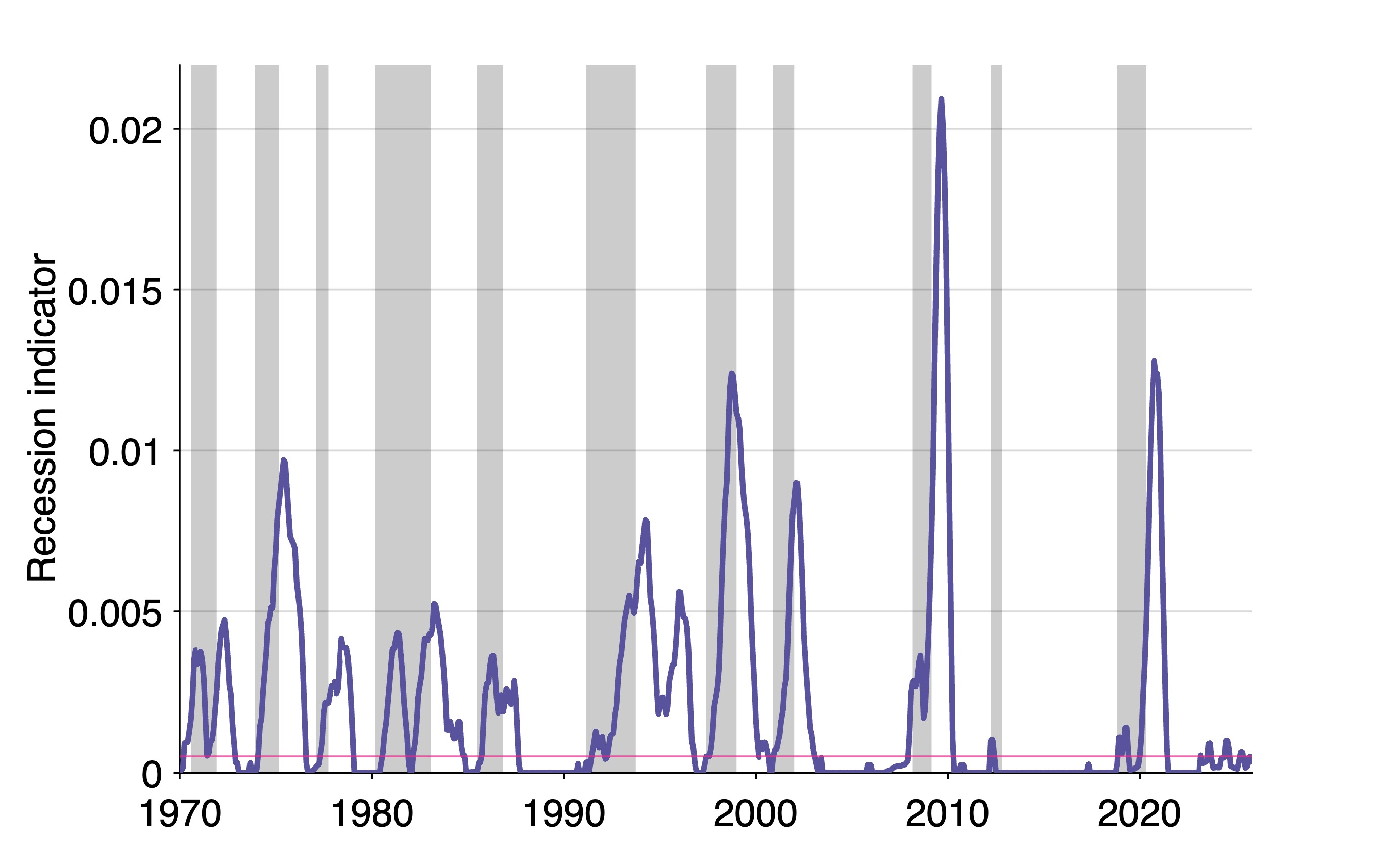}
    \caption{Classifier 1: Threshold 0.0005}
    \label{fig:ensemble1}

    \vspace{0.3em}
    \begin{minipage}{0.95\linewidth}
    \footnotesize
    \textit{Notes:} The classifier specification corresponds to the first classifier reported in Table~\ref{tab:classifier-ensemble}. The purple line represents the recession indicator underlying the classifier, while the pink line represents the corresponding recession threshold. Gray shaded regions indicate recessions dated by the Economic and Social Research Institute (ESRI).
    \end{minipage}
\end{figure}

\begin{figure}[H]
    \centering
    \includegraphics[width=0.8\linewidth]{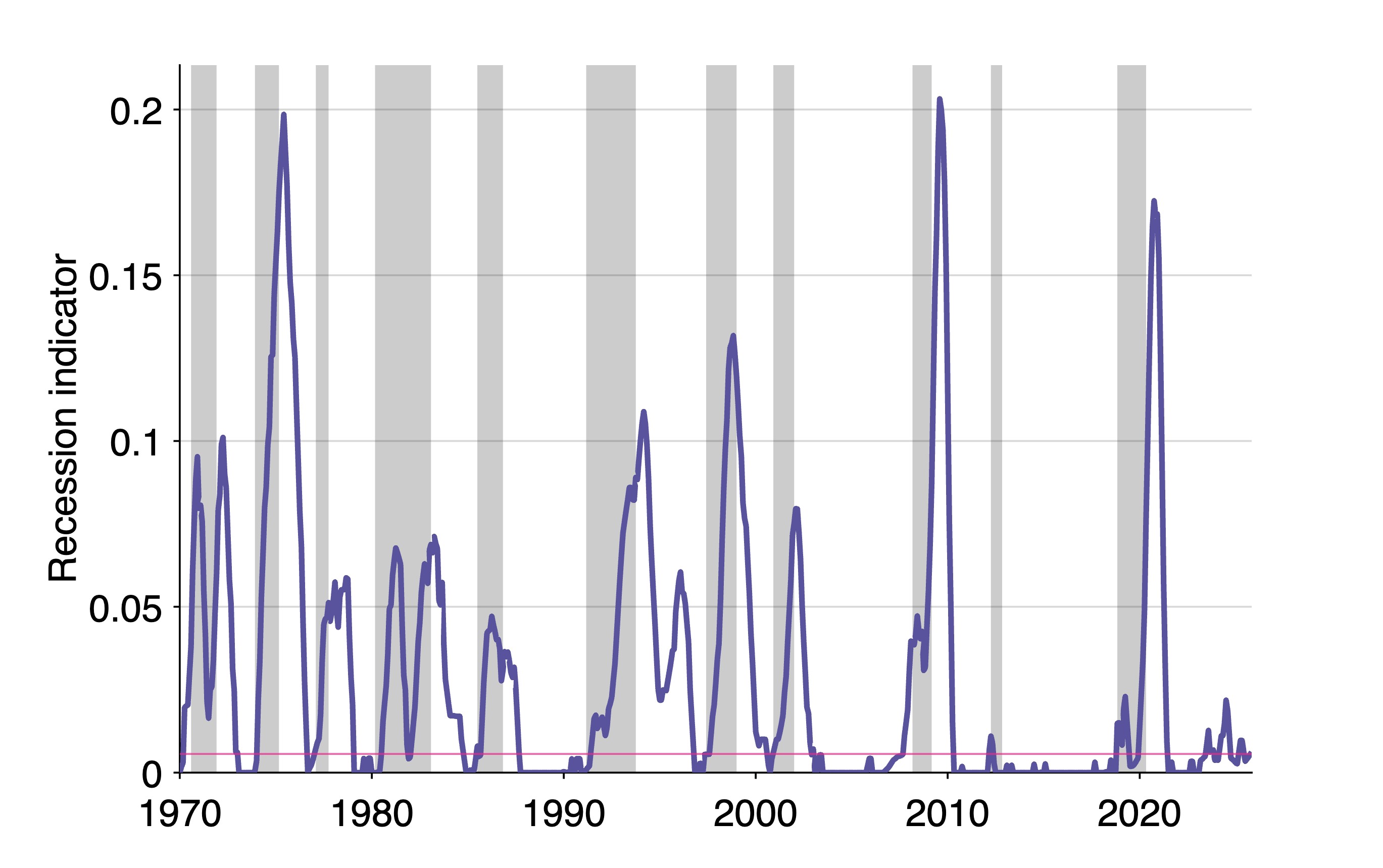}
    \caption{Classifier 2: Threshold 0.0056}
    \label{fig:ensemble2}

    \vspace{0.3em}
    \begin{minipage}{0.95\linewidth}
    \footnotesize
    \textit{Notes:} The classifier specification corresponds to the second classifier reported in Table~\ref{tab:classifier-ensemble}. The purple line represents the recession indicator underlying the classifier, while the pink line represents the corresponding recession threshold. Gray shaded regions indicate recessions dated by the Economic and Social Research Institute (ESRI).
    \end{minipage}
\end{figure}

\begin{figure}[H]
    \centering
    \includegraphics[width=0.8\linewidth]{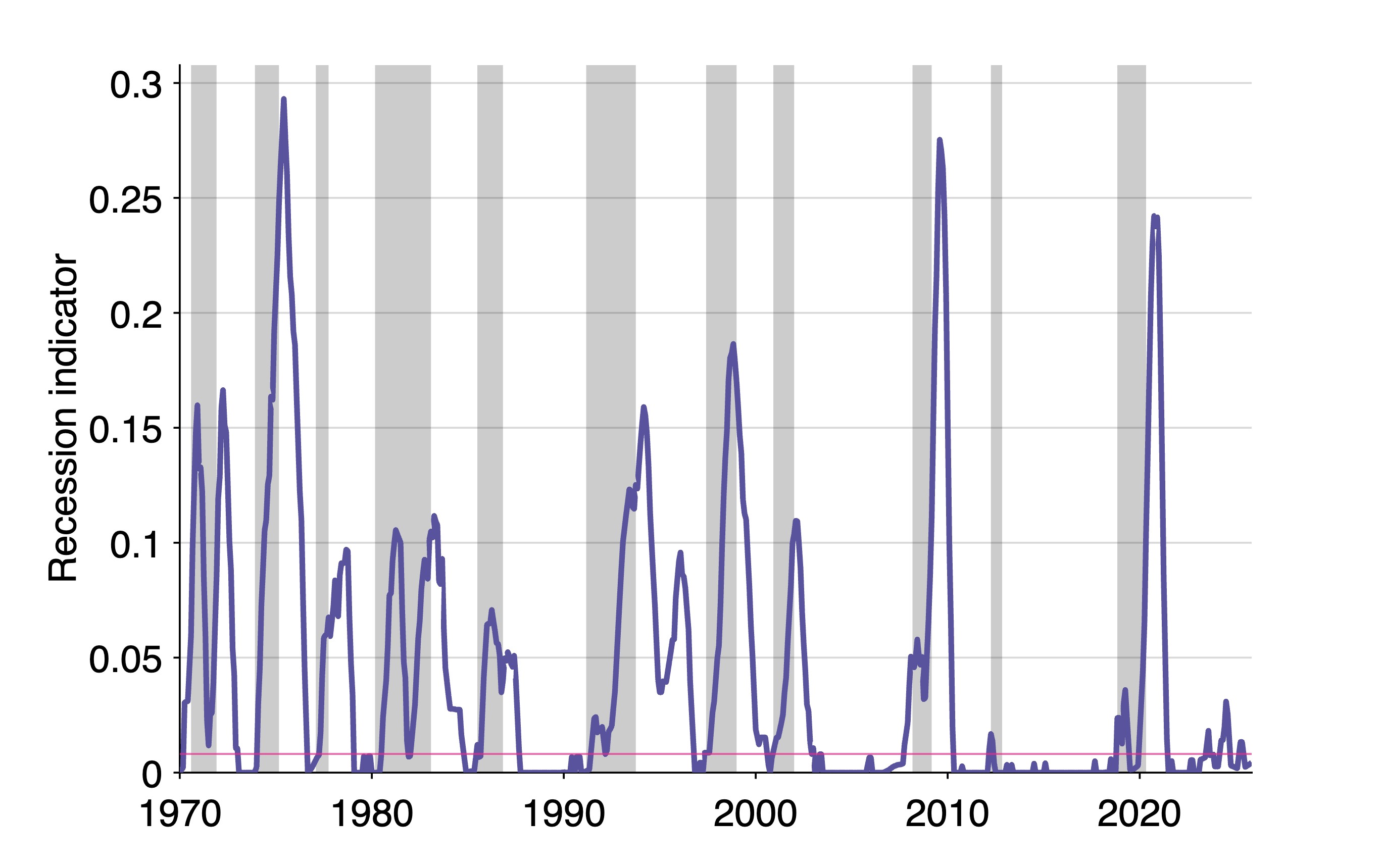}
    \caption{Classifier 3: Threshold 0.0081}
    \label{fig:ensemble3}

    \vspace{0.3em}
    \begin{minipage}{0.95\linewidth}
    \footnotesize
    \textit{Notes:} The classifier specification corresponds to the third classifier reported in Table~\ref{tab:classifier-ensemble}. The purple line represents the recession indicator underlying the classifier, while the pink line represents the corresponding recession threshold. Gray shaded regions indicate recessions dated by the Economic and Social Research Institute (ESRI).
    \end{minipage}
\end{figure}

\begin{figure}[H]
    \centering
    \includegraphics[width=0.8\linewidth]{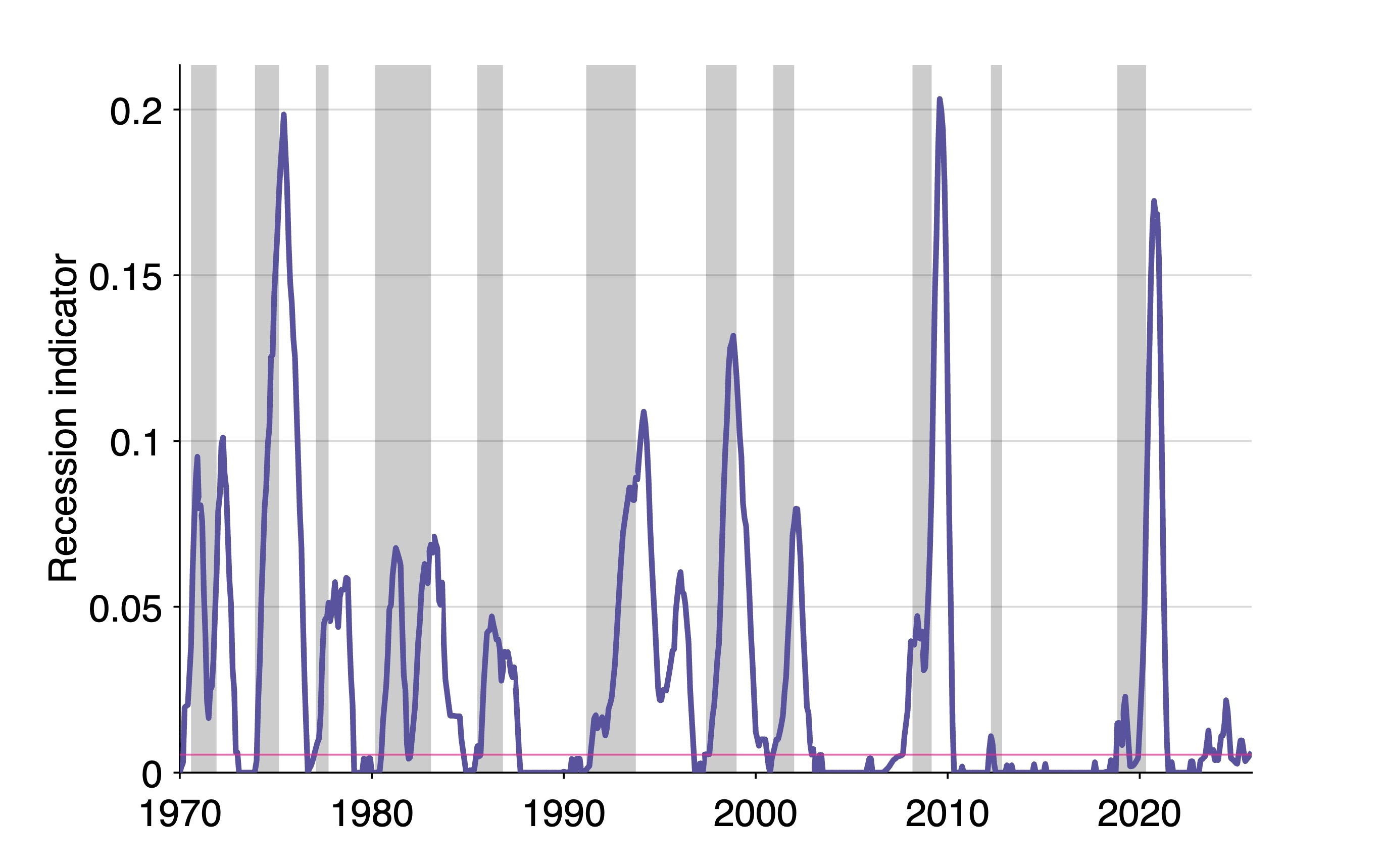}
    \caption{Classifier 4: Threshold 0.0054}
    \label{fig:ensemble4}

    \vspace{0.3em}
    \begin{minipage}{0.95\linewidth}
    \footnotesize
    \textit{Notes:} The classifier specification corresponds to the fourth classifier reported in Table~\ref{tab:classifier-ensemble}. The purple line represents the recession indicator underlying the classifier, while the pink line represents the corresponding recession threshold. Gray shaded regions indicate recessions dated by the Economic and Social Research Institute (ESRI).
    \end{minipage}
\end{figure}

\begin{figure}[H]
    \centering
    \includegraphics[width=0.9\linewidth]{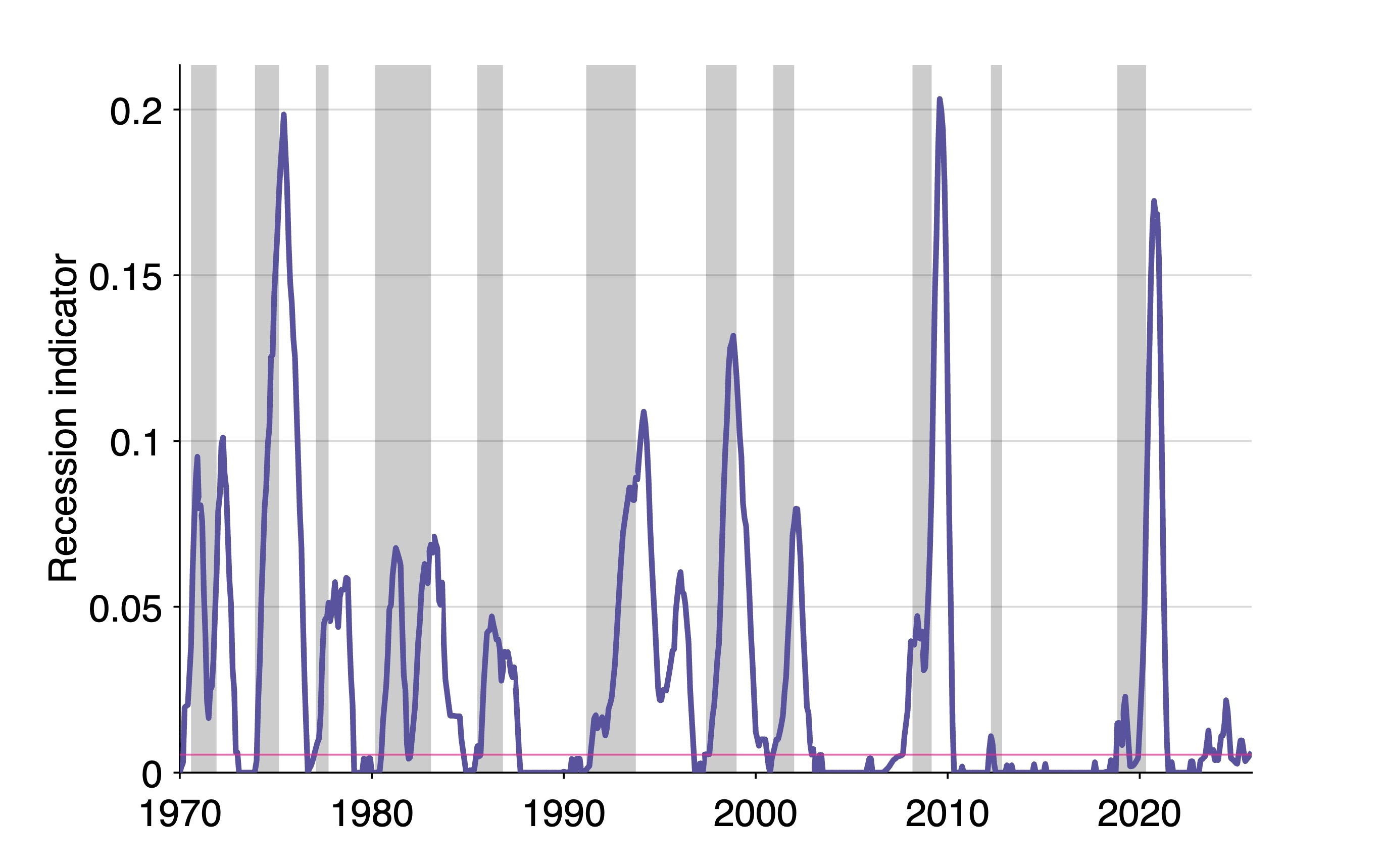}
    \caption{Classifier 5: Threshold 0.0073}
    \label{fig:ensemble5}

    \vspace{0.3em}
    \begin{minipage}{0.95\linewidth}
    \footnotesize
    \textit{Notes:} The classifier specification corresponds to the fifth classifier reported in Table~\ref{tab:classifier-ensemble}. The purple line represents the recession indicator underlying the classifier, while the pink line represents the corresponding recession threshold. Gray shaded regions indicate recessions dated by the Economic and Social Research Institute (ESRI).
    \end{minipage}
\end{figure}

\begin{figure}[H]
    \centering
    \includegraphics[width=0.9\linewidth]{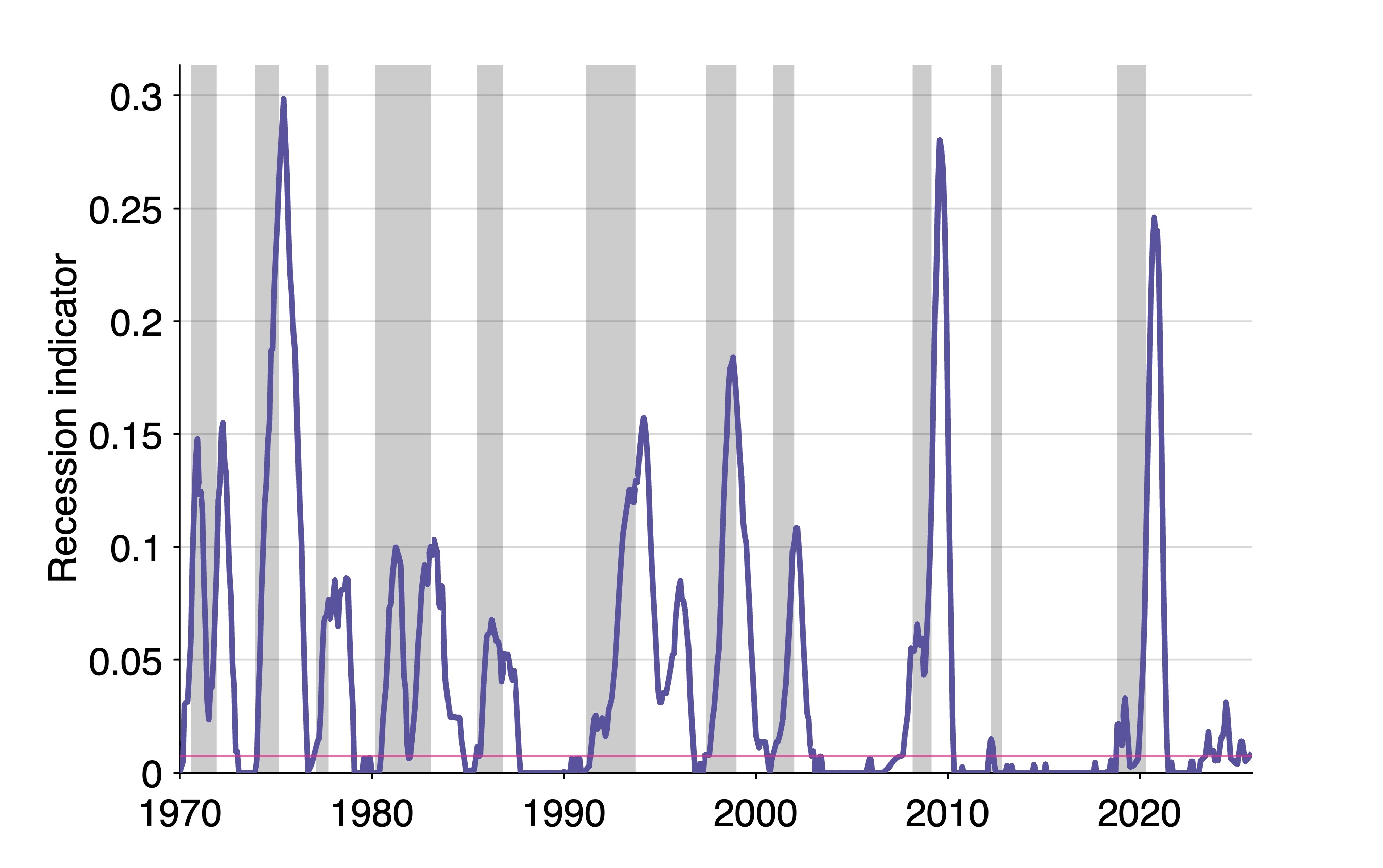}
    \caption{Classifier 6: Threshold 0.0072}
    \label{fig:ensemble6}

    \vspace{0.3em}
    \begin{minipage}{0.95\linewidth}
    \footnotesize
    \textit{Notes:} The classifier specification corresponds to the sixth classifier reported in Table~\ref{tab:classifier-ensemble}. The purple line represents the recession indicator underlying the classifier, while the pink line represents the corresponding recession threshold. Gray shaded regions indicate recessions dated by the Economic and Social Research Institute (ESRI).
    \end{minipage}
\end{figure}

\begin{figure}[H]
    \centering
    \includegraphics[width=0.9\linewidth]{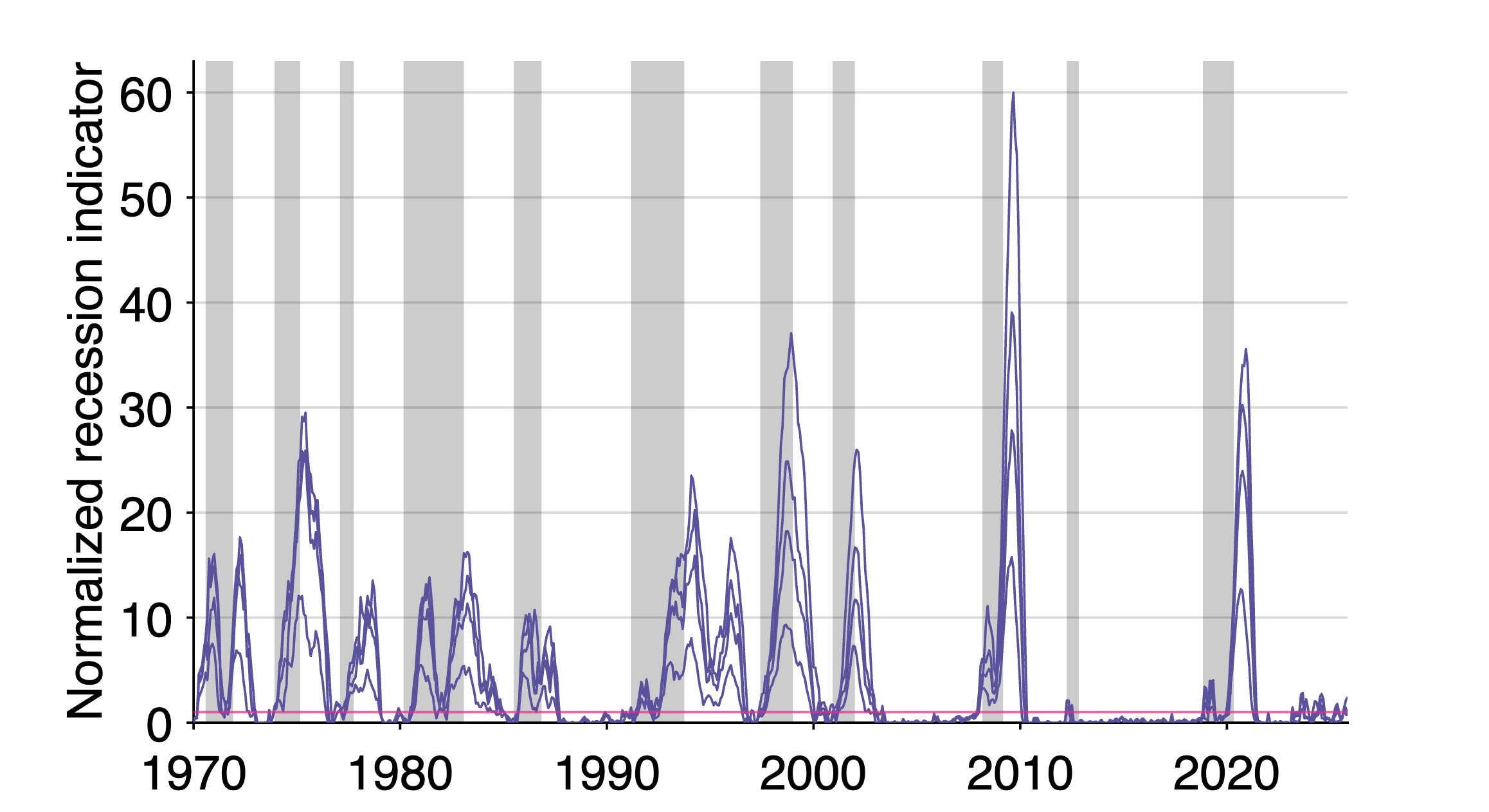}
    \caption{Full Set of All Six Recession Classifiers}
    \label{fig:ensemble_all}

    \vspace{0.3em}
    \begin{minipage}{0.95\linewidth}
    \footnotesize
    \textit{Notes:} The figure plots all classifiers reported in Table~\ref{tab:classifier-ensemble}. The purple lines represent the recession indicators underlying the classifiers, while the pink lines represent the corresponding recession thresholds. Gray shaded regions indicate recessions dated by the Economic and Social Research Institute (ESRI).
    \end{minipage}
\end{figure}

\clearpage
\section{Detecting Recessions}

To build the recession detection algorithm, we follow \cite{michaillat2025} to aggregate the detection signals produced by the classifier ensemble. We then use the aggregated signals to compute the current recession risk. 

\subsection{Empirical Construction of Recession Probabilities}

We follow \cite{michaillat2025} exactly. For each classifier $k$ and recession $j$ in the training sample, we compute the detection error
\[
\varepsilon_{k,j} = d_{k,j} - s_j,
\]
where $d_{k,j}$ is the classifier's detection date and $s_j$ is the ESRI Japanese recession start date as shown in Table ~\ref{tab:japan_recession_start_1970}. For each classifier, these historical errors are summarized by their sample mean $\mu_k$ and standard deviation $\sigma_k$, which capture systematic anticipation or delay and timing precision, respectively.

When classifier $k$ issues a real time detection at date $d_k$, we treat the historical detection errors as an empirical forecast-error sample and approximate their distribution by a normal distribution with mean $\mu_k$ and variance $\sigma_k^2$, exactly as in \citet{michaillat2025}. Letting
\[
\varepsilon_k = d_k - s,
\]
where $s$ is the unknown start date of the current recession, the event $\{s \le t\}$ is equivalent to $\{\varepsilon_k \ge d_k - t\}$. Therefore,
\[
\Pr(s \le t \mid d_k) = \Pr(\varepsilon_k \ge d_k - t).
\]
This probability is computed using only information from the training sample.\\

Whenever an individual recession indicator crosses its threshold, we infer the probability that the recession has already started from the distribution of the detection error. If classifier $k$ is on average exactly on time and the detection error is symmetric, then the recession has started with probability $0.5$ when the classifier is activated. If the classifier tends to detect recessions early on average, this probability is below $0.5$; if the classifier tends to detect recessions late on average, it is above $0.5$. In the months following detection, this probability converges to one according to the cumulative distribution function of the detection error.

More formally, suppose that classifier $k$ detects the $J$ recessions in the training sample. Each detection $j$ yields a detection date $d(k,j)$ and a detection error $\varepsilon(k,j)$. As in \citet{michaillat2025}, we assume that the detection error $\varepsilon(k)$ is normally distributed with mean $\mu(k)$ and standard deviation $\sigma(k)$:
\[
\varepsilon(k) \sim \mathcal{N}\bigl(\mu(k), \sigma^2(k)\bigr).
\]
Then, for any date $t \ge d(k)$, the probability that the start date $s$ of the new recession occurred before $t$, conditional on classifier $k$ having detected a recession at date $d(k)$, is
\[
P(k,t)
= \Pr\bigl(s < t \mid d(k), \mu(k), \sigma(k)\bigr)
= \Pr\bigl(d(k)-s > d(k)-t \mid \mu(k), \sigma(k)\bigr)
= \Pr\bigl(\varepsilon(k) > d(k)-t \mid \mu(k), \sigma(k)\bigr).
\]
Under the normal approximation, this becomes
\[
P(k,t)
= 1 - \Phi\!\left(\frac{d(k)-t-\mu(k)}{\sigma(k)}\right),
\]
where $\Phi(\cdot)$ denotes the cumulative distribution function of the standard normal distribution.

\subsection{Recession Probability from Classifier Ensemble}

\begin{figure}[t!]
    \centering
    \includegraphics[width=0.8\linewidth]{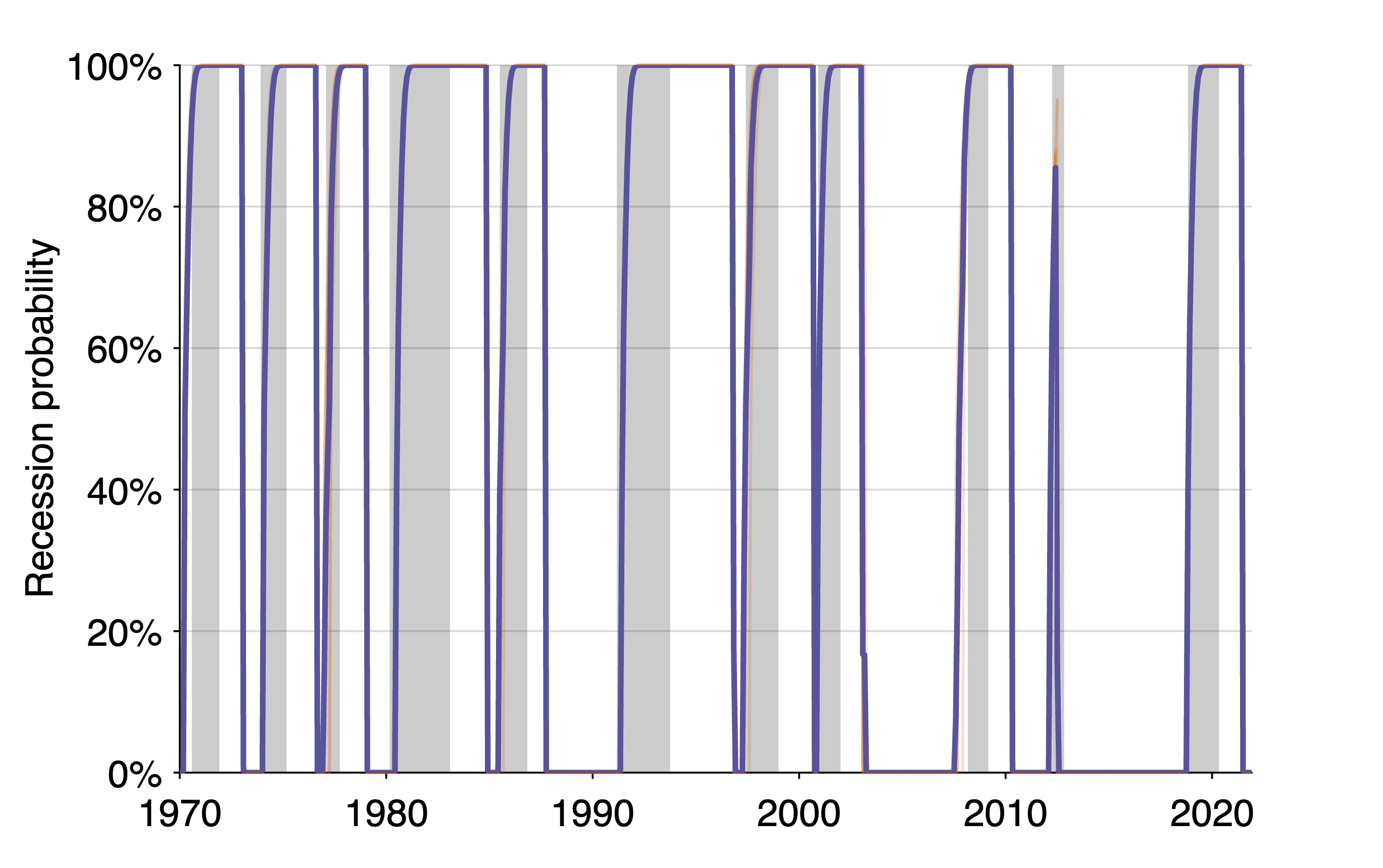}
    \caption{In-sample recession probability, January 1970--December 2021. The figure plots the estimated probability of a recession over time.}
    \label{fig:recession_probability}

    \vspace{0.3em}
  
    \footnotesize
    \textit{Notes:} The thick purple line shows the average probability across the six classifiers in the selected ensemble. The thin orange lines show the probability from each classifier separately. Shaded areas indicate recessions dated by the ESRI.

\end{figure}

\begin{figure}[H]
    \centering
    \includegraphics[width=0.8\linewidth]{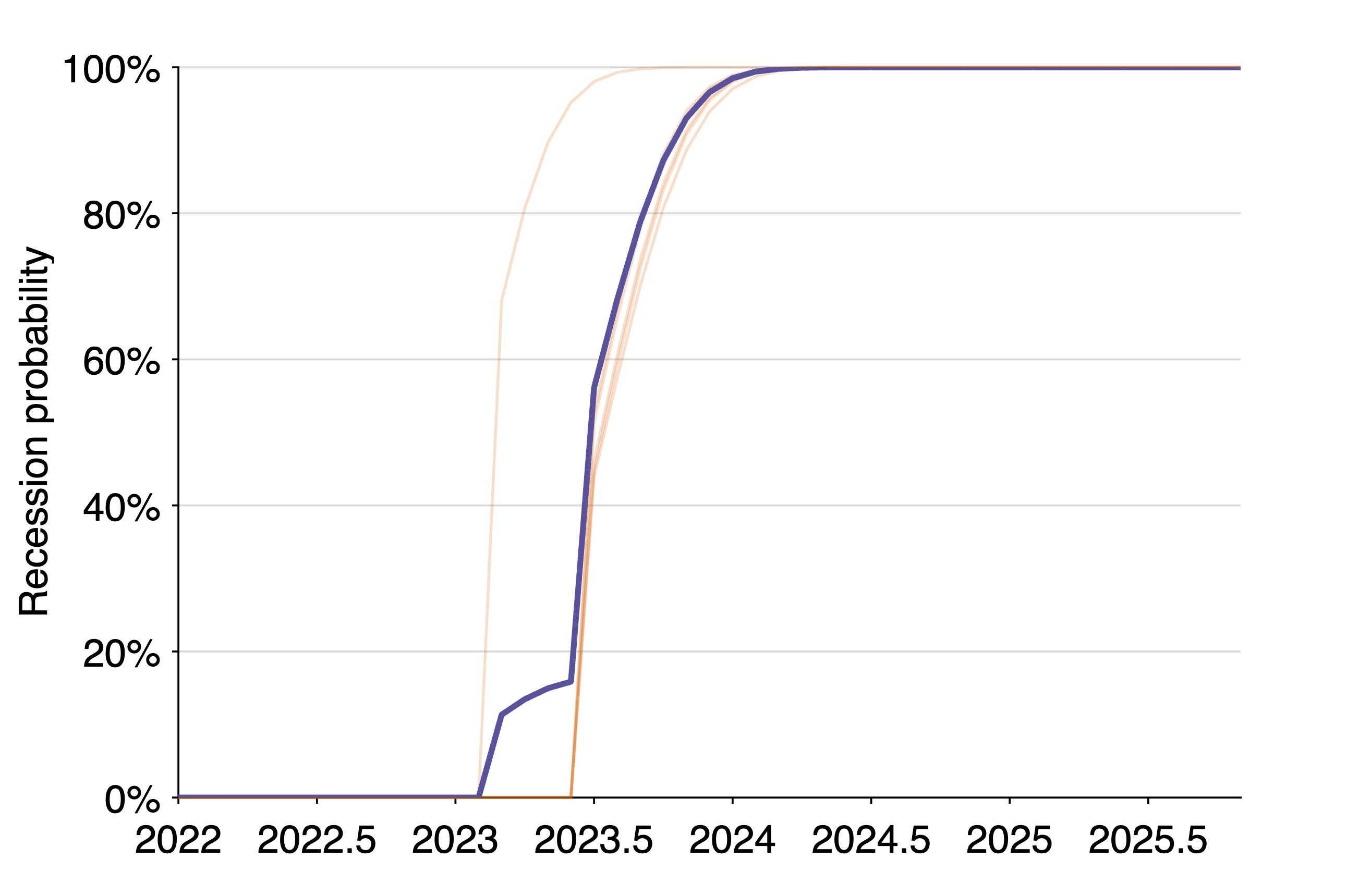}
    \caption{Out-of-sample recession probability, January 2022--December 2025}
    \label{fig:placeholder-2}
\end{figure}

Lastly, we summarize the information from the $K=6$ detection classifiers by constructing an ensemble recession probability. Each classifier produces a probability that a recession has started, and we average these probabilities to obtain a single aggregate measure. Figure ~\ref{fig:recession_probability} shows the resulting ensemble probability for the 1970--2021 training sample. For all 11 recessions, the probability rises in advance of the official start date, reaches 1 at the onset of the recession, and remains there until the recession has ended.

\subsection{Application of Recession Detection to Current Data}

Next, we apply the classifier estimated on the 1970-2021 training sample to the out-of-sample period 2022-2025 to assess current recession risk in the Japanese economy. The estimated recession probability first rises in 2023 and increases sharply after mid-2023, peaking in 2024 before easing somewhat. It rises again around mid-2024 and remains elevated through the end of the year. More recently, the probability begins to increase again around March 2025.

\section{Evaluating the recession detection algorithm with backtests}
To examine the reliability of the algorithm, we perform a series of out-of-sample backtests. In each exercise, the training sample is shortened and the algorithm is evaluated on recessions occurring after the training period. The results indicate that the detection algorithm performs robustly across different training windows. Moreover, the classifier ensembles obtained from these backtests consistently imply a meaningful probability that the Japanese economy entered a recession in 2025. 

\begin{table}[!htbp]
\centering
\caption{Performance of the algorithm on backtests}
\label{tab:backtest_performance}
\begin{tabular}{lccc}
\toprule
Backtest & 2015 & 2005 & 1995 \\
\midrule

Number of classifiers & 10 & 14 & 9 \\

\multicolumn{4}{l}{\textit{Training sample detection error (months)}} \\
Mean & 0.9 & 0.4 & -0.3  \\
Standard deviation & 2.7 & 2.4 & 2.1  \\
Minimum & -4.2 & -4.1 & -3.2  \\
Maximum & 4.6 & 4.1 & 3.1  \\

\multicolumn{4}{l}{\textit{Testing sample detection error (months)}} \\
Mean & 0.7 & -3.9 & 23.1  \\
Standard deviation & 0.0 & 5.0 & 25.5  \\
Minimum & 0.7 & -10.6 & -4.9  \\
Maximum & 0.7 & 0.6 & 63.8 \\
False positives & 0 & 1 & 2  \\
False negatives & 0 & 0 & 0 \\

Training recessions & 10 & 8 & 6 \\
Testing recessions & 1 & 3 & 5  \\

\bottomrule
\end{tabular}

\vspace{0.3cm}

\begin{minipage}{0.92\linewidth}
\footnotesize
The table reports the performance of the selected classifier ensembles in each backtest exercise. 
The training samples are Jan 1970--Dec 1994, 
Jan 1970--Dec 2004, and Jan 1970--Dec 2014. 
The testing periods are January 1995--December 2025, 
January 2005--December 2025, and January 2015--December 2025. 
Detection errors are measured in months relative to recession onset dates.
\end{minipage}

\end{table}

\subsection{Backtesting from 2015}

We train the algorithm using data through December 2014. The training sample contains 10 recessions, as summarized in Table 5. This procedure selects 10 classifiers from the high-precision segment of the anticipation–precision frontier, reported in Table A2. We then evaluate these classifiers out of sample using data from January 2015 through December 2025. During this test period, all 10 classifiers detect the 2020 recession with no false positives. The detection is also timely. On average, the recession is signaled only 0.7 months after its official onset.

\begin{figure}[H]
    \centering
    \includegraphics[width=0.8\linewidth]{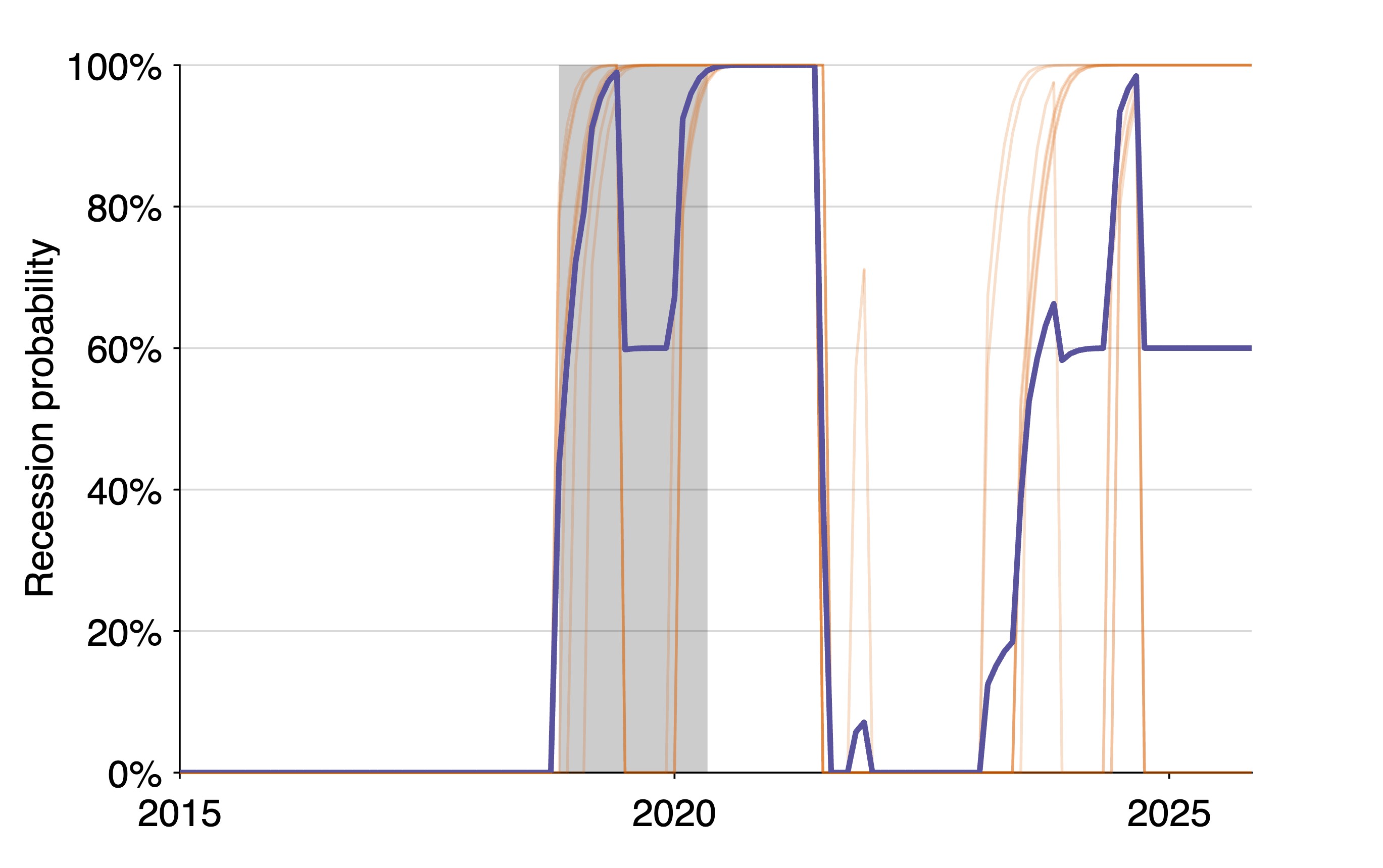}
    \caption{Out-of-sample Japan Recession Probability, January 2015--December 2025}
    \label{fig:placeholder-3}
\end{figure}

\subsection{Backtesting from 2005}

We train the algorithm using data through December 2004. The training sample contains eight recessions, as summarized in Table 5. This procedure selects 14 classifiers from the high-precision segment of the anticipation–precision frontier, reported in Table A3. We then evaluate these classifiers out of sample using data from January 2005 through December 2025.

The out-of-sample performance is particularly striking, as the classifiers exhibit anticipatory detection since the mean detection error is  -3.9 months, implying that, on average, the classifiers signal recessions nearly four months before the official recession onset dates. The standard deviation of the detection error is approximately five months, which remains reasonably small given the difficulty of predicting business-cycle turning points in real time.

\begin{figure}[H]
    \centering
    \includegraphics[width=0.8\linewidth]{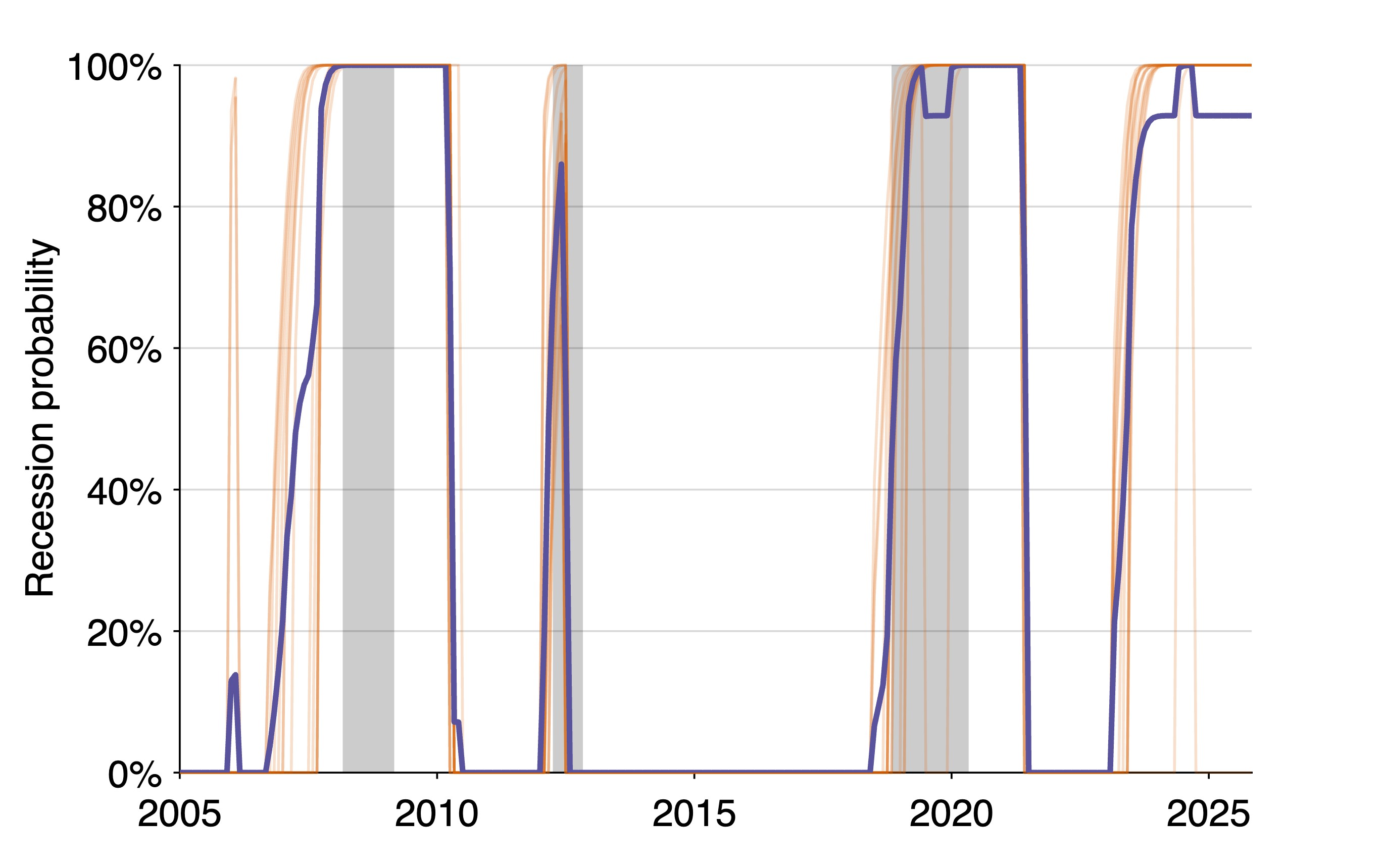}
    \caption{Out-of-sample Japan Recession Probability, January 2005--December 2025}
    \label{fig:placeholder-4}
\end{figure}

\subsection{Backtesting from 1995}

The backtesting exercise beginning in 1995 exhibits substantial noise, as shown in Figure 23, although the algorithm still successfully captures all recession episodes. However, the out-of-sample performance deteriorates considerably, with the mean detection error rising to 23.1 months, implying that recessions are identified nearly two years after their official onset on average. The exceptionally large standard deviation further suggests that the backtest calibrated through 1995 becomes unstable and loses predictive reliability. Therefore, we conclude the backtesting analysis at this point.

\begin{figure}[H]
    \centering
    \includegraphics[width=0.8\linewidth]{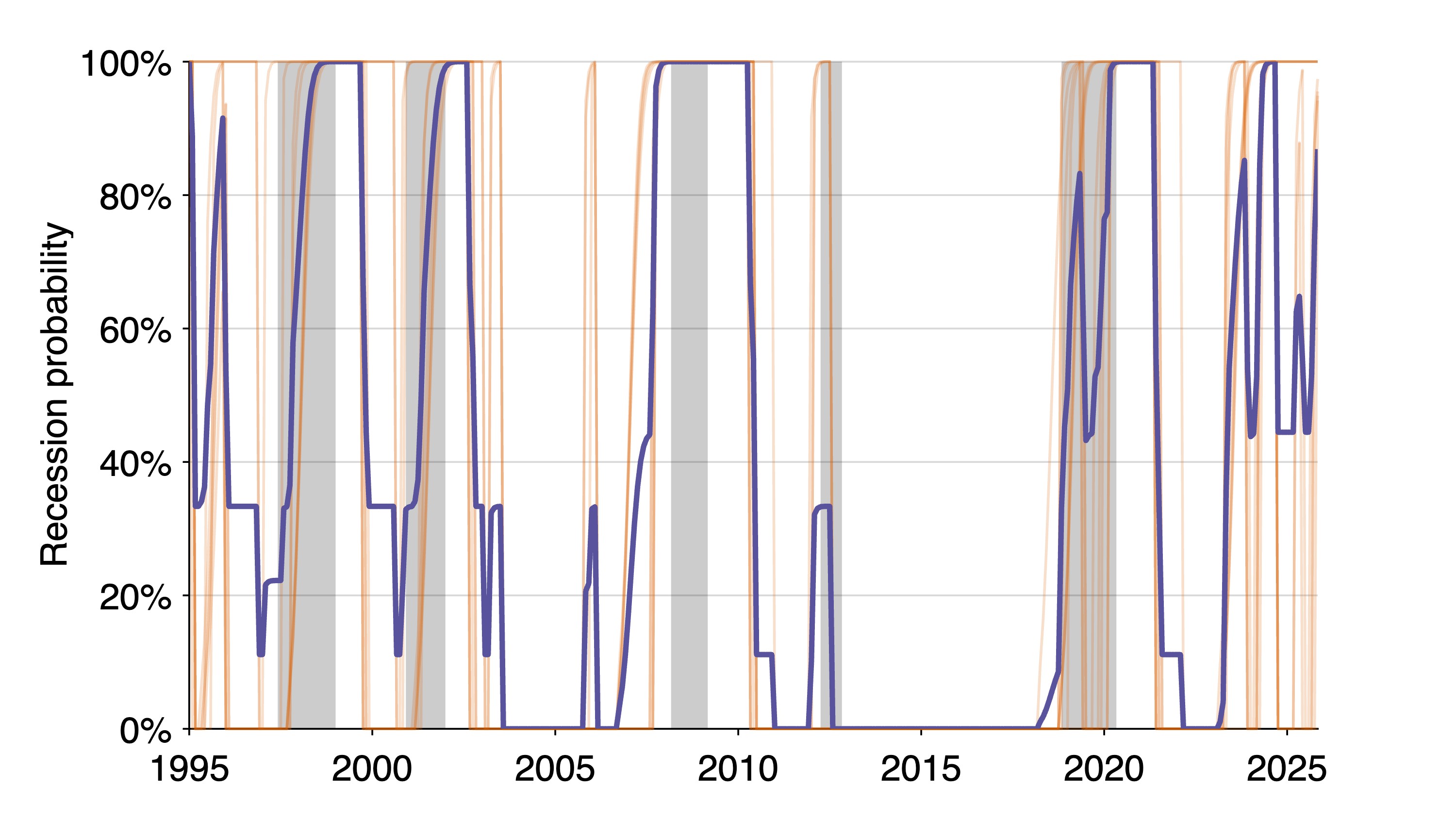}
    \caption{Out-of-sample Japan Recession Probability, January 1995--December 2025}
    \label{fig:placeholder-5}
\end{figure}

\subsection{Sensitivity Test: Excluding the 2012 recession episode}

We conduct an additional robustness exercise excluding the 2012 recession episode because its underlying economic dynamics may differ from those of more standard cyclical recessions. The slowdown occurred in the aftermath of the Great East Japan Earthquake and the post-Lehman recovery period, and therefore reflects a combination of reconstruction effects, temporary supply disruptions, and broader cyclical weakening. Since the recession detection framework is designed to identify labor-market based signals associated with typical business-cycle downturns rather than disaster-related shocks, the 2012 episode may represent a structurally atypical recession episode in the Japanese context. Excluding this episode allows us to examine the extent to which the detection results depend on this unusual historical event.

Consistent with this interpretation, even when the 2012 episode is included in the training sample, the model does not assign a near-100\% recession probability to that episode.

Nevertheless, excluding the 2012 recession substantially worsens the overall detection performance. The average detection timing deteriorates from nearly contemporaneous detection in the baseline specification, with an average delay of only 0.06 months after onset, to roughly two months after recession onset. In addition, the predicted probability of a recession in 2025 declines from nearly 100\% to around 80\%.

Overall, these findings suggest that, despite its atypical origin, the 2012 episode still provides economically meaningful information about recession dynamics in Japan and helps the model identify weaker and more gradual downturns that are characteristic of the Japanese business cycle.

The exercise of excluding 2012 asks whether the results are stable to excluding a weak and ambiguous official recession. Performance weakens, not because the labor-market signal disappears, but because excluding 2012 changes the exact-count classifier-selection problem. The selected classifiers are re-optimized to generate 10 signals instead of 11, producing a different frontier. This shows that the method is informative but sensitive to how ambiguous recession episodes are treated.
\begin{figure}[H]
    \centering
    \includegraphics[width=0.8\linewidth]{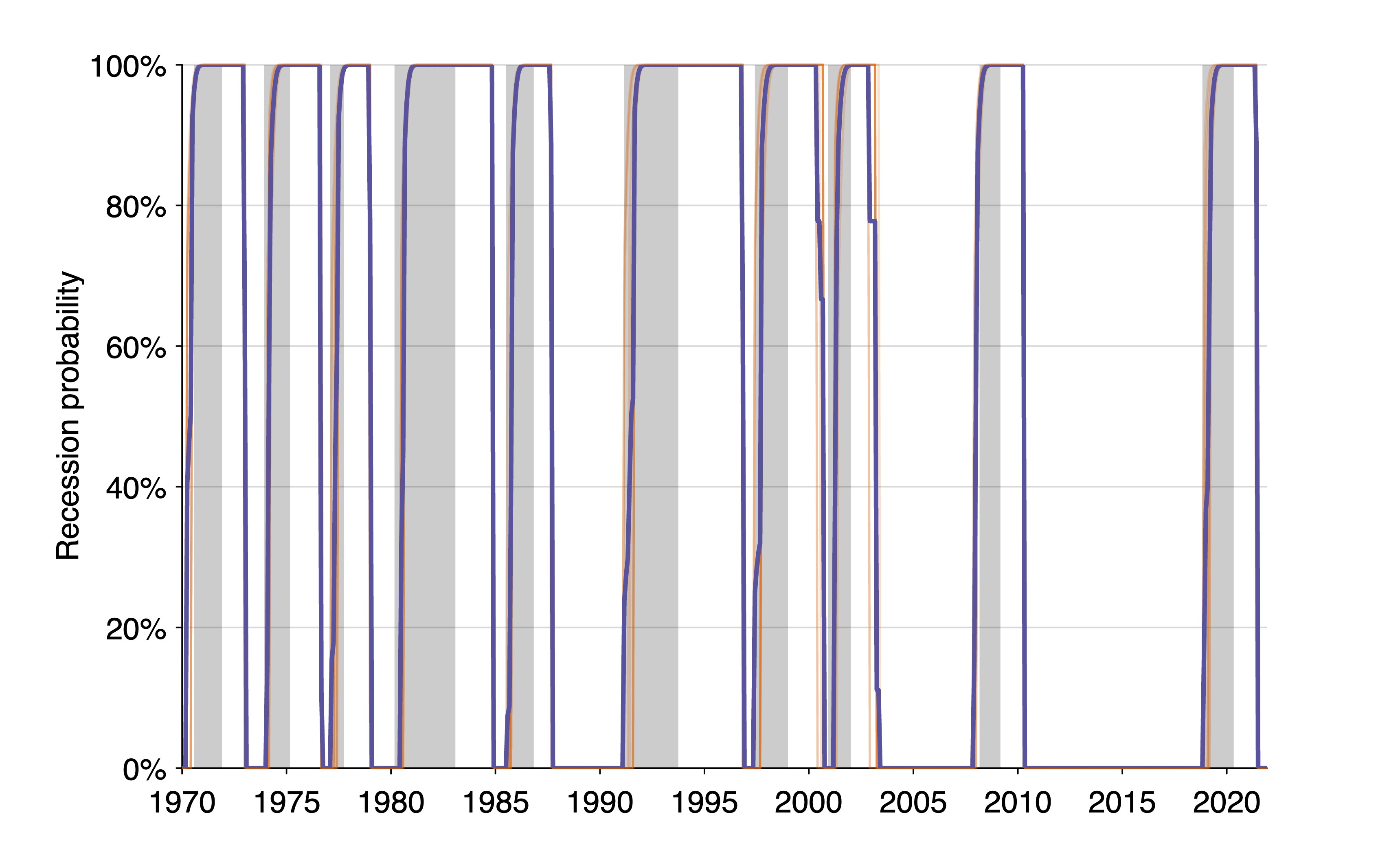}
    \caption{In-sample recession probability, January 1970--December 2021}
    \label{fig:placeholder-6}
\end{figure}

\begin{figure}[H]
    \centering
    \includegraphics[width=0.8\linewidth]{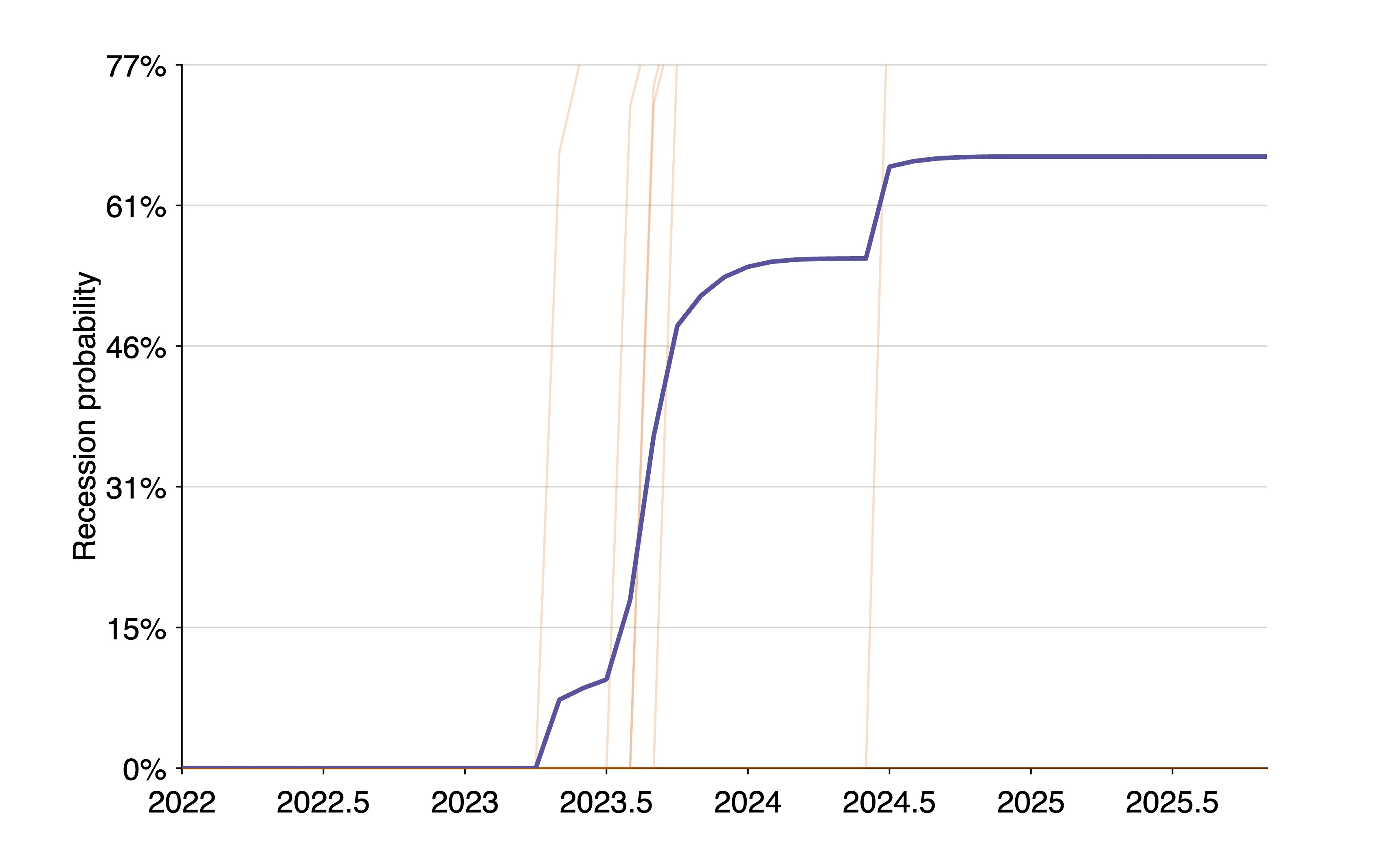}
    \caption{Out-of-sample recession probability, January 2022--December 2025}
    \label{fig:placeholder-7}
\end{figure}

\begin{figure}[H]
    \centering
    \includegraphics[width=0.8\linewidth]{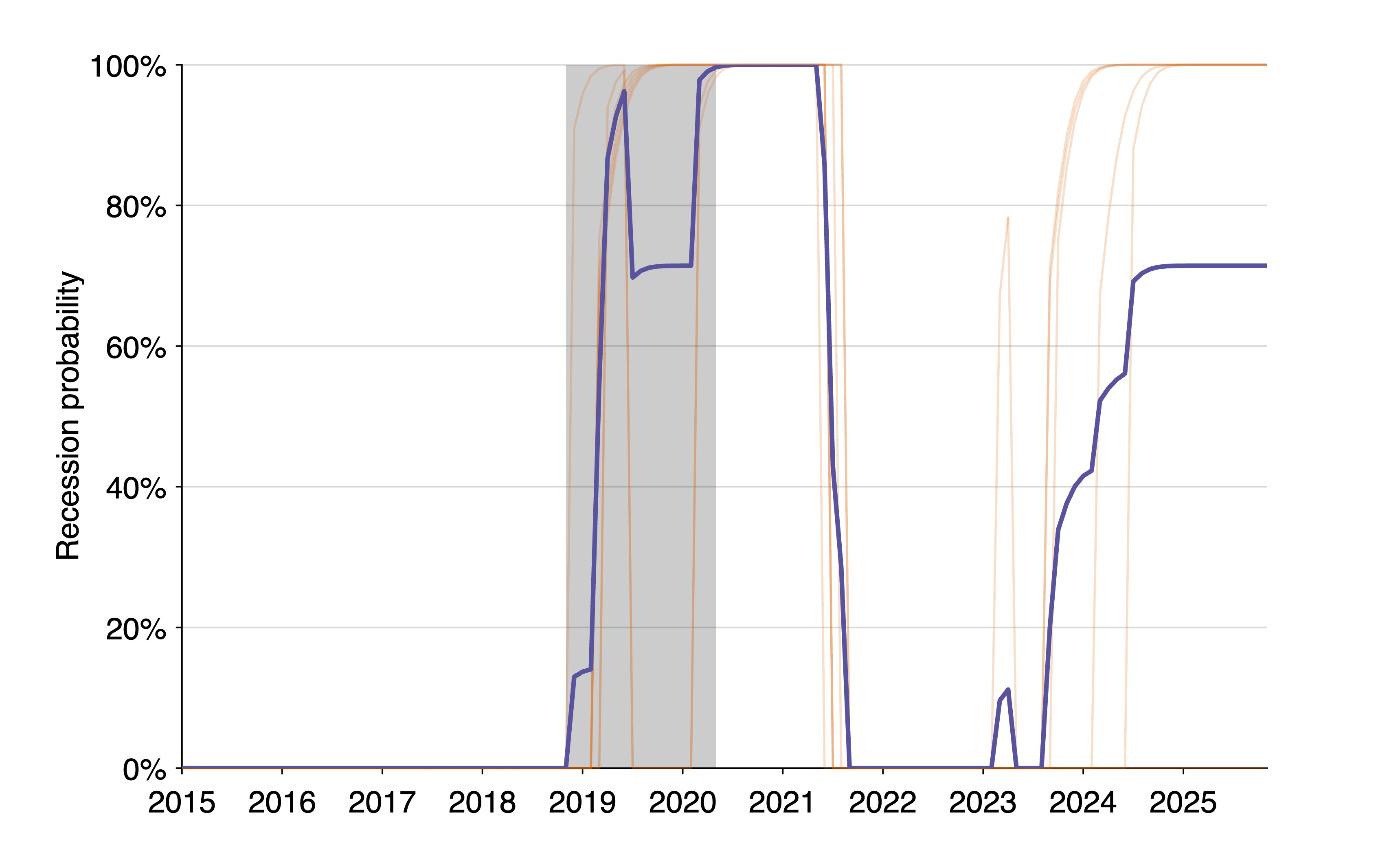}
    \caption{Out-of-sample recession probability, January 2015--December 2025}
    \label{fig:placeholder-8}
\end{figure}

\begin{figure}[H]
    \centering
    \includegraphics[width=0.8\linewidth]{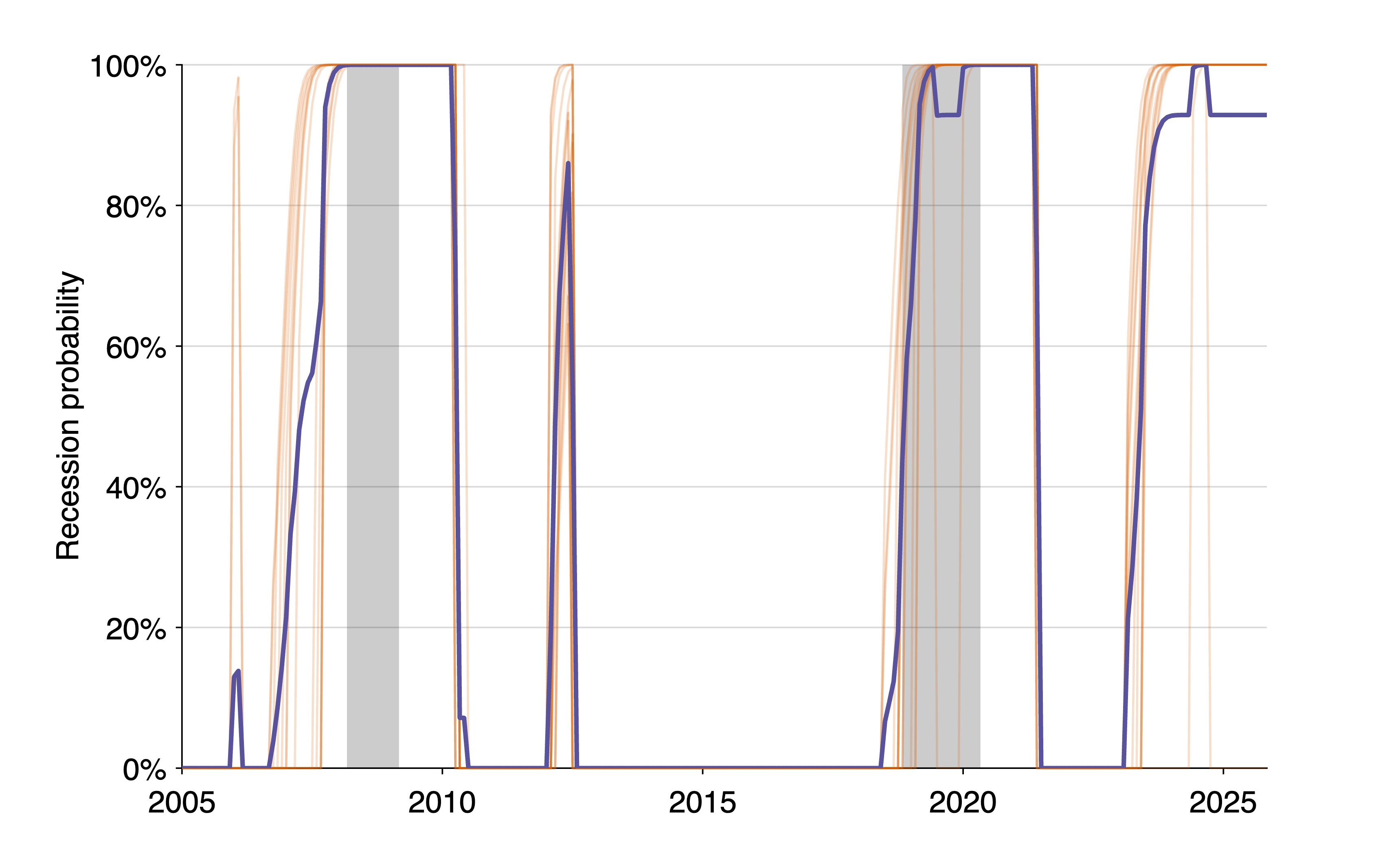}
    \caption{Out-of-sample recession probability, January 2005--December 2025}
    \label{fig:placeholder-9}
\end{figure}

\begin{figure}[H]
    \centering
    \includegraphics[width=0.8\linewidth]{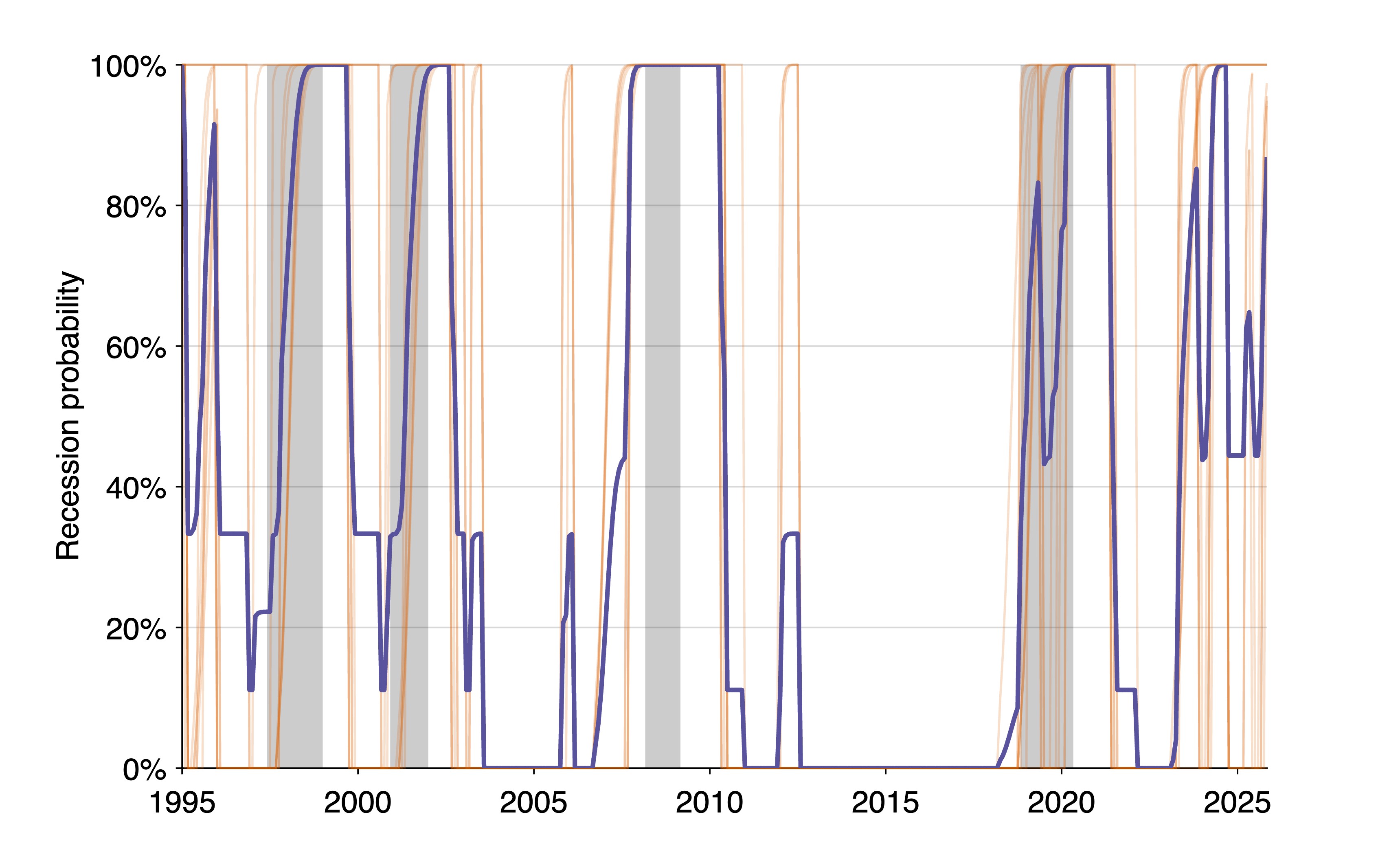}
    \caption{Out-of-sample recession probability, January 1995--December 2025}
    \label{fig:placeholder-10}
\end{figure}

\section{Conclusion}

This paper extends the algorithm developed by \citet{michaillat2025} to Japanese data. These results suggest that the algorithm provides a simple and effective framework for real time recession detection in Japan. The algorithm detects recessions both more accurately and far earlier than official ESRI
announcements, and it outperforms simpler benchmark rules such as the Sahm rule and the Michez rule. The algorithm is trained on data from January 1970 to December 2021, a period that includes 11 recessions. On the training sample, it performs very well. On average, the selected classifier ensemble signals recessions 0.06 months after their true onset, with a standard deviation of detection errors of 2.72 months.

These findings are also consistent with the evidence for the United States reported by \citet{michaillat2025}, where the classifier ensemble substantially improves upon official recession dating delays. Between 1979 and 2021, the classifier ensemble identified recession starts on average 1.2 months after onset, compared to 6.3 months for the National Bureau of Economic Research (NBER) Business Cycle Dating Committee. Taken together, these results suggest that classifier-based recession detection provides a robust and generalizable framework for identifying business-cycle turning points across economies.

An important avenue for future research is to extend the sample further back in time, which would allow the analysis to incorporate additional recession episodes and provide a more stringent test of the algorithm’s robustness. A longer historical sample may also improve the precision of the estimated classifiers by exposing them to a wider range of macroeconomic conditions.

More broadly, our results show that this classifier-based approach can be applied to other countries by calibrating thresholds and smoothing parameters to local labor market conditions, providing a practical framework for real time recession detection across different economies.

\bibliography{\bib}

\appendix
\section{Additional Selected Ensemble Detection Results}
\begin{table}[H]
\centering
\begin{threeparttable}

\caption{Detected Recession Starts by Threshold for the Main Ensemble (continued)}
\label{tab:mainensemblethresholds2}

\scriptsize
\setlength{\tabcolsep}{4pt}
\renewcommand{\arraystretch}{1.0}

\begin{tabular}{ccccccc}
\toprule
\makecell{Ensemble\\Member} & Threshold & Recession & \makecell{Official\\Start} & \makecell{Official\\Trough} & \makecell{Detection\\Start} & \makecell{Error\\(Months)} \\
\midrule

\multicolumn{7}{c}{\textit{Threshold = 0.0054}} \\
4 & 0.0054 & 1  & 1970.58 & 1971.92 & 1970.25 & -4 \\
4 & 0.0054 & 2  & 1973.92 & 1975.17 & 1974.08 & 2 \\
4 & 0.0054 & 3  & 1977.08 & 1977.75 & 1977.08 & 0 \\
4 & 0.0054 & 4  & 1980.17 & 1983.08 & 1980.50 & 4 \\
4 & 0.0054 & 5  & 1985.50 & 1986.83 & 1985.50 & 0 \\
4 & 0.0054 & 6  & 1991.17 & 1993.75 & 1991.42 & 3 \\
4 & 0.0054 & 7  & 1997.42 & 1999.00 & 1997.33 & -1 \\
4 & 0.0054 & 8  & 2000.92 & 2002.00 & 2000.92 & 0 \\
4 & 0.0054 & 9  & 2008.17 & 2009.17 & 2007.67 & -6 \\
4 & 0.0054 & 10 & 2012.25 & 2012.83 & 2012.17 & -1 \\
4 & 0.0054 & 11 & 2018.83 & 2020.33 & 2018.83 & 0 \\

\midrule
\multicolumn{7}{c}{\textit{Threshold = 0.0073}} \\
5 & 0.0073 & 1  & 1970.58 & 1971.92 & 1970.25 & -4 \\
5 & 0.0073 & 2  & 1973.92 & 1975.17 & 1974.08 & 2 \\
5 & 0.0073 & 3  & 1977.08 & 1977.75 & 1977.00 & -1 \\
5 & 0.0073 & 4  & 1980.17 & 1983.08 & 1980.50 & 4 \\
5 & 0.0073 & 5  & 1985.50 & 1986.83 & 1985.50 & 0 \\
5 & 0.0073 & 6  & 1991.17 & 1993.75 & 1991.42 & 3 \\
5 & 0.0073 & 7  & 1997.42 & 1999.00 & 1997.33 & -1 \\
5 & 0.0073 & 8  & 2000.92 & 2002.00 & 2000.92 & 0 \\
5 & 0.0073 & 9  & 2008.17 & 2009.17 & 2007.67 & -6 \\
5 & 0.0073 & 10 & 2012.25 & 2012.83 & 2012.17 & -1 \\
5 & 0.0073 & 11 & 2018.83 & 2020.33 & 2018.83 & 0 \\

\midrule
\multicolumn{7}{c}{\textit{Threshold = 0.0072}} \\
6 & 0.0072 & 1  & 1970.58 & 1971.92 & 1970.25 & -4 \\
6 & 0.0072 & 2  & 1973.92 & 1975.17 & 1974.08 & 2 \\
6 & 0.0072 & 3  & 1977.08 & 1977.75 & 1977.00 & -1 \\
6 & 0.0072 & 4  & 1980.17 & 1983.08 & 1980.50 & 4 \\
6 & 0.0072 & 5  & 1985.50 & 1986.83 & 1985.50 & 0 \\
6 & 0.0072 & 6  & 1991.17 & 1993.75 & 1991.42 & 3 \\
6 & 0.0072 & 7  & 1997.42 & 1999.00 & 1997.33 & -1 \\
6 & 0.0072 & 8  & 2000.92 & 2002.00 & 2000.92 & 0 \\
6 & 0.0072 & 9  & 2008.17 & 2009.17 & 2007.58 & -7 \\
6 & 0.0072 & 10 & 2012.25 & 2012.83 & 2012.17 & -1 \\
6 & 0.0072 & 11 & 2018.83 & 2020.33 & 2018.83 & 0 \\

\bottomrule
\end{tabular}

\begin{tablenotes}[flushleft]
\scriptsize
\item \textit{Notes:} Continued from Table~\ref{tab:mainensemblethresholds1}. The table reports recession detection dates for Japan under alternative ensemble thresholds. Official start dates correspond to the benchmark recession chronology dated by the Economic and Social Research Institute (ESRI). Detection start denotes the first month in which the recession indicator crosses the specified threshold. Detection errors are measured as the difference, in months, between the detected and official recession start dates. Negative values indicate early detection, while positive values indicate delayed detection.
\end{tablenotes}

\end{threeparttable}
\end{table}

\section{Additional Results for 1985-2015 Backtests}
\begin{figure}[H]
    \centering
    \includegraphics[width=0.8\linewidth]{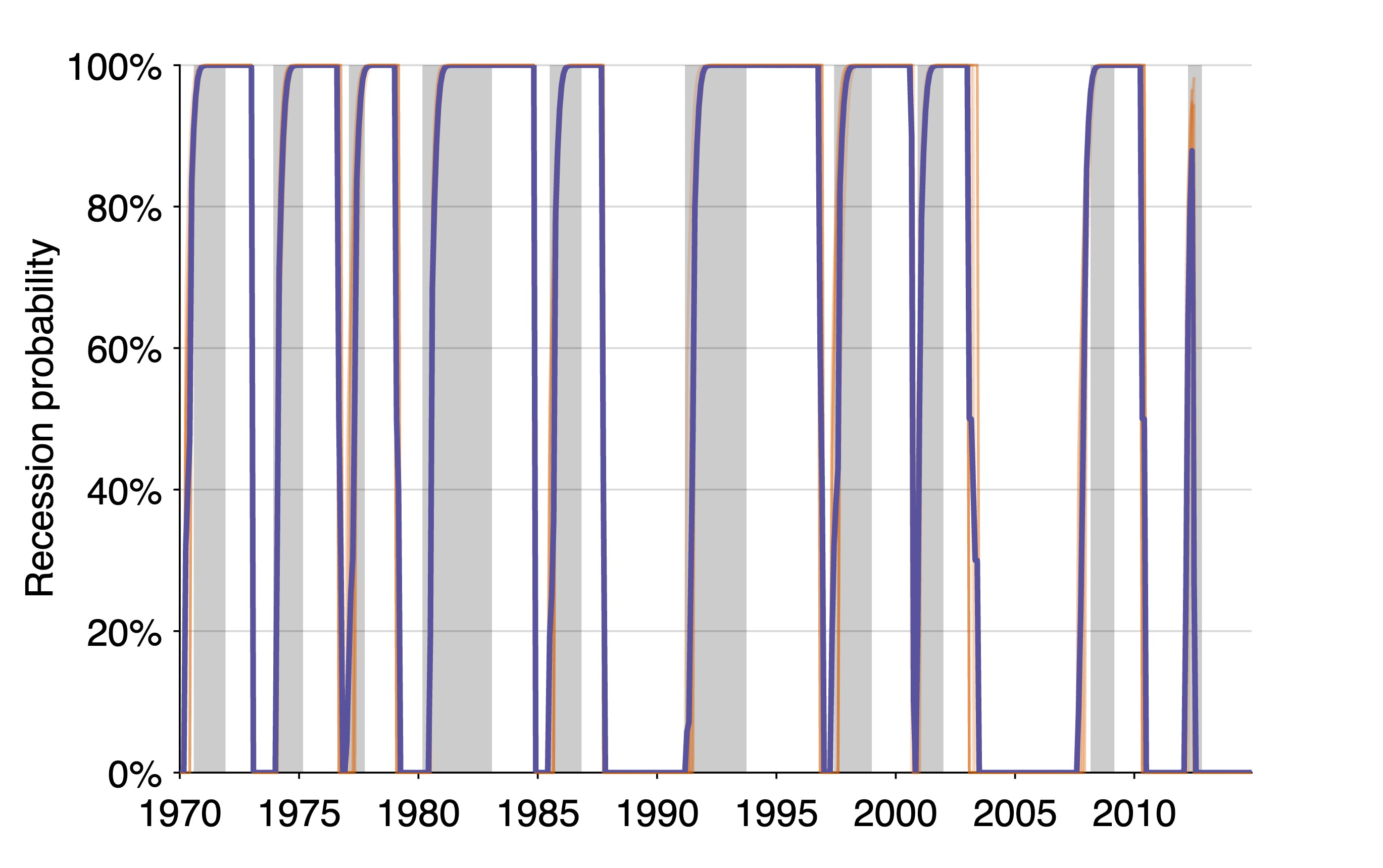}
    \caption{Out-of-sample Japan recession probability, January 2015--December 2025}
    \label{fig:placeholder-11}
\end{figure}

\begin{figure}[H]
    \centering
    \includegraphics[width=0.8\linewidth]{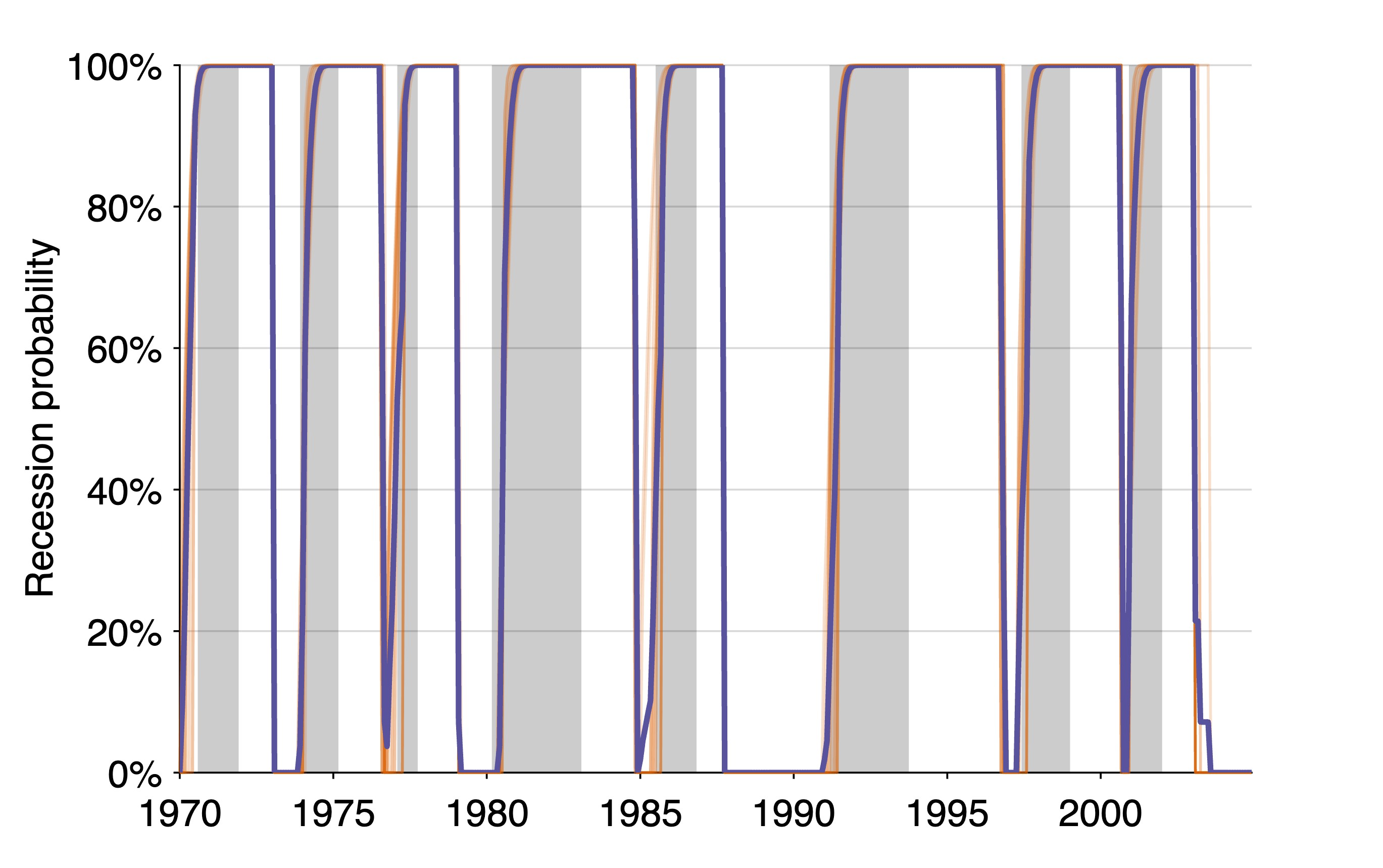}
    \caption{Out-of-sample Japan recession probability, January 2005--December 2025}
    \label{fig:placeholder-12}
\end{figure}

\begin{figure}[H]
    \centering
    \includegraphics[width=0.8\linewidth]{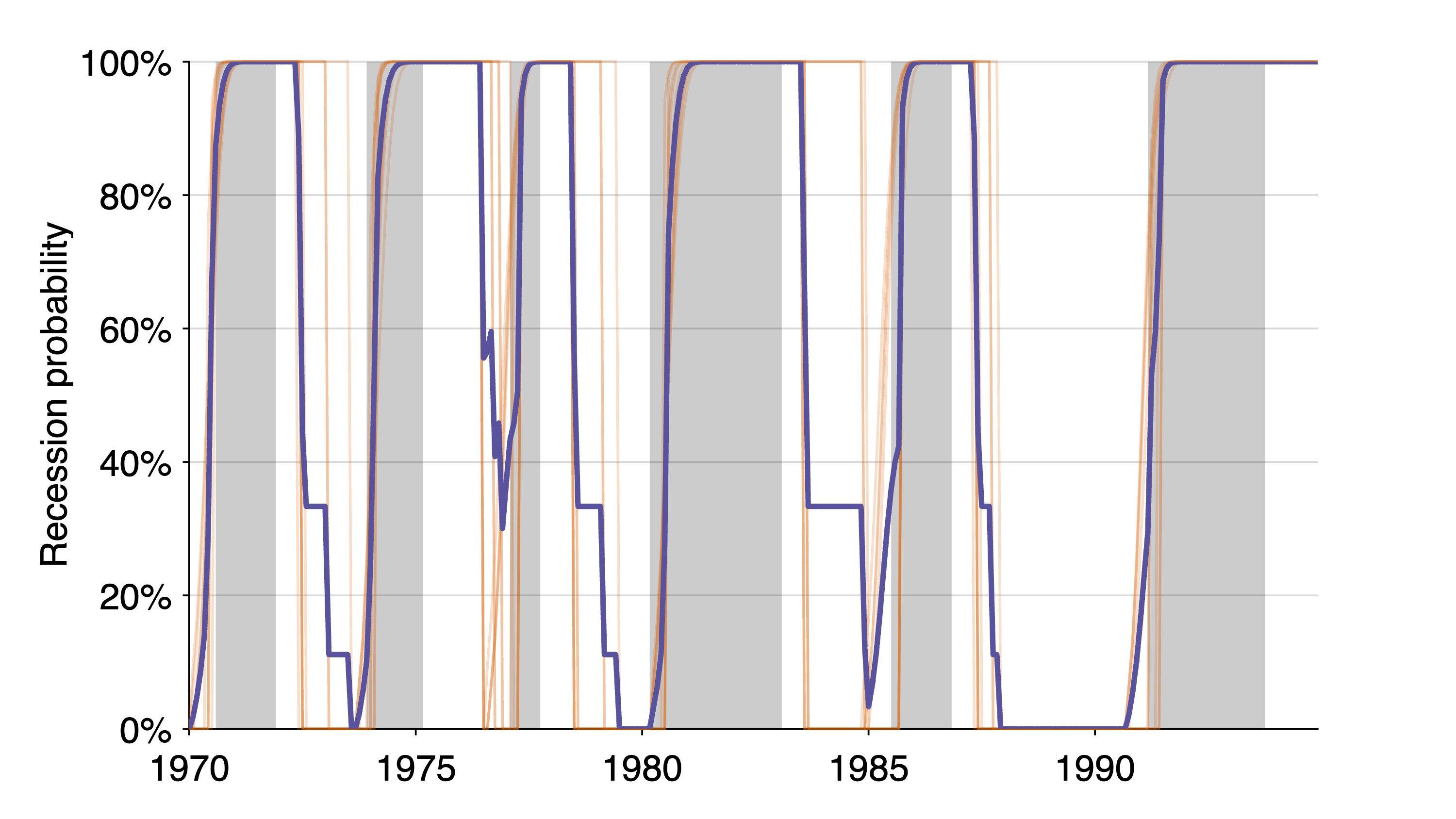}
    \caption{Out-of-sample Japan recession probability, January 1995--December 2025}
    \label{fig:placeholder-13}
\end{figure}

\begin{figure}[H]
    \centering
    \includegraphics[width=0.8\linewidth]{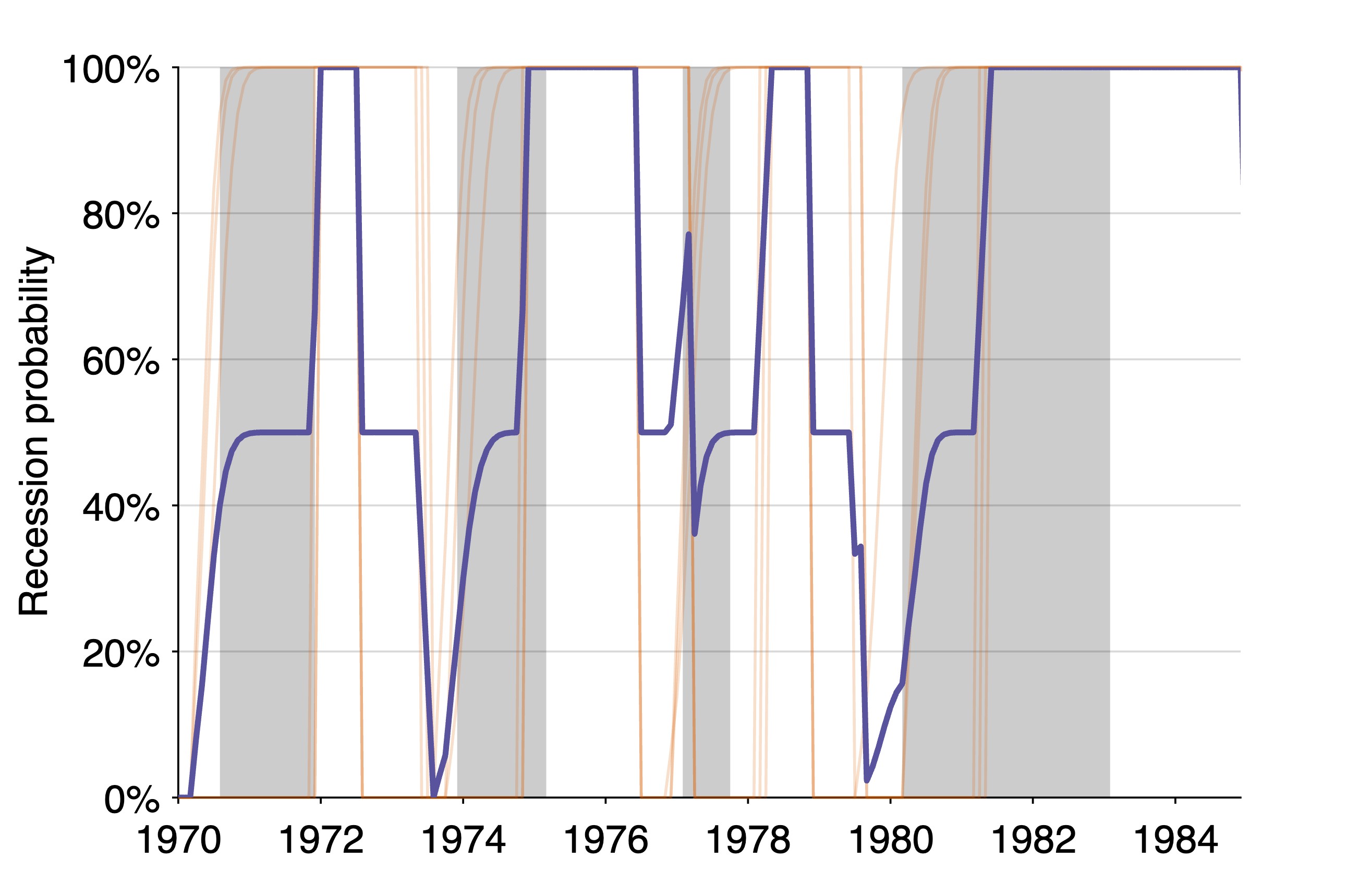}
    \caption{Out-of-sample Japan recession probability, January 1985--December 2025}
    \label{fig:placeholder-14}
\end{figure}

\begin{table}[h]
\centering
\footnotesize
\setlength{\tabcolsep}{5pt}
\renewcommand{\arraystretch}{1.1}

\caption{Classifier Ensemble Selected in the 2015 Backtest}
\label{tab:ensemble2015}

\begin{tabular}{cccccccccc}
\toprule

 & Smoothing & Smoothing & Curving & Turning & Mixing & Mixing & Threshold & Standard & Mean \\
 & method & parameter & parameter & parameter & method & parameter & value & error & error \\

\midrule

(1)  & SMA & 5 & 0.5 & 11 & linear & 1.0 & 0.0034 & 2.315 & 2.2 \\
(2)  & SMA & 5 & 0.6 & 11 & linear & 1.0 & 0.0023 & 2.508 & 2.1 \\
(3)  & SMA & 5 & 0.7 & 11 & linear & 1.0 & 0.0014 & 2.530 & 2.0 \\
(4)  & SMA & 4 & 0.3 & 11 & linear & 1.0 & 0.0080 & 2.617 & 1.5 \\
(5)  & SMA & 6 & 0.8 & 8  & linear & 0.9 & 0.0005 & 2.638 & 1.2 \\
(6)  & SMA & 7 & 0.0 & 9  & linear & 0.9 & 0.0043 & 2.693 & 0.5 \\
(7)  & SMA & 5 & 0.1 & 9  & linear & 0.8 & 0.0056 & 2.821 & 0.2 \\
(8)  & SMA & 5 & 0.0 & 9  & linear & 0.9 & 0.0081 & 2.844 & 0.1 \\
(9)  & SMA & 5 & 0.1 & 9  & linear & 0.8 & 0.0054 & 2.865 & -0.3 \\
(10) & SMA & 5 & 0.0 & 9  & linear & 0.8 & 0.0073 & 2.871 & -0.4 \\

\bottomrule
\end{tabular}

\vspace{0.2cm}

\begin{minipage}{0.92\linewidth}
\footnotesize
\textit{Notes:} The table reports the classifier ensemble selected in the 2015 backtest. The ensemble is restricted to classifiers satisfying a high-precision criterion based on the standard deviation of detection errors. Detection-error statistics are reported in months. \texttt{SMA} denotes simple moving average smoothing. The Box--Cox parameter $\gamma$ governs the transformation of the recession indicator, the turning parameter $\beta$ determines the turning horizon, and indicators are combined linearly using weight $\delta$. A recession is identified when the resulting indicator crosses the threshold $\zeta$ from below. Detection errors are defined as the difference, in months, between estimated and official recession start dates dated by the Economic and Social Research Institute (ESRI). 
\end{minipage}

\end{table}

\begin{table}[h]
\centering
\footnotesize
\setlength{\tabcolsep}{5pt}
\renewcommand{\arraystretch}{1.1}

\caption{Classifier Ensemble Selected in the 2005 Backtest}
\label{tab:ensemble2005}

\begin{tabular}{cccccccccc}
\toprule

 & Smoothing & Smoothing & Curving & Turning & Mixing & Mixing & Threshold & Standard & Mean \\
 & method & parameter & parameter & parameter & method & parameter & value & error & error \\

\midrule

(1)  & SMA & 4 & 0.6 & 11 & linear & 1.0 & 0.0026 & 1.728 & 2.625 \\
(2)  & SMA & 4 & 0.8 & 8  & linear & 0.8 & 0.0012 & 1.732 & 2.500 \\
(3)  & SMA & 4 & 0.7 & 8  & linear & 0.8 & 0.0016 & 1.996 & 2.375 \\
(4)  & SMA & 4 & 0.6 & 8  & linear & 0.8 & 0.0024 & 2.278 & 2.250 \\
(5)  & SMA & 4 & 1.0 & 8  & minmax & 0.9 & 0.0001 & 2.288 & 0.625 \\
(6)  & SMA & 5 & 0.1 & 9  & linear & 0.8 & 0.0054 & 2.345 & 0.500 \\
(7)  & SMA & 5 & 0.0 & 9  & linear & 0.8 & 0.0072 & 2.395 & 0.375 \\
(8)  & SMA & 6 & 0.5 & 8  & linear & 0.7 & 0.0009 & 2.421 & 0.125 \\
(9)  & SMA & 6 & 0.3 & 8  & linear & 0.6 & 0.0018 & 2.449 & 0.000 \\
(10) & SMA & 5 & 0.0 & 9  & linear & 0.6 & 0.0054 & 2.643 & -0.375 \\
(11) & SMA & 6 & 0.5 & 8  & linear & 0.4 & 0.0005 & 2.739 & -0.500 \\
(12) & SMA & 5 & 0.0 & 9  & minmax & 0.9 & 0.0009 & 2.781 & -0.625 \\
(13) & SMA & 5 & 0.0 & 9  & linear & 0.3 & 0.0027 & 2.817 & -1.750 \\
(14) & SMA & 5 & 0.0 & 9  & linear & 0.2 & 0.0019 & 2.958 & -2.000 \\

\bottomrule
\end{tabular}

\vspace{0.2cm}

\begin{minipage}{0.92\linewidth}
\footnotesize
\textit{Notes:} The ensemble comprises the classifiers selected from the anticipation--precision frontier in the 2005 backtest. The ensemble is restricted to classifiers satisfying a high-precision criterion based on the standard deviation of detection errors. Detection-error statistics are reported in months. \texttt{SMA} denotes simple moving average smoothing. The Box--Cox parameter $\gamma$ governs the transformation of the recession indicator, the turning parameter $\beta$ determines the turning horizon, and indicators are combined using either linear weighting or the minimum--maximum (\texttt{minmax}) method with weight $\delta$. A recession is identified when the resulting indicator crosses the threshold $\zeta$ from below. Detection errors are defined as the difference, in months, between estimated and official recession start dates dated by the Economic and Social Research Institute (ESRI).
\end{minipage}

\end{table}

\section{Additional results for the 1985–2015 backtests excluding the 2012 recession episode}

\begin{figure}[H]
    \centering
    \includegraphics[width=0.8\linewidth]{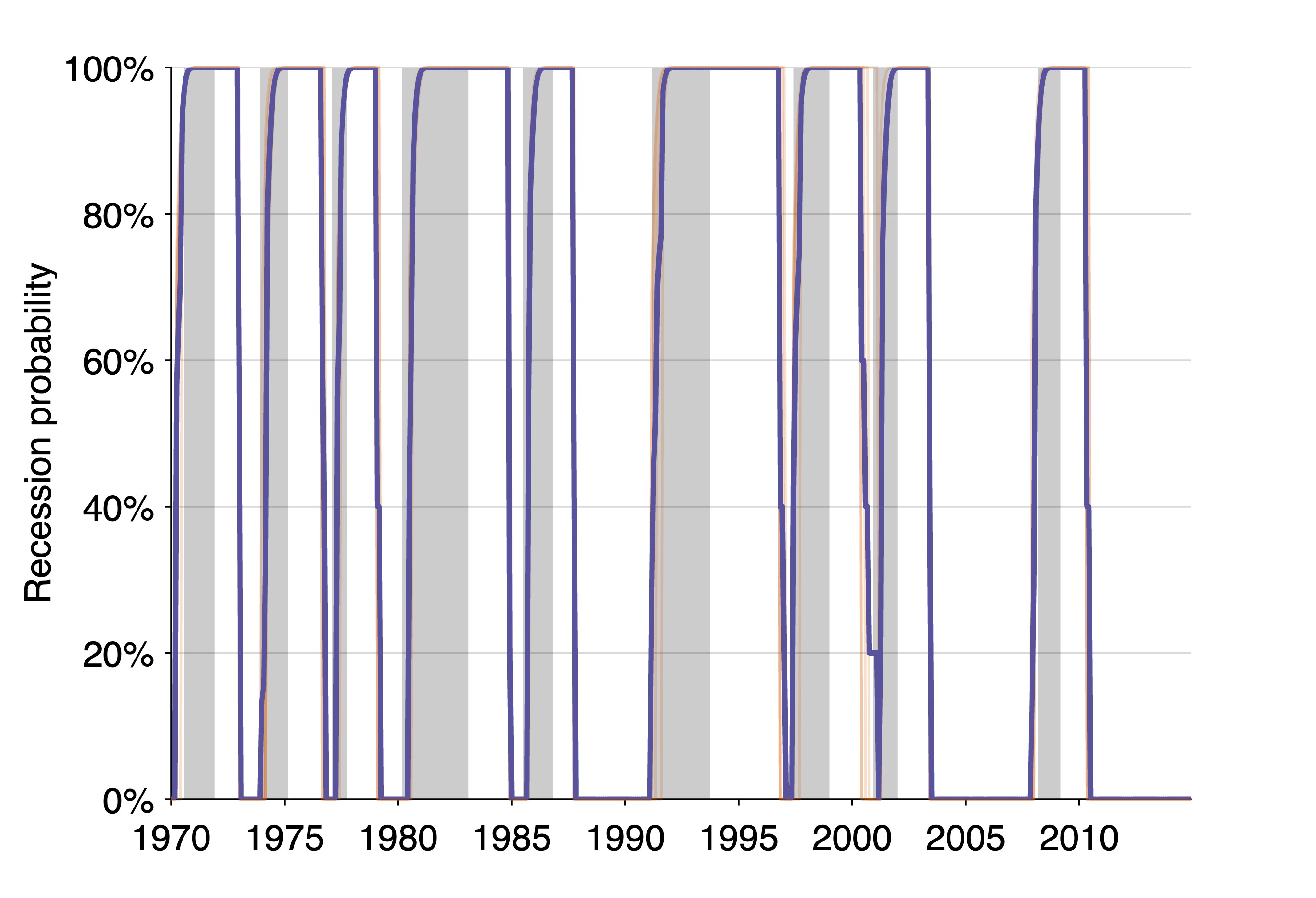}
    \caption{In-sample recession probability, January 1970--December 2014}
    \label{fig:placeholder-15}
\end{figure}
\begin{figure}[H]
    \centering
    \includegraphics[width=0.8\linewidth]{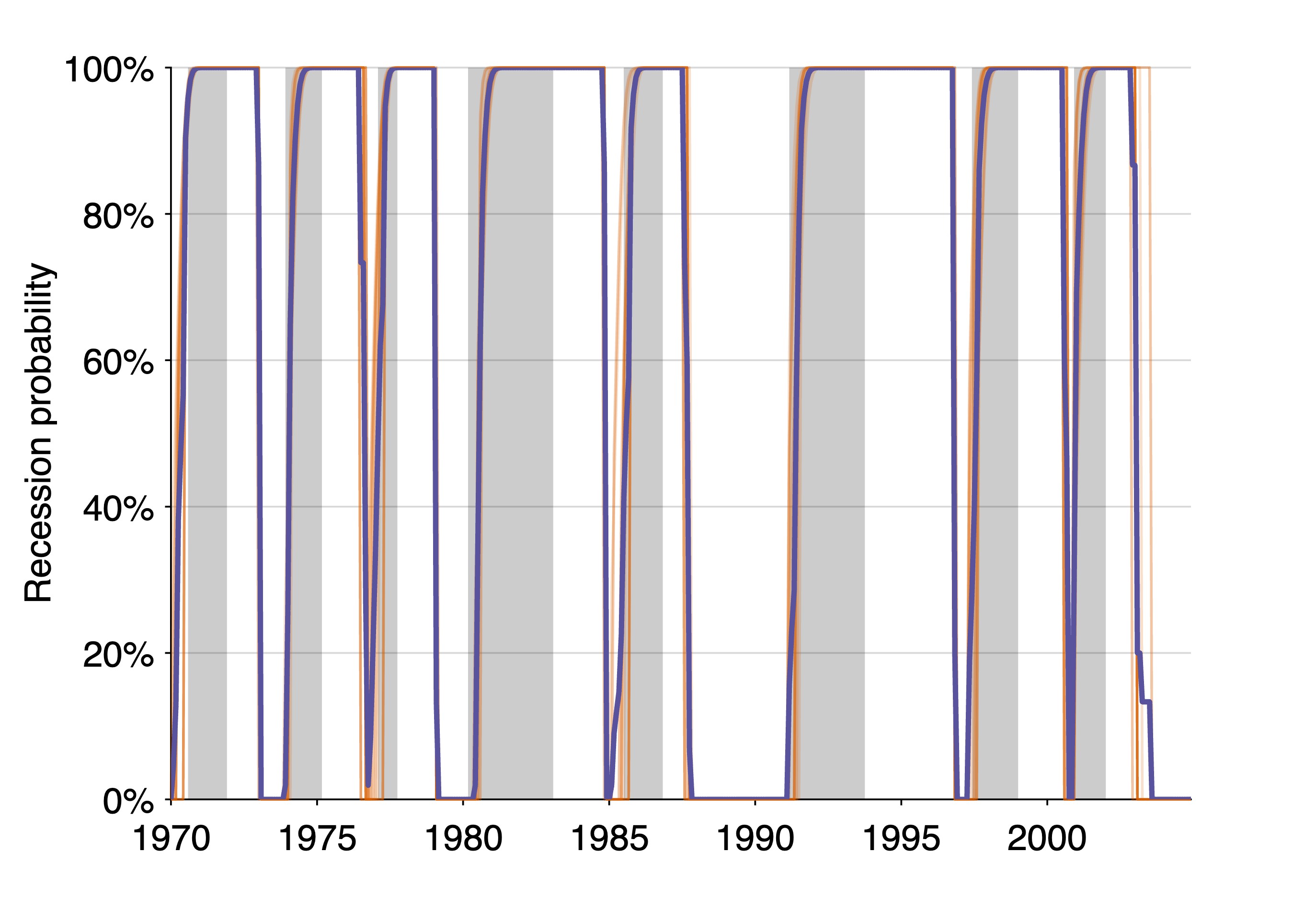}
    \caption{In-sample recession probability, January 1970--December 2004}
    \label{fig:placeholder-16}
\end{figure}

\begin{figure}[H]
    \centering
    \includegraphics[width=0.8\linewidth]{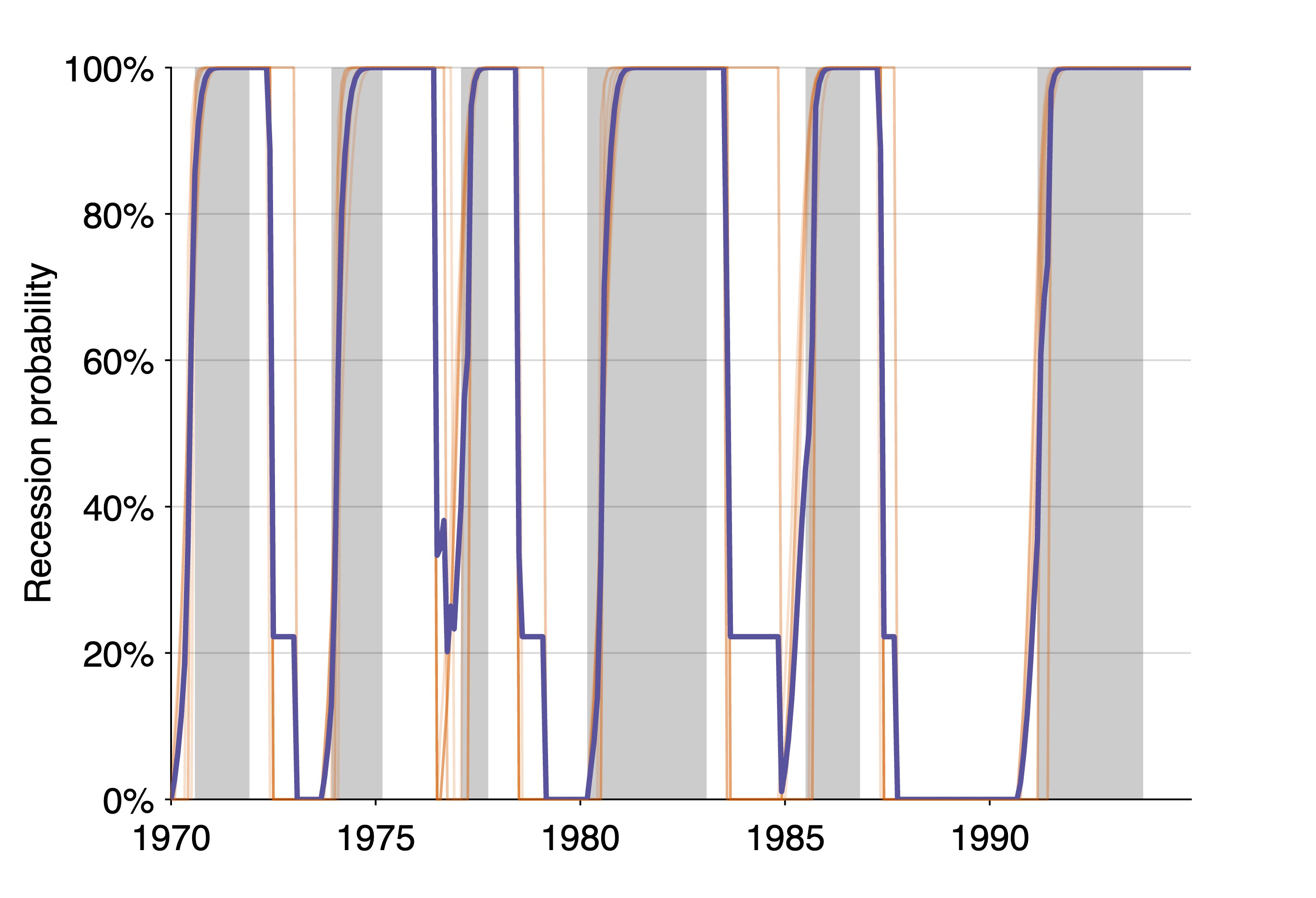}
    \caption{In-sample recession probability, January 1970--December 1994}
    \label{fig:placeholder-17}
\end{figure}

\bibliographystyle{paper}

\end{document}